# Transport anomalies and quantum criticality in electron-doped cuprate superconductors

Xu Zhang[1], Heshan Yu[1], Ge He[1], Wei Hu[1], Jie Yuan[1], Beiyi Zhu[1], Kui Jin[1,2]*

[1] Beijing National Laboratory for Condensed Matter Physics, Institute of Physics, Chinese Academy of Sciences, Beijing 100190, China
[2] Collaborative Innovation Center of Quantum Matter, Beijing, 100190, China

**Abstract**

Superconductivity research is like running a marathon. Three decades after the discovery of high-$T_c$ cuprates, there have been mass data generated from transport measurements, which bring fruitful information. In this review, we give a brief summary of the intriguing phenomena reported in electron-doped cuprates from the aspect of electrical transport as well as the complementary thermal transport. We attempt to sort out common features of the electron-doped family, e.g. the strange metal, negative magnetoresistance, multiple sign reversals of Hall in mixed state, abnormal Nernst signal, complex quantum criticality. Most of them have been challenging the existing theories, nevertheless, a unified diagram certainly helps to approach the nature of electron-doped cuprates.

# 1. Introduction

In last several decades, the developments in advanced scientific instruments have brought great convenience to condensed matter physics. One paradigm is probing the electronic states and electronic structures of strongly correlated electron systems. Remarkably in high-$T_c$ superconductors, tools such as scanning tunneling microscope (STM) [1] and angle-resolved photoemission spectroscopy (ARPES) [2] have been exhibiting the power to discern complex density states and topology of Fermi surface. Nevertheless as an utmost used method, transport probe is unique for discovering new materials and novel properties, as well as a necessary complement to advanced probes in unraveling electron correlations, phase diagrams and so on. For instance, a panoply of discoveries, such as superconductivity [3], Kondo effect [4], integer and fractional quantum Hall effects [5, 6] and giant magnetoresistance effect [7, 8] were first witnessed by transport measurements.

Since the discovery of first superconductor, i.e. the element mercury in 1911 [3], the milestones of searching for new materials in this field leastwise include the heavy fermion superconductor $CeCu_2Si_2$ in 1978 [9], the organic superconductor $(TMTSF)_2PF_6$ in 1980 [10], the copper-oxide perovskite superconductor (cuprates) $La_{2-x}Ba_xCuO_4$ in 1986 [11], the iron-based superconductor LaOFeP in 2006 [12, 13]. The cuprates keeping the record of $T_c$ at ambient pressure (~134 K) have been of greatest concern to the superconductivity community. For the cuprates, there is a common feature in crystal structures, that is, the copper-oxygen blocks separated by charge reservoir blocks which donate charge carriers to the $CuO_2$ planes. Nominally, the cuprate superconductors can be categorized into types of hole doping and electron doping according to the sign of doped carriers. Soon after the discovery of hole-doped $La_{2-x}Ba_xCuO_4$, the first electron-doped $Nd_{2-x}Ce_xCuO_4$ was reported in 1989 [14, 15].

The distinction between these "214-type" $La_{2-x}Ba_xCuO_4$ and $Nd_{2-x}Ce_xCuO_4$ is the apical oxygen, where one copper atom and six oxygen atoms form a $CuO_6$ octahedron in the former but only a Cu-O plane in the latter as shown in Fig. 1. For convenience, the community abbreviates the hole- and electron-doped 214 types as T and T', respectively. There are only two branches in electron-doped family: the aforementioned T' superconductor (point group $D_{4h}^{17}$, space group *I4/mmm*) and infinite-layer superconductor (point group $D_{4h}^{1}$, space group *P4/mmm*). Owing to a limited number of electron-doped cuprates and their complicated synthesis procedures compared to the hole-doped ones, heretofore, researches were addressed mostly on the hole-doped family and rarely on electron-doped counterparts. However, it is undoubtedly that exploring the nature of electron-doped cuprates is indispensable for approaching the mechanism of high-$T_c$ superconductors.

Not expected to recall the whole achievements on electron-doped cuprates in last 27 years, instead this short review centers on intriguing transport anomalies and quantum criticality. To provide a profile of electron-doped cuprates from the aspect of transport, we select the following topics, i.e. electrical transport anomalies (Section 2), two-band feature in both normal and mixed states (Section 3), the complementary thermal transport behavior (Section 4), and quantum phenomena in extreme conditions (Section 5). One can refer to other nice reviews published recently for an overall view on structures, properties and applications [16-18].

## 2. Electrical transport anomalies

A characteristic of all superconductors is zero electrical resistance below the critical superconducting transition temperature ($T_c$) and fully expulsion of magnetic field known as Meissner effect. For type-I superconductors, transition width of *R*(*T*) curve, i.e. the temperature from normal state to Meissner state, is typical of 0.1 K or less. For type-II layered cuprate superconductors (high-$T_c$ cuprates), the transition is usually broadened by an order of magnitude, due to Kosterlitz-Thouless transition where vortex pairs with opposite sign unbind with lifting up the temperature. When applying magnetic field, there is a mixed state located between the normal state and the Meissner state. In this state, vortices with normal core coexist with the superconducting area. Consequently, the resistance behavior becomes more complicated, since both intrinsic properties of the vortex and pinning effects play roles in fruitful vortex states [19]. From the aspect of electrical transport, once entering the mixed state rich phenomena can be observed in Hall signal (reviewed in Section 3), compared to the rare from resistance signal. However, a numbers of well-known anomalies were first uncovered from the resistance measurements in the normal state when tuning chemical doping, defects, temperature, magnetic field, and so on. Fig. 2 exhibits a typical Hall-bar configuration to measure voltages of both Hall (**V** // $y$, **I** // $x$, **B** // $z$) and resistance (**V** // **I** // $x$, **B** // $z$).

In this section, we hash over resistance anomalies in electron-doped cuprates, e.g. low temperature metal-insulator transitions, linear-in-temperature resistivity (the 'strange metal' behavior), negative magnetoresistance, anisotropic in-plane angular dependent magnetoresistance (AMR), and linear-in-field magnetoresistance. Although these intriguing phenomena are present in the normal state, their underlying physics is crucial to the understanding of high-$T_c$ superconductivity.

**2.1. Metal-insulator transitions**

Metal-insulator transitions (MITs) mean huge change in resistivity, by even tens of orders of magnitude, which have been widely observed in correlated electron systems [20]. On the basis of different driving forces, the MITs are sorted into several types and named after a few memorable physicists like Wilson, Peierls, Mott, and Anderson. In this sense, unveiling the nature of MITs has profound influence on condensed matters. In electron-doped cuprates, MITs have been inevitably observed by tuning chemical doping [21-25], sample annealing process (adjusting oxygen concentration in the samples) [26-29], magnetic field [30] and disorder [31-34]. Acquainted with the MITs in electron-doped cuprates, we first look through two key elements, i.e. crossover from metallic- to insulating behavior by tuning temperature and superconductor -insulator transitions by tuning nonthermal parameters.

1) **Crossover from metallic- to insulating-behavior.** In $Ln_{2-x}Ce_xCuO_{4\pm\delta}$ (Ln = Nd, Pr, La…), the slightly Ce-doped or heavily oxygen-off-stoichiometric samples show insulating (or semiconducting) behavior with the residual resistivity in the range from $m\Omega\cdot cm$ to $\Omega\cdot cm$. In contrast, the optimally- or over-doped sample has a residual resistivity of tens of $\mu\Omega\cdot cm$. Most of the time, the $R(T)$ curve displays a crossover from metallic behavior (higher $T$) to insulating-like behavior (lower $T$) as seen in Fig. 3. In this case, the ground state is not exactly an insulating (or semiconducting) state, whereas literature still prefers to use MIT (we will not stick to this issue in the following part). The origin of crossover from metallic- to insulating-behavior, (i.e. upturn of resistivity) is still in debate, which may be subject to two-dimensional (2D) weak localization [35, 36], Kondo-like scattering [37], additional scattering by magnetic droplets trapped at impurity sites [38, 39], or a link to antiferromagnetism [40].

2) **Superconductor-insulator transitions (SITs).** For an electron-doped cuprate superconductor in the underdoped region or in the condition far from oxygen optimization, the upturn of resistivity usually happens at temperature $T_{up}$ above $T_c$ ('upturn' is frequently used in the community, which emphasizes the violation of metallic behavior at low temperature). The $T_{up}$ will be gradually suppressed as a function of doping [41], usually coming across the superconducting transition temperature at the optimal doping level and terminating at a slightly overdoping. After the superconductivity is killed by applying magnetic field, the upturn underneath the superconducting dome can be seen as shown in Fig. 4.

In early 90's, Tanda *et al.* reported a SIT in $Nd_{2-x}Ce_xCuO_{4\pm\delta}$ thin films by tuning magnetic fields [26, 27]. They found that the sheet resistance $R_\square$ (=$\rho_{ab}/d$) at the SIT was close to the critical value $h/(2e)^2$ (= 6.45 kΩ *per* $CuO_2$ plane), suggesting a Bose-insulator state before entering into the Fermi insulator(Fig. 5). Here, $\rho_{ab}$ is the residual resistivity and $d$ is the distance between adjacent $CuO_2$ planes. The Bose-insulator state is a quantum phenomenon, where Cooper pairs are localized in 2D superconductors and rendered immobile by disorder. In field-tuned SITs, the resistivity should satisfy a scaling theory given by Fisher [35],

$$\rho(B,T) = \frac{h}{4e^2} f\left[\frac{c_0(B-B_c)}{T^{1/(\nu z)}}\right], \tag{1}$$

where $f$ is a dimensionless scaling function, $c_0$ is a non-universal constant, $B_c$ is the critical magnetic field characterizing the SIT, $\nu$ and $z$ are the correlation length critical exponent and the dynamical critical exponent, respectively.

In $Nd_{2-x}Ce_xCuO_{4\pm\delta}$ thin films, Tanda *et al.* got $\nu z$ = 1.2. Very recently, Bollinger *et al.* [42] reported a SIT at the pair quantum resistance $h/(2e)^2$ and $\nu z$ = 1.5 in ultrathin $La_{2-x}Sr_xCuO_4$ films by tuning charge carrier concentration via ionic liquid gating method (electric double layer transistor, abbreviated as EDLT). Leng et al. [43] carried out similar experiments on ultrathin YBa$_2$Cu$_3$O$_{7-x}$ films and found $vz$=2.2 (Fig. 6). In EDLT experiments, the correlation length diverges upon approaching the critical carrier concentration rather than the critical magnetic field, which may result in different $vz$.

Sawa et al. [44] found that in La$_{2-x}$Ce$_x$CuO$_4$ thin film with $x$=0.08, the $R_\square$ is about 32 kΩ, by 5 times larger than the value of $h/(2e)^2$. Jin et al. [30] did field and doping dependent resistance measurements on La$_{2-x}$Ce$_x$CuO$_4$ thin films. They found that in slightly overdoped La$_{2-x}$Ce$_x$CuO$_4$ thin film with $x$ = 0.12, the $R_\square$ is about 1.43 kΩ and $vz$ = 0.75. However, in underdoped La$_{2-x}$Ce$_x$CuO$_4$ with $x$ = 0.09, the $R_\square$ is found to be temperature dependent. That is, the isothermal $R(B)$ curves do not cross at a fixed point (see Section 2.3). Recently, Zeng et al. studied the resistance behavior of ultrathin Pr$_{2-x}$Ce$_x$CuO$_4$ films on Pr$_2$CuO$_4$ buffer layer using EDLT device. They arrived at $R_\square$ = 2.88 kΩ and $vz$ = 2.4 [45].

Theoretically, different values of $vz$ signify different universality classes, e.g. 7/3 in quantum percolation model [46], 4/3 in classic percolation model [47]. Certainly, the application of quantum scaling theory can reveal underlying physics of SITs which confirms that these values of the critical exponent are intrinsic. Nevertheless, the non-universal critical sheet resistance requires more careful work on issues like sample quality, finite temperature influence and Griffiths effects [48].

Now it is clear that once superconductivity is stripped away, the MITs can be observed with doping, magnetic field, electric field and disorder/oxygen. Next we will turn to physics behind the metallic state, the upturn, and the magnetoresistance.

## 2.2. Temperature dependence of resistivity in metallic state

In ordinary metals, the Landau Fermi liquid theory can well describe low temperature dependence of resistivity, which obeys $\rho \sim T^2$ [49]. At high temperature, resistance mainly comes from electron-phonon scattering, which results in $\rho \sim T$ at $T > \Theta_D$ and $\rho \sim T^5$ at $T < \Theta_D$, where $\Theta_D$ is the Debye temperature. At low temperature, the electron-phonon scattering becomes weak and electron-electron scattering starts to dominate the transport.

Restricted to the Pauli exclusion principle, two scattered electrons should go to unoccupied states in a range of ~ $k_B T$ to the Fermi level, in that the resistivity follows a $T^2$ relationship.

1) **The strange metal**. In cuprate superconductors, the temperature dependence of resistance in metal regime is very intriguing. In 1987, Gurvitch and Fiory found that the resistivity of optimally doped $YBa_2Cu_3O_{7-x}$ and $La_{2-x}Sr_xCuO_4$ is surprisingly linear in temperature, i.e. $\rho \sim T$, which can be held from tens of Kelvin just above $T_c$ up to hundreds of Kelvin [50] as seen in Fig. 7. Thereafter, the linear-in-temperature behavior has been widely observed in organic [51], heavy-fermion [52], cuprates [53, 54] and iron-based superconductors [55], which earned it a widespread reputation, i.e. 'strange metal'.

2) **Violation of MIR limit**. In hole-doped cuprates, the strange is not only the linear-in-$T$ resistivity far below the Debye temperature, but also the unsaturated resistivity up to 1000 K violating the Mott-Ioffe-Regel (MIR) limit around 100-1000 μΩ·cm ($\rho_{MIR} = 3\pi^2 \hbar / e^2 k_F^2 l$) in the framework of Bloch Grüneisen theory, on the basis of the criterion that the mean free path cannot be shorter than the crystals' interatomic spacing [56]. The unsaturation of resistivity up to 1000 K was also observed in electron-doped $Nd_{2-x}Ce_xCuO_4$ and $Pr_{2-x}Ce_xCuO_4$ [57].

3) **Crossover from Fermi liquid to strange metal**. Unlike the hole-doped cuprates in which the linear-in-$T$ resistivity persists from right above $T_c$ to hundreds of Kelvin, a nearly $T^2$ dependence of $\rho_{ab}$ is reported in $Nd_{2-x}Ce_xCuO_4$ with $x \geq 0.13$ [21]. Similar behavior has been observed in $La_{2-x}Ce_xCuO_4$ and a 2D Fermi liquid theory was employed to fit $\rho(T)$ of $x$ = 0.10 -0.20 as well as Co-doped samples [58, 59]. For slightly overdoped $Pr_{2-x}Ce_xCuO_4$ with $x$=0.17, Fournier *et al.* observed that the linearity could persist from 40 mK to 10 K, then there is a crossover from $T$ to $T^2$ near 40 K as seen in Fig. 8(a) and (b) [60]. On the contrary, the underdoped $HgBa_2CuO_{4+\delta}$ shows a linear-resistivity regime from 400 K to 280 K but Fermi liquid behavior from 170 K to 91 K as shown in Fig. 8(c) and (d) [61]. Hussey *et al.* [62] claimed that the normal state transport of overdoped $La_{2-x}Sr_xCuO_4$ actually contained two regimes in which the electrical resistivity varies approximately linearly with temperature. Therefore, the one at higher $T$ should correspond to the regime from 400 K to 280 K in $HgBa_2CuO_{4+\delta}$, and the other one at low $T$ matches the regime from 40 mK to 10 K in $Pr_{2-x}Ce_xCuO_4$.

4) **Relation between strange metal and superconductivity.** Interestingly, in $La_{2-x}Ce_xCuO_4$ with $x$ from 0.11 to 0.17, there is a regime where the linear resistivity persists down to 20 mK once the superconductivity is suppressed. The Fermi liquid behavior is recovered in non-superconducting samples at $x > 0.19$ [54] (see Fig. 9). The best linearity of $\rho(T)$ can span over three orders of magnitude. Using the formula $\rho(T) = \rho_0 + A_1(x)T$ to fit their data, Jin et al. found that $A_1(x)$ decreased with decreasing doping (x) and displayed a positive correlation with $T_c$. The scaling of $A_1$ with $T_c$ also works for $Pr_{2-x}Ce_xCuO_4$ as shown in Fig. 10, indicating intimate relation between linear resistivity and superconductivity. Such relation has been also confirmed in unconventional superconductors $(TMTSF)_2PF_6$, $YBa_2Cu_3O_{7-x}$, $La_{2-x}Sr_xCuO_4$, $Ba(Fe_{1-x}Co_x)_2As_2$, thus a unifying rule is concluded [51, 63].

5) **The origin of strange metal**. Fournier et al. tried to bridge it over two-band feature of electron-doped cuprates [60]. They assumed the temperature dependence of relaxation times of electron and hole bands as $1/\tau_{el} \sim T^2$ and $1/\tau_{hole} \sim T$, respectively. Since hole carriers dominate the transport at low temperature, then the behavior of holes could be consistent with electron-electron scattering in a 2D disordered metal [64]. Moriya et al. pointed out that the generic linear-in-$T$ resistivity is the typical feature of 2D antiferromagnetism (AFM) quantum critical point (QCP), and the linear-temperature scattering arise from 2D antiferromagnetic spin fluctuations [65]. Rosch considered an AFM QCP in 3D disorder system, where a linear-temperature dependence of resistivity could also be achieved by anisotropic scattering from critical spin fluctuations [66]. Abrahams et al. studied quasi-two-dimensional metals with small-angle elastic scattering and angle-independent inelastic scattering. They suggested that linear temperature resistivity behavior has a relation to the marginal Fermi liquid [67]. Our theoretical colleagues have been pushing forward the phenomenology theory, considering such as a flat band pinned to the Fermi surface [68], Umklapp scattering vertex [69] and higher order of spin-fermion coupling [70]. However, clarifying the micro mechanism of the linear-temperature resistivity down to mK is still a big challenge.

## 2.3. Negative magnetoresistance

Magnetoresistance is the change in electrical resistance of a material when a magnetic field is applied. In conventional metals, the ordinary magnetoresistance is positive and the isotherms subject to the Kohler plot, that is, a plot of $\Delta\rho/\rho_0$ vs. $(B/\rho_0)^2$ should fall on a straight line with a slope that is independent of temperature. Here, $\Delta\rho = \rho(B) - \rho_0$. The underlying picture is that the mean free path becomes shorter in magnetic field due to Lorentz force. In the framework of Boltzmann equation, the magnetoresistance is proportional to $B^2\mu^2$ assuming single type of carriers. The mobility satisfies $\mu \sim \rho_0^{-1}$ in Drude model, so we get $\Delta\rho/\rho_0 \sim (B/\rho_0)^2$.

1) **Negative to positive magnetoresistance.** In electron-doped cuprates, the insulating behavior or the upturn can be suppressed in magnetic field as seen in Fig. 11, which means a negative magnetoresistance (n-MR). With increasing doping the n-MR can turn to positive (p-MR) as seen in Fig. 12(a). The phenomenon of n-MR to p-MR has been also obtained by tuning oxygen/disorder (Fig. 12 (b)) [28, 29] or temperature (Fig. 12(c)) [24, 40].

2) **Crossing points of magnetoresistance isotherms.** As mentioned in Section 2.1, if the critical sheet resistance is temperature independent in the superconductor-insulator transition, the magnetoresistance isotherms will cross at a fixed point and obey the scaling theory. In many cuprate superconductors, the magnetoresistance isotherms have one crossing point. Two things should be pointed out. First, there are two crossing points in La$_{2-x}$Ce$_x$CuO$_4$ thin films with $x$=0.12, the first crossing point occurs before entering the normal state, whereas the second crossing point shows up in the regime of n-MR as seen in Fig. 13 [30]. Second, the magnetoresistance isotherms do not always cross at a fixed critical field, e.g. in underdoped La$_{2-x}$Ce$_x$CuO$_4$ thin films with $x$=0.09 and underdoped Pr$_{2-x}$Ce$_x$CuO$_4$ thin films with $x$=0.12 [38] as shown in Fig. 14.

3) **The origin of negative magnetoresistance.** The n-MR is usually accompanied with the upturn. Tanda *et al.* [26] fitted the n-MR of Nd$_{2-x}$Ce$_x$CuO$_4$ thin films to the 2D weak localization theory. The conductivity obeys the following formula [71].

$$\Delta\sigma(B) = \sigma(B) - \sigma(0) = \frac{-\alpha e^2}{2\pi^2\hbar}\left[\psi\left(\frac{1}{2} + \frac{1}{a\tau}\right) - \psi\left(\frac{1}{2} + \frac{1}{a\tau_\varepsilon}\right) - ln\left(\frac{\tau_\varepsilon}{\tau}\right)\right], \qquad (2)$$

where $\alpha$ is constant, $\tau$ is the relaxation time due to normal impurity scattering, $\tau_\varepsilon$ is the inelastic scattering time, and $a = 4DeB/\hbar$ with $D$ the diffusion coefficient. In this situation, spatially localized states by quantum interference result in a quantum correction to Drude conductivity. The magnetic field destroys the quantum interference and leads to enhanced conductivity, i.e. n-MR. The 2D weak localization also requires a log$T$ dependence of resistivity, which is observed in underdoped $Nd_{2-x}Ce_xCuO_4$ with $x$ = 0.10 [36]. Sekitani et al. [37] carried out electrical transport study on underdoped $La_{2-x}Ce_xCuO_4$, $Pr_{2-x}Ce_xCuO_4$ and $Nd_{2-x}Ce_xCuO_4$ thin films. They found a deviation from log $T$ behavior towards the lowest temperature and attributed the n-MR to suppression of Kondo scattering off $Cu^{2+}$ spins. Dagan *et al.* [40] studied MR of $Pr_{2-x}Ce_xCuO_4$ from $x$ = 0.11 to $x$ = 0.19 and found that the spin-related MR vanished near the boundary of AFM ($x$ = 0.16). Therefore, they linked the n-MR and upturn to AFM correlation. Finkelman *et al.* [38] found the spin-related MR was linear in field, inconsistent with the Kondo scattering which gives a log $B$ dependence. They favors the picture of antiferromagnetic magnetic droplets [39]. Recently, Naito group [72] got superconductivity in parent compounds, and the upturn could be suppressed after a two-step 'protect annealing'. Since upturn and n-MR are twinborn, clarifying what happened in different annealing processes will be very instructive.

## 2.4. Anisotropic in-plane angular dependent magnetoresistance

Probing the in-plane AMR is another widely used method to unveil broken symmetry and phase boundary in unconventional superconductors, since anisotropic scattering processes can be manifested as order forms. For instance, fourfold AMR has been commonly observed in electron-doped cuprates [73-76], whereas twofold AMR mostly appears in hole-doped cuprates [77, 78], iron-based superconductors [79], as well as the spinel oxides superconductor [80].

Lavrov *et al.* [73] reported a fourfold AMR in highly underdoped, antiferromagnetic $Pr_{1.29}La_{0.7}Ce_{0.01}CuO_4$ crystals. They found that the anisotropy was caused by the anisotropic spin-flop field. In this system, the Cu spins are arranged in a non-collinear configuration (Fig. 15). It is easier to flip the non-collinear structure to a collinear structure with field along the Cu-Cu direction than that along the Cu-O-Cu direction. Such fourfold AMR has also been observed in $Nd_{2-x}Ce_xCuO_4$ [75, 76] and $Pr_{2-x}Ce_xCuO_4$ [74, 81] (Fig. 16(a)). In $Pr_{2-x}Ce_xCuO_4$, the temperature at which the fourfold AMR vanishes seems consistent with the static AFM ordering temperature. However, Jin et al. [24] found a twofold AMR in electron-doped $La_{2-x}Ce_xCuO_4$ thin films as shown in Fig. 16(b). The onset temperature of twofold symmetry tracks the AFM correlations [82, 83]. Jovanovic *et al.* [25] also found a twofold symmetry in infinite-layer $Sr_{1-x}La_xCuO_2$ thin films, following the explanation used in $La_{2-x}Ce_xCuO_4$. Besides, the twofold AMR has also been observed in $YBa_2Cu_3O_{7-x}$ [77], $La_{2-x}Sr_xCuO_4$ [78], $LiTi_2O_4$ [80] and $BaFe_{2-x}Co_xAs_2$ [79].

The hole-doped cuprates have a collinear spin structure, that may be the reason why the symmetry of AMR is twofold rather than fourfold. For electron-doped $La_{2-x}Ce_xCuO_4$ and $Sr_{1-x}La_xCuO_2$, since only films are of high quality, information on magnetic structure is absent. To clarify this issue, we need more details on these two systems.

## 2.5. Linear-in-field magnetoresistance

Linear magnetoresistance is first reported in non-magnetic $Ag_2Te$ [84]. The pristine sample exhibits negligible magnetoresistance, whereas slightly doping leads to a linear positive magnetoresistance. Successively, the linear-in-field magnetoresistance has been widely seen in high-$T_c$ cuprates [24, 85, 86], Graphene [87], topological insulators [88], Dirac and Weyl semi-metals [89, 90].

In electron doped cuprates, Sckitani *et al.* [85] reported a negative linear magnetoresistance in $Nd_{2-x}Ce_xCuO_4$ thin films. Finkelman *et al.* [38] found the negative spin-related MR was linear in field in underdoped $Pr_{2-x}Ce_xCuO_4$ thin films with $x$ = 0.12. A linear negative magnetoresistance in $La_{2-x}Sr_xCuO_4$ is also argued to be a spin source [91]. Jin *et al.* [24] also found the negative linear magnetoresistance in underdoped $La_{2-x}Ce_xCuO_4$ thin films with $x$ = 0.06 (Fig. 12(a)). Interestingly, it will become positive at $x$ = 0.10. Li et al. also [92] found a positive linear MR in $Pr_{2-x}Ce_xCuO_4$ but with the field normal to the $CuO_2$ plane (Fig. 17).

Theoretically, there exist both classic and quantum approaches to a linear positive magnetoresistance. The classic one is based on the importance of phase inhomogeneities. Herring [93] obtained a linear positive magnetoresistance by numerical calculations on an 'impedance network'. Guttal and Stroud [94] extended it to 2D disordered semiconducting film and reproduced the linear positive magnetoresistance. Bulgadaev and Kusmartsev deduced explicit expressions for magnetoresistance of strongly inhomogeneous planar and layered systems, and also obtained large linear magnetoresistance [95].

In the quantum approach, Abrikosov [96] proposed a model on the basis of the assumption of a gapless spectrum with a linear momentum dependence (the limiting quantum case with electrons only in one Landau band). In this case, $\rho = N_i H / \pi n^2 ec$, where $N_i$ is the density of scattering centers. Fenton *et al.* [97] suggested that linear magnetoresistance could be observed at a simple density-wave QCP where the Fermi surface is reconstructed and shows a local radius of curvature, i.e. cusp. Consequently, the magnetotransport is dominated by a fraction of quasiparitcles (~ $ev_F B\tau$) deflected around the cusp, leading to a nonanalytic response of linear magnetoresistivity. The origin of positive/ negative linear magnetoresistance in electron doped cuprates is still not confirmed. The negative linear MR seems to be a common behavior in underdoped samples. It is worthy of checking whether the positive linear MR is an accident event or not.

## 3. Two band phenomena

MgB$_2$, the $T_c$ record holder among conventional superconductors at ambient pressure, is a multiband superconductor [98]. The hole-doped cuprates YBa$_2$Cu$_3$O$_y$ and YBa$_2$Cu$_4$O$_8$ contain two types of charge carriers in underdoped regime, which has been verified from the Hall coefficient ($R_H$) and Seebeck coefficient [99-101]. Almost all the iron based superconductors are known to be multiband superconductors, possibly except the one unit cell FeSe thin film [102-105]. Therefore, it turns out that multiband feature is essential to achieving a high-$T_c$.

The electron-doped cuprates, not unexpectedly, also belong to the multiband family. Hitherto, the powerful ARPES has observed the coexistence of electron- and hole- Fermi surfaces in Nd$_{2-x}$Ce$_x$CuO$_4$, Pr$_{2-x}$Ce$_x$CuO$_4$, Pr$_{1-x}$LaCe$_x$CuO$_4$, Sm$_{1-x}$Ce$_x$CuO$_4$, and Sr$_{1-x}$La$_x$CuO$_2$ near the optimal doping [106-111]. As a function of Ce doping, these electron-doped cuprates arrive at a unified picture, i.e. as the doping increases electron pockets first come across the Fermi level near ($\pi$, 0) and (0, $\pi$) in the momentum space, then a hole pocket emerges at ($\pi/2$, $\pi/2$) near the optimal doping, and finally a large hole FS forms. Perhaps not coincidentally, the ARPES study on $Pr_{1.3-x}La_{0.7}Ce_xCuO_4$ showed a similar evolution of FS with removing oxygen via annealing process[109] (Fig. 18).

In this section, we will go over the two band feature and its impact on the normal state, the mixed state, and the correlation to superconductivity on the basis of transport studies.

### 3.1. Two band feature in the normal state

Soon after the discovery of electron-doped $Nd_{2-x}Ce_xCuO_{4\pm\delta}$, Jiang *et al*. [28] found that the Hall coefficient in optimal doped $Nd_{2-x}Ce_xCuO_{4\pm\delta}$ ($x$=0.15) changed from negative to positive with removing the oxygen content. Combined with the thermoelectric transport measurements, they attributed such phenomenon to the coexistence of electron and hole carriers, aforementioned, ARPES studies on $Nd_{2-x}Ce_xCuO_4$ later confirmed this speculation [106, 107]. Similar behavior of the Hall coefficient was also observed in series of oxygen tuned $Pr_{2-x}Ce_xCuO_{4\pm\delta}$ with $x$ = 0.17 [112]. Interestingly, Ce substitution gave a quite similar Hall behavior as seen in $Pr_{2-x}Ce_xCuO_4$ and $La_{2-x}Ce_xCuO_4$ [113, 114] (Fig. 19). Therefore, it seems once again that oxygen and doping (Ce) play roughly the same role in the evolution of band structure in the normal state.

Great efforts have been made to understand the origin of the band evolution. Dagan *et al.* measured the doping dependent Hall coefficient ($R_H$) down to 350 mK in $Pr_{2-x}Ce_xCuO_4$ [113] and found a 'kink' in the $R_H$ near a critical concentration, $x_c$ ~ 0.165, which happens to be the doping where the electron and hole pockets merge together as revealed by ARPES, slightly higher than the optimal doping level $x$ = 0.15 for this system. This critical doping was also notified on the same system by other transport measurements such as the spin-related magnetoresistance [40], the AMR [81], Nernst [115, 116], thermopower [117], as well as spectrum probes like tunneling [118] and infrared [119]. Assuming that a commensurate ($\pi$, $\pi$) spin density wave (SDW) order occurs for $x < x_c$, Lin and Millis were able to capture the 'kink' with t-t'-t''-J model [120]. In this picture, when $x > x_c$ a large hole FS centered at (π, π), but once passing the critical point, the SDW (or AFM) steps in. Consequently, a magnetic unit cell equals to two lattice unit cells in the real space, and the magnetic Brillouin zone will be reduced by a half in the momentum space. Then the large hole FS will be cut by the boundary of magnetic Brillouin zone and open folding gap at the cutting points (i.e. hotspots). Therefore, the 'kink' is regarded as a result of FS reconstruction by the SDW or AFM. Since driven by a nonthermal quantity, the transition to AFM is a quantum phenomenon. As mentioned in Section 2, a plausible explanation for the strange metal behavior is based on the AFM quantum criticality [65]. Yet this interpretation has been commonly adopted, there are still drawbacks in that solely considering the role of *J* (i.e. the AFM exchange coupling) is not enough to describe all the experimental details. In the framework of t-t'-t''-*U* model with *U* the on-site Coulomb repulsion and density wave gap contained in the dispersion, Kusko et al. [121] and Tremblay's group [122] were able to reproduce the ARPES results by taking the self-consistent renormalization and the dynamical mean-field theory calculations, respectively. Instead of choosing an adjustable Mott gap, Xiang *et al*. [123] considered an effective t-*U'*-J model where the effective *U'* represents the Coulomb repulsion between O 2*p* and Cu 3*d* electrons. The essential difference among these models is how to treat the contribution of oxygen 2*p* orbitals.

It is not easy to distinguish between the AFM and the Coulomb repulsion that which one is more important to the two band feature. Nevertheless, as passing the critical point, the scenario of FS reconstruction should result in anti-correlation between the concentration of hole and electron carriers, i.e., one decreases as the other increases, whereas in Xiang's model the interplay between Cu 3*d* and O 2*p* bands can give a positive correlation between the two type carriers. Obviously, the physics behind two band feature is awaiting more reliable experimental results.

### 3.2. Manifestation of two bands in mixed state

Now we move to the mixed state. Once entering the mixed state, rich phenomena come out in Hall signal [124]. Among them, the most intriguing one is the sign reversal with temperature or magnetic field. One-time sign reversal was observed in samples such as Nb films[125], $\alpha$-Mo$_3$Si films [126], YBa$_2$Cu$_3$O$_{7-\delta}$ single crystals [127], YBa$_2$Cu$_3$O$_y$/PrBa$_2$Cu$_3$O$_y$ superlattices [128], and Nd$_{1.85}$Ce$_{0.15}$CuO$_4$ single crystals [129]. A double sign reversal was found in highly anisotropic cuprates, such as Bi$_2$Sr$_2$CaCu$_2$O$_x$ [130], Tl$_2$Ba$_2$CaCu$_2$O$_8$ [131, 132] and HgBa$_2$CaCu$_2$O$_{6+\delta}$ [133]. Besides, in twinned YBa$_2$Cu$_3$O$_{7-\delta}$ thin films, Göb et al. reported a double sign reversal with the applied magnetic fields parallel to the crystallographic c axis and to the twin boundaries [134].

In the mixed state, the Hall conductivity can be expressed as $\sigma = \sigma_n + \sigma_f$, where $\sigma_n$ originates from the normal carriers in the vortex cores, $\sigma_f$ comes from the transverse motion of the vortices according to Faraday's law $E = -\frac{v_L \times H}{c}$ [135]. Since $\sigma_n$ always has the same sign as that in the normal state, $\sigma_f$ is the key point to investigate the anomalous Hall effect, e.g. the sign reversal. When the vortices move anti-parallel to the supercurrent, the sign of $\sigma_f$ and $\sigma_n$ should be opposite and results in Hall anomaly. Related to this transverse motion, various models have been proposed.

The early work to understand flux flow is based on the standard Bardeen-Stephen (BS) model [136]. In traditional BS model, the intrinsic transverse motion of vortices is always in the same direction with the superfluid flow. Therefore, it requires extrinsic factors, such as pinning force [137, 138], thermal fluctuation [139], and vortex-vortex interaction [140], to give an anti-parallel vortex motion to the superfluid flow. However this unusual motion has never been observed in any other fluid and cannot be explained in the framework of classical hydrodynamic theory.

On the basis of time-dependent Ginzburg-Landau equation, the intrinsic force exerted on a single vortex has been reinvestigated from a micro perspective by some groups [141-145]. They argued that the anomalous Hall effect can be intrinsic, relying on the electronic structure of the normal state. However, the vortex motion is unavoidably influenced by the extrinsic factors mentioned above, so the difficulty is how to extract the intrinsic information.

Besides, there is also a model employing two bands to explain the Hall anomaly [146]. The Hall anomaly is naturally attributed to the change of predominant type of charge carrier from the normal state to vortex state, while the theoretical work is based on the BS model. At the early stage, few multiband superconductors had been recognized but the Hall anomaly seemed general for superconductors. Hence, two-band feature had not been widely considered.

For electron-doped cuprates, the study on $Nd_{1.85}Ce_{0.15}CuO_{4-y}$ by Hagen *et al.* [147] supports that the Hall anomaly originates from the intrinsic motion of vortex. In their work, they compared different systems and found that the value of $l/\xi_0$ was very important to the appearance of sign reversal. Here, $l$ is the length of mean free path and $\xi_0$ is the BCS coherence length. Such finding stimulates a series of theoretical studies closely related to that quantity, $l/\xi_0$ [142, 144, 145, 148].

Charikova *et al.* reconsidered the two-band model to describe the Hall anomaly in $Nd_{2-x}Ce_xCuO_4$ [149]. To explain their data at doping with $x$ = 0.14 and 0.15, the authors assumed that the electron and hole bands dominated the transport in the normal state and in mixed state, respectively, i.e. the two types of carriers have different pairing strengths.

Actually, a weakly coupled two-gap model has been proposed to explain the unusual temperature dependence of superfluid density $\rho_s(T)$ in electron-doped cuprates [150] (Fig. 20). The model requires different pairing strengths of electrons and holes in electron-doped cuprates, which is also used to ascribe the feature in Raman scattering on $Nd_{2-x}Ce_xCuO_4$ and $Pr_{2-x}Ce_xCuO_4$ [151].

However, the observation of one-time sign reversal cannot pin down the manifestation of two-band feature. Recently, a double sign reversal has been observed in the mixed state of $Pr_{1.85}Ce_{0.15}CuO_4$ (Fig. 21), and the Hall anomaly can be tuned by the EDLT method. Compared with traditional chemical substitutions, the tuning of carrier concentration by electrostatic doping will not bring more disorder or pinning centers into the system [152]. Thus, such double sign reversal urges the consideration of two band feature in mixed state [153].

### 3.3. Multiband superconductivity

The blooming multiband feature to superconductivity deserves careful study. The electron-doped cuprate superconductor has a common two band feature and a relatively small upper critical field $H_{c2}$ (~10 T), thus it would be a good candidate. After the birth of the iron based superconductor, the multiband superconductivity becomes flourishing [102, 104, 154-158]. Before going to the details of $H_{c2}$ in multiband superconductors, we first stop by the issue of determining $H_{c2}$ from the transport measurements.

In conventional superconductors and some iron based superconductors, the magnetoresistance is negligible. So the most convenient method is to pick up critical fields at 90%, 50% and 10% percentages of normal-state resistance ($\rho_n$) of the magnetoresistance isotherms [102, 154]. However, the magnetoresistance isotherms in electron-doped cuprates are complex, e.g. the crossing point at SIT, the negative or positive unsaturated MR. One has to define the $\rho_n$ for each isotherm, the error bar is big and the value is not so reliable to do analysis [159]. The above method is thus not applicable in electron-doped ones. A scaling of the fluctuation conductivity $\sigma_{flu}$ ($H$, $T$) has been used to extract $H_{c2}(T)$ [160-163]. In this method, the $\sigma_{flu}$ was obtained by subtracting the extrapolated normal state conductivity from the total conductivity. However, this method also suffers the anomalies such as the upturn.

Balci *et al.* [164] used Nernst signal to determine the $H_{c2}(T)$ in $Pr_{2-x}Ce_xCuO_4$. They discerned a valley-like behavior in the isotherms (Fig. 22) so the minimum is defined as $H_{c2}$(T). As we will discuss in Section 5, this method relies on the remarkable two-band Nernst signal, which overcomes the 'long-tail' influence from fluctuations. By coincidence, Jin *et al.* [114] extracted $H_{c2}(T)$ from the derivative of magnetoresistance isotherms. They differentiated the magnetoresistance isotherm of $La_{2-x}Ce_xCuO_4$ (i.e. $\rho^{'}(H)$ = d$\rho$/d$H$), and found that the peak of $\rho^{'}(H)$ first moved to low field with increasing temperature, and then moved up once the superconductivity is destroyed. This behavior implies the competitive contributions between vortex motion and the two-type carriers. The advantage of these two methods is to use an explicit criterion to pin down the normal state resistance, reducing the uncertainty to a bearable degree.

In electron-doped $La_{2-x}Ce_xCuO_4$ and $Pr_{2-x}Ce_xCuO_4$, the $H_{c2}(T)$ from the differential method exhibits an unusual upward feature (Fig. 23(a)), mimicking the behavior of superfluid density [150, 165], which signifies a multiband superconductivity. The upward curvature has also been widely observed in iron based superconductors (Fig. 23(b)). On the basis of the multiband BCS model, the $H_{c2}$ of a two-gap superconductor in the dirty limit is derived by Gurevich [166], which can account for the upward curvature [102, 154-156].

## 4. Thermal transport properties

For cuprates, thermal transport is complementary and indispensable to the electrical transport in clarifying such as the multiband feature [28, 29], superconducting fluctuations [167, 168], and phase transitions [117, 169]. The thermal transport signals, Nernst and thermopower [170, 171], can in some sense be regarded as thermally driven Hall signal and resistivity, respectively. As shown in Fig. 24, when a steady temperature gradient $\nabla_x T$ is applied to a material, the thermopower, i.e. the Seebeck coefficient, is defined as $S = -\frac{E_x}{\nabla_x T}$, and in presence of a perpendicular magnetic field $H_z$, the Nernst signal can be extracted from the transverse electric field $E_y$, as $N = \frac{E_y}{\nabla_x T}$.

In superconductors, the Nernst signal is contributed by mobile charge carriers and superconducting fluctuations [172, 173]. Referring to the mobile carriers, $N$ is generally small in ordinary metals with a single carrier type due to the Sondheimer cancellation [174], whereas it can be large in multiband metals, e.g. the electron-doped cuprates [115, 116]. In mixed state, Nernst signal in cuprates is greatly enhanced [167, 175] compared to the organic [176, 177] and heavy fermion systems [178, 179], signifying strong superconducting fluctuations. The Seebeck signal can leastwise provide information on evolution of carriers and phase transitions due to its high sensitivity to the topology of Fermi surface [101, 117]. Nevertheless, the thermal transport has been suffering challenges of high-precision signal collection and data analysis. In this section, we will skim over the abnormal Nernst signal in the normal state, superconducting fluctuations, and the Fermi surface reconstruction under survey by thermopower in electron-doped cuprates.

### 4.1. Abnormal Nernst signal in the normal state

In semi-classic transport theory [180], the charge current density $J_e$, the electrical conductivity tensor $\bar{\sigma}$, and the thermoelectric (Peltier) tensor $\bar{\alpha}$ satisfy

$$\boldsymbol{J_e} = \bar{\sigma}\boldsymbol{E} - \bar{\alpha}|\nabla T|. \tag{3}$$

The steady state yields $\boldsymbol{J_e}$ = 0, therefore neglecting small temperature gradient along the transverse direction, the Nernst signal can be written as

$$N = \frac{\alpha_{xy}\sigma_{xx} - \alpha_{xx}\sigma_{xy}}{\sigma_{xx}^2 + \sigma_{xy}^2}. \tag{4}$$

When $\sigma_{xy} \ll \sigma_{xx}$, the above Eq. (4) is further simplified as

$$N = \frac{\alpha_{xy}}{\sigma_{xx}} - S\tan\theta_H = S(\tan\theta_T - \tan\theta_H). \tag{5}$$

Here, $S = \frac{\alpha_{xx}}{\sigma_{xx}}$. $\tan\theta_T$ and $\tan\theta_H$ are thermal and electric Hall angles, respectively. From two-dimensional system like cuprates, $\alpha_{ij} = -\frac{\pi^2 k_B^2 T}{3e}\frac{\partial \sigma_{ij}}{\partial \epsilon}\Big|_{\epsilon=E_F}$, then the Nernst signal is

$$N = -\frac{\pi^2 k_B^2 T}{3e}\frac{\partial \tan\theta_H}{\partial \epsilon}\Big|_{\epsilon=E_F}. \tag{6}$$

If the Hall angle is only weakly dependent on energy in the vicinity of the Fermi energy, then the Nernst signal is negligible in systems where only one type of charge carriers dominate the transport such as in hole-doped $Tl_2Ba_2CaCuO_8$ and $La_{2-x}Sr_xCuO_4$ [181, 182], as well as in the slightly underdoped and heavily overdoped regimes of electron-doped cuprates (e.g. tens of nV/K). In other words, a single metal gives

$$\alpha_{xy}\sigma_{xx} = \alpha_{xx}\sigma_{xy}. \tag{7}$$

For a two band system, the Eq. (4) should be rewritten as

$$N = \frac{(\alpha_{xy}^h + \alpha_{xy}^e)(\sigma_{xx}^h + \sigma_{xx}^e) - (\alpha_{xx}^h + \alpha_{xx}^e)(\sigma_{xy}^h + \sigma_{xy}^e)}{(\sigma_{xx}^h + \sigma_{xx}^e)^2 + (\sigma_{xy}^h + \sigma_{xy}^e)^2}. \tag{8}$$

The superscripts $h$ and $e$ stand for hole and electron, respectively. Since $\alpha_{xx}^h$ and $\alpha_{xx}^e$ are expected to have different signs, Eq.(7) implies the same signs of $\alpha_{xy}^h$ and $\alpha_{xy}^e$ [183]. Simply for a compensated system, i.e. the case of electron-doped cuprates near the optimal doping, the first term of Eq. (8) is a non-zero value but the second term is zero for $\sigma_{xy}^h = -\sigma_{xy}^e$. Therefore, Nernst signal is obviously enhanced in a two-band system compared to one-band system, by one or two orders of magnitude.

Fournier *et al.* [29] discovered a distinct Nernst signal in $Nd_{2-x}Ce_xCuO_4$ thin films near the optimal doping. Li et al. [115] found that the Nernst signal of optimally doped $Pr_{2-x}Ce_xCuO_4$ was several times larger than the under- and over-doped samples as seen in Fig 25(a). The optimally doped $La_{2-x}Ce_xCuO_4$ also shows a large *N* of the same order of magnitude (i.e. several μV/K in Fig 25(b)).

In addition, based on the two-band theory magnetoresistance can be written as $\frac{\Delta\rho_{xx}}{\rho_0} = \frac{(\sigma_{xx}^h R_H^h - \sigma_{xx}^e R_H^e)^2 \sigma_{xx}^h \sigma_{xx}^e B^2}{(\sigma_{xx}^h + \sigma_{xx}^e)^2}$ for compensated metals. The Nernst signal in Eq. (8) is rewritten as

$$N = \frac{N^h \sigma_{xx}^h + N^e \sigma_{xx}^e}{\sigma_{xx}^h + \sigma_{xx}^e} + \frac{\sigma_{xx}^h \sigma_{xx}^e (\sigma_{xx}^h R_H^h - \sigma_{xx}^e R_H^e)(S^h - S^e) B}{(\sigma_{xx}^h + \sigma_{xx}^e)^2} \quad . \tag{9}$$

Here, $N^i$ and $S^i$ are Nernst signal and thermopower for the *i* band (*i*= h, e), respectively. The factor $(\sigma_{xx}^h R_H^h - \sigma_{xx}^e R_H^e)$ can be found in both formulas, which indicates that a maximum of the magnetoresistance is likely to coincide with a maximum of the Nernst coefficient. Note that $S^e < 0$, so $(S^h - S^e)$ is always positive. This speculation has been validated in $Nd_{2-x}Ce_xCuO_4$ and $Pr_{2-x}Ce_xCuO_4$ [29, 115], once again pointing to the two-band feature.

## 4.2. Superconducting fluctuations

In hole-doped cuprates, a large Nernst signal has been observed in an extended region above $T_c$ [167, 175]. As mentioned above, the Nernst signal in the normal state of hole-doped cuprates is small because of the single type carriers, except for the case of Fermi surface reconstruction [101, 169]. Therefore, such abnormal signal persisting far beyond $T_c$ has been attracting considerable attention and suffering hot debate on its origin, i.e., phase fluctuations vs. amplitude fluctuations. Superconducting order parameter is comprised of phase $e^{i\theta}$ and amplitude $|\Psi|$. Fluctuating either one can get the Nernst signal enhanced.

1) **Phase fluctuations.** The superconducting phase fluctuation scenario is stimulated by the theoretical model of Emery and Kivelson [172]. In conventional superconductors, the superfluid density is pretty large so that electron pairing and long-range-order phase coherence occur simultaneously. In cuprates superconductors, owing to a small superfluid density, the long-range phase coherence is destroyed above $T_c$ whereas the local Cooper pairing amplitude remains sizable. In underdoped region, $T_c$ is decided by the phase coherence temperature $T_\theta^{max}$, which is proportional to the superfluid density over the effective electron mass, whereas in overdoped side the phase coherence becomes stronger so $T_c$ is the onset temperature of Cooper pairing, following the mean-field transition temperature $T^{MF}$ predicted by BCS-Elishberg theory as shown in Fig. 26. These two characteristic temperatures shape $T_c$ to be a dome, and thus there is an extended regime of phase fluctuations in underdoped region. Empirically, Uemura et al. [184] had concluded such relation between the $T_c$ and the superfluid density based on the μSR experimental results on a series of hole-doped cuprates, i.e. $T_c \propto \sigma(T\rightarrow 0) \propto 1/\lambda^2 \propto n_s/m^*$ holds up to optimal doping but $T_c$ is suppressed with further increasing carrier doping. Here, $n_s$ is the superconducting carrier density. In the mixed state of type-II superconductors, the large Nernst signal is due to the motion of vortices [185]. Consequently, the extended regime of large Nernst signal was attributed to short-lived vortex excitations above $T_c$ [186, 187].

2) **Amplitude fluctuations**. Alternatively, the superconducting amplitude fluctuation scenario lies upon the Aslamazov-Larkin (AL) theory [188], where the fluctuations are limited by the coherence length of Cooper pairs. Ussishkin *et al.* [173] calculated thermoelectric transport based on the Gaussian amplitude fluctuations, and found that this AL-type fluctuations were responsible for the optimally doped and overdoped samples in $La_{2-x}Sr_xCuO_4$ system [189]. In this picture, the lifetime of Cooper pairs diffusing toward the cold end of the sample is longer than those to the hot end, so the thermal gradient gives rise to a net drift of Cooper pairs towards the cold end, and then a Nernst signal is generated by the perpendicular magnetic field. Pourret *et al.* [190] showed the evidence that the larger Nernst signal above $T_c$ came from the superconducting amplitude fluctuations in amorphous films of $Nb_xSi_{1-x}$.

In electron-doped cuprates, the superconducting fluctuations are not so strong compared to the hole-doped ones. Li *et al.* [115] found that in $Pr_{2-x}Ce_xCuO_4$ the onset temperature of notable vortex Nernst signal was slightly higher than $T_c$, i.e. by less than 4 K. While, there are two peaks in the temperature dependence of the Nernst signal, which are associated with evolution of two-band feature by AFM in the normal state and the vortex motion in mixed state, respectively. Moreover, the overdoped samples with $x = 0.17$ still have discernable peak in the normal state which seems inconsistent with the picture of a large full Fermi surface for the ARPES. Similar two-peak feature is also found in $La_{2-x}Sr_xCuO_4$, where the one in the normal state is linked to the stripe order [169] as shown in Fig. 27.

Tafti *et al* [116] carried out similar Nernst experiments on $Pr_{2-x}Ce_xCuO_4$, and identified that the superconducting Nernst signal from underdoped (*x* = 0.13) to overdoped (*x* = 0.17) was quantitatively consistent with theory of Guassian fluctuations in a dirty 2D superconductor by Ussishkin *et al.* [173].

Before concluding this subsection, we would like to point out two things. First, the Guassian fluctuations cannot fully account for the large Nernst signal in underdoped $La_{2-x}Sr_xCuO_4$ [173], where the physics of pseudogap inevitably get involved in the contention [191, 192]. Secondly, so far our understanding of normal-state large Nernst signal relies on a lot of assumptions from Boltzmann transport theory; obviously, it is oversimplified for the correlated systems, even not suitable for a system with anisotropic scattering.

### 4.3. Functions of thermopower

In Boltzmann theory, $S = -\frac{\pi^2 k_B^2 T}{3e} \frac{\partial ln\sigma}{\partial \epsilon}\Big|_{\epsilon_F}$ [193]. In zero-temperature limit, $\sigma$ is proportional to energy in the vicinity of the Fermi energy [194]. Therefore, we can simplify the expression in case of free electron gas,

$$S = -\frac{\pi^2 k_B^2 T}{3e} \frac{1}{\epsilon_F}. \qquad (10)$$

From the above equation, we have $S/T \propto E_F^{-1} \propto k_F^{-2} \propto n^{-1} \propto R_H$ for a two dimensional system, linking the Seebeck coefficient to the Hall coefficient.

Li. *et al* found that when the superconductivity is killed by magnetic field, the doping dependence of *S/T* at 2 K followed the behavior of $R_H(x)$ in $Pr_{2-x}Ce_xCuO_4$ (Fig. 28). The kink in Hall coefficient implies a quantum critical doping at *x* = 0.16 as discussed in Section 3.1. In the same sense, the Seebeck signal can provide useful information on the Fermi surface reconstruction. The dramatic change in temperature dependence of *S/T* has been also used to catch the onset temperature of stripe order in hole-doped cuprates [101, 195].

In addition, by thermopower measurements, Jiang *et al.* [196] reported that an orbital effect led to a large magneto-thermopower due to the anisotropic scattering; Xu *et al.* [197] studied the extra oxygen introduced impurity scattering without changing the carrier density in $Nd_{2-x}Ce_xCuO_4$ films; Budhani *et al.* [198] investigated the weak localization on the Cu-O planes in combination with the electrical transport.

## 5. Quantum phenomena in extreme conditions

Although superconductivity itself is a macroscopic quantum phenomenon, approaching the nature of unconventional superconductivity, e.g. in heavy fermion, cuprates and pnictides, relies upon the understanding of its concomitant phenomena characterized by quantum fluctuations and criticality, which are prominent in the extreme conditions, such as ultralow temperature down to millikelvin and strong magnetic field up to hundred Tesla. In previous sections, some of these phenomena have been insinuated about the electron-doped cuprates, e.g. the linear-in-*T* resistance persists down to 40 mK [60], the 'kink' behavior in doping dependence of Hall coefficient at 350 mK [113], the magnetic-field induced SIT occurring at the critical sheet resistance $h/(2e)^2$ [27]. In this section, we will overlook quantum oscillations, quantum phase transitions and controversy over QCPs in the electron-doped cuprates.

### 5.1. Quantum oscillations

In the semi-classical theory [199], quantum oscillations are caused by the Landau quantization of energy levels, which is considered as a signature of Fermi liquid behavior. When the magnetic field increases, the density of states has a discontinuous change as the Landau levels pass over the closed Fermi surface one after one.

The oscillations of transport quantities, i.e. Shubnikov-de Haas effect, can provide following information. First, the cross-section area, $A_F$, of Fermi surface normal to the applied magnetic field can be calculated from the oscillation frequency *f* through the Onsager relation $f = \frac{\Phi_0}{2\pi^2} A_F$, where $\Phi_0 = 2.07 \times 10^{-15}$ T·m² is the flux quantum. Second, for a quasi-two dimensional Fermi surface like in the cuprates, the oscillating component of the magnetoresistance is described as

$$\rho_{osc} \propto B^{1/2} R_T R_D \sin(2\pi f/B + \gamma) \ , \tag{11}$$

where $R_T = \frac{2\pi^2 k_B T/\hbar\omega_c}{\sinh(2\pi^2 k_B T/\hbar\omega_c)}$ is the thermal damping factor, $R_D = e^{-\pi/(\omega_c \tau_D)}$ is the Dingle factor, and $\gamma$ is the Onsager phase. The effective mass $m^* = \frac{eB}{\omega_c}$ and the mean free path $l_D \sim \hbar(A_F/\pi)^{1/2} \tau_D/m_c$ can be calculated from the temperature and scattering damping factors $R_T$ and $R_D$, respectively.

The quantum oscillations in cuprates were first observed from the c-axis transport study on underdoped $YBa_2Cu_3O_{6.5}$ in 2007 with $f$ the order of magnitude of $10^2$ Tesla [200] (Fig. 29(a)). Subsequently, quite a few experiments verified the oscillations from various measurements such as the magnetization (i.e. de Haas-van Alphen) [201, 202], the thermopower [195], specific heat [203], and thermal conductivity [204] of $YBa_2Cu_3O_{6.5}$, as well as the in-plane magnetoresistance of $HgBa_2CuO_{4+\delta}$ [205]. The oscillations were also observed from the c-axis transport and magnetic torque in overdoped $Tl_2Ba_2CuO_{6+\delta}$ with $f$ the order of magnitude of $10^4$ Tesla [206].

As expected, the quantum oscillations were soon reported by Helm *et al.* [207] in 2009, from the c-axis transport in electron-doped $Nd_{2-x}Ce_xCuO_4$ with $x$ = 0.15, 0.16, and 0.17, where the $f$ changes from ~ 300 to $10^4$ Tesla with increasing doping. As shown in Fig.30(c), there is a slow oscillation frequency probed in $x$ = 0.15 and $x$ = 0.16, whereas a fast one observed in $x$ = 0.17. Since the frequency of quantum oscillations yields the cross-section area of Fermi surface normal to the applied magnetic field (*H* // c-axis), the huge change in frequency thus signifies the Fermi surface reconstruction between $x$ = 0.16 and $x$ = 0.17. Recently, the in-plane transport on superconducting $Pr_2CuO_{4-\delta}$ also showed oscillations above 60 Tesla, with $f$ ~ 300 Tesla [208].

The above experiments convey very important information: 1) closed Fermi surface existing in the certain underdoped regime, whether it is induced by magnetic field or not, is under debate for hole-doped cuprates [205]; 2) Fermi surface reconstruction occurring with increasing doping from underdoped to overdoped in both hole-doped (Fig. 29) and electron-doped cuprates (Fig.

30), consistent with the ARPES results; 3) a comparable Fermi surface between the optimally doped and the new superconducting parent samples in electron-doped cuprates.

## 5.2. Quantum phase transitions

We have mentioned that in $Pr_{2-x}Ce_xCuO_4$ thin films, a critical doping at $x \sim 0.165$ has been verified by different transport measurements, e.g. Hall coefficient [113], spin-related magnetoresistance [40], AMR [81], Nernst [115, 116], and thermopower [117], as well as the spectrum probes like tunneling [118] and infrared [119]. The aforementioned quantum oscillations in electron-doped $Nd_{2-x}Ce_xCuO_4$ single crystals point to the same critical doping between $x = 0.16$ and $x = 0.17$, also in coincidence with the ARPES results [107]. As the Ce dopants increase, this critical point in $Pr_{2-x}Ce_xCuO_4$ and $Nd_{2-x}Ce_xCuO_4$ has been commonly accepted as a quantum phase transition from the antiferromagnetism to the Fermi liquid at zero temperature [16].

As shown in Fig. 31, a continuous quantum phase transition undergoes two different ground states at zero temperature by tuning nonthermal parameter like doping, magnetic field, or pressure [209]. Consequently, there is a 'fan-shaped' quantum critical regime above the QCP at finite temperature, where the quantum fluctuations remain dominant. Since the correlations at a QCP are characterized by scale invariance in space and time, quantum critical scaling functions can be used to describe the divergence upon approaching the critical boundary [52]. In Section 2.1, we have introduced the quantum critical scaling function by Fisher [35], which is used to describe the superconductor-insulator quantum phase transition.

Butch *et al.* [210] reported quantum critical scaling plots of $\Delta\rho/(A_2T^2)$ vs. $f(\Delta B^\gamma/T)$ at the edge of Fermi liquid state in electron-doped $La_{2-x}Ce_xCuO_4$. In Fig. 32(a), a single power-law exponent ($n$ < 2) can describe the resistivity behavior in the quantum critical regime, i.e. $\rho \sim T^n$ in the non-Fermi liquid region. Here, the quasiparticle-quasiparticle scattering coefficient $A_2$ can be achieved by fitting the Fermi liquid region with $\rho = \rho_0 + A_2T^2$. They deduced a simple relation, $\gamma = \alpha\,(2-n)$, among the scaling exponent $\gamma$, the power-law exponent $n$, and the critical exponent α obtained from the divergence of $A_2$ as the critical field is approached from the Fermi liquid region. The critical exponent α is constant for different doping as seen in Fig. 32(b). This relation reflects that the competition between two energy scales, i.e. by magnetic field and temperature, drives the quantum disordered state (Fermi liquid) to the quantum critical region (non-Fermi liquid). In order to reach the quantum critical region, smaller magnetic field is needed to overcome the weaker thermal fluctuations as $T \rightarrow 0$.

Surprisingly, they found that for $La_{2-x}Ce_xCuO_4$ with $x$ = 0.15, the scaling exponent $γ$ = 0.4 since the power-law exponent $n$ = 1 (like the strange metal). While for $x$ = 0.17, $γ$ = 1 since $n$ = 1.6. Different values of scaling exponent imply different types of quantum fluctuations of the ordered state. That is, the linear-in-$T$ resistance is linked to the antiferromagnetic fluctuations [54]. However, the origin of quantum fluctuations for $n$ = 1.6, which is also observed above the Fermi liquid regime in $La_{2-x}Sr_xCuO_4$ [86, 211] remains to be clarified in future.

Besides, quantum scaling functions of $ω/T$ are commonly used to describe the spectra function in the quantum criticality region, e.g. describing the quantum critical behavior in hole-doped $Bi_2Sr_2Ca_{0.92}Y_{0.08}Cu_2O_{8+δ}$ by scaling the optical spectra [212], verifying the continuous antiferromagnetic phase transition in Ce-doped $Nd_{2-x}Ce_xCuO_4$ and oxygen-doped $Pr_{0.88}LaCe_{0.12}CuO_4$ by scaling the inelastic neutron scattering spectra [213].

Obviously, although quantum phase transition occurs at zero temperature, the quantum scaling functions at finite temperature can be used to verify the QCPs in cuprates. However, the scaling from quantum disordered state does not tell us what the ordered state is. For instance, we don't know which ground state is responsible for n = 1.6 power law [54]. The strange metal in different unconventional systems has been attributed to different origins by different theoretical models [214].

## 5.3. Controversy over quantum critical points

There is much controversy over QCPs: the number of QCPs, the accurate locations of these QCPs, and the origin of the QCPs. For the hole-doped cuprates in Fig. 33(a), there are multiple critical points. However, owing to the composite competing orders, not all of them have been verified as QCPs.

In electron-doped cuprates as seen in Fig. 33(b), there seems to be at least two QCPs. One is at the edge of the Fermi liquid state, which has been verified in $La_{2-x}Ce_xCuO_4$ [210], as well as claimed in $Nd_{2-x}Ce_xCuO_4$ [215]. The origin of this QCP is still unclear. Another truncates the superconducting dome near the optimal doping such as in $Pr_{2-x}Ce_xCuO_4$, $Nd_{2-x}Ce_xCuO_4$ and $La_{2-x}Ce_xCuO_4$, where the Fermi surface reconstruction happens. However, the origin of the Fermi surface reconstruction is still under debate, yet much transport evidence points to the antiferromagnetic order.

In $Nd_{2-x}Ce_xCuO_{4\pm\delta}$, Yamada *et al.* [216] reported that the transition between the AFM and superconductivity was first order and the AFM QCP does not exist, also supported by few experimental results [217, 218]. However, Motoyama *et al.* [219] reported that the long range AFM order terminated at $x \sim 0.13$, whereas the superconductivity appeared beyond this doping. Mang *et al.* [220] proposed that the non-superconducting $Nd_{2-x}Ce_xCuO_{4\pm\delta}$ might display a ground state with 2D antiferromagnetic order.

Similar controversy also exists in $Pr_{1-x}LaCe_xCuO_{4\pm\delta}$. Wilson *et al.* [221] reported that the high-energy spin and charge excitations could be observed in $x = 0.12$. Furthermore, Ishii *et al.* [222] probed them up to the highest doping level of superconductivity. Fujita *et al.* [223] reported that there exist low-energy spin fluctuations over doping level of superconductivity. Besides, by annealing the $Pr_{0.88}LaCe_{0.12}CuO_{4-\delta}$ samples, the long-ranged antiferromagnetic order vanishes when the superconductivity appears [213].

Consequently, the neutron scattering measurements provide quite conflicting information on the boundary of AFM. Alternatively, the aforementioned transport measurements arrive at a roughly consistent QCP, i.e., $x \sim 0.16$ in both $Pr_{2-x}Ce_xCuO_4$ and $Nd_{2-x}Ce_xCuO_4$, in agreement with the results of ARPES and infrared optical measurements.

For the unique $La_{2-x}Ce_xCuO_{4\pm\delta}$ with optimal doping at $x = 0.10$, the controversy exists as well. As shown in Fig. 34, the $\mu$SR probe [224] revealed that the long-range antiferromagnetic order vanishes at $x \sim 0.08$. However, the angular magnetoresistance [24] and the low-temperature

Hall resistance [114] reported a magnetic QCP locates at $x \sim 0.14$. Very recently, Yu et al. [82] built up a multidimensional phase diagram of $La_{2-x}Ce_xCuO_{4\pm\delta}$ as a function of Ce, oxygen and the magnetic field. These new results revealed that in $La_{2-x}Ce_xCuO_4$ the long-rang AFM vanishes at $x_{c1} \sim 0.08$, whereas 2D AFM correlations can persist up to a QCP, $x_{FS} \sim 0.14$. Besides, the upturn of resistivity signifies the formation of 3D AFM, which becomes prominent once the superconductivity is stripped away. Undoubtedly, the quantum criticality plays a significant role in approaching the nature of the superconductivity in electron-doped cuprates.

## 6. Concluding remarks

The transport anomalies and quantum criticality in electron-doped cuprates have been briefly summarized. By seeking the correlations among various transport phenomena, a general phase diagram has been sketched out to manifest the common features, such as two-band structure, superconducting fluctuations and quantum criticality. In this way, a profile of the intrinsic electron structure and its evolution gradually emerges out of the intricate phenomena, yet some of them like the Nernst signal in mixed state and the positive linear magnetoresisitance are still lack of explicit description. In order to stride forward the nature of high-$T_c$ superconductivity, it is essential to reveal more details about the electronic states as a function of different tuning parameters, i.e. urging a multidimensional phase diagram. Being versatile and flexible, transport probes are easy to integrate with these new techniques. Some advanced techniques, such as the electric double-layer transistors (EDLTs) [225] and combinatorial syntheses [226], have been applied to tune carrier density and chemical composition in films, respectively. Therefore, there is plenty room for the transport to catch the essence of high-$T_c$ superconductors.

*Finally, the transport anomalies and quantum criticality in electron-doped cuprate superconductors are summarized in a form of phase diagram as seen in* Fig. 35.

**Acknowledgments**

The corresponding author would like to give special thanks to R.L. Greene for his guidance and fruitful discussions. The authors would like to take this opportunity to thank all the collaborators with whom the researches on electron-doped cuprates have been conducted, including J. Paglione, F. V. Kusmartsev, T. Xiang, Y.F. Yang, Y. Dagan, R.F. Kiefl, P. Abbamonte, J. Qi, J.F. Wang, L. Li, J. Lian, X. Zhang, P. Bach, N.P. Butch, K. Kirshenbaum, Y. Jiang, I. Takeuchi, S. Smadici, J. Vanacken, F. Herlach, V.V. Moshchalkov, B. Leridon, L. Zhao, H. Wu, B.R. Zhao, H.B. Wang, and T. Hatano. This work was supported by the National Key Basic Research Program of China (Grant No. 2015CB921000), the National Natural Science Foundation of China (Grant No. 11474338), the Open Research Foundation of Wuhan National High Magnetic Field Center (Grant No. PHMFF2015008), and the Strategic Priority Research Program (B) of the Chinese Academy of Sciences (Grant No. XDB07020100) .

# Reference

[1] Ø. Fischer, M. Kugler, I. Maggio-Aprile, C. Berthod, C. Renner, Scanning tunneling spectroscopy of high-temperature superconductors, Rev. Mod. Phys. 79 (2007) 353.

[2] A. Damascelli, Z. Hussain, Z.-X. Shen, Angle-resolved photoemission studies of the cuprate superconductors, Rev. Mod. Phys. 75 (2003) 473.

[3] H.K. Onnes, The superconductivity of mercury, Commun. Phys. Lab. Univ. Leiden 122 (1911).

[4] J. Kondo, Resistance minimum in dilute magnetic alloys, Prog. Theor. Phys. 32 (1964) 37.

[5] K. Vonklitzing, G. Dorda, M. Pepper, New method for high-accuracy determination of the fine-structure constant based on quantized Hall resistance, Phys. Rev. Lett. 45 (1980) 494.

[6] D.C. Tsui, H.L. Stormer, A.C. Gossard, Two-dimensional magnetotransport in the extreme quantum limit, Phys. Rev. Lett. 48 (1982) 1559.

[7] M.N. Baibich, J.M. Broto, A. Fert, F. Nguyen Van Dau, F. Petroff, P. Etienne, G. Creuzet, A. Friederich, J. Chazelas, Giant magnetoresistance of (001)Fe/(001)Cr magnetic superlattices, Phys. Rev. Lett. 61 (1988) 2472.

[8] G. Binasch, P. Grünberg, F. Saurenbach, W. Zinn, Enhanced magnetoresistance in layered magnetic structures with antiferromagnetic interlayer exchange, Phys. Rev. B 39 (1989) 4828.

[9] F. Steglich, J. Aarts, C.D. Bredl, W. Lieke, D. Meschede, W. Franz, H. Schäfer, Superconductivity in the presence of strong Pauli paramagnetism: $CeCu_2Si_2$, Phys. Rev. Lett. 43 (1979) 1892.

[10] D. Jérome, A. Mazaud, M. Ribault, K. Bechgaard, Superconductivity in a synthetic organic conductor $(TMTSF)_2PF_6$, J. Phys. Lett. 41 (1980) 95.

[11] J.G. Bednorz, K.A. Müller, Possible high $T_c$ superconductivity in the Ba-La-Cu-O system, Z.Phys.B 64 (1986) 189.

[12] Y. Kamihara, H. Hiramatsu, M. Hirano, R. Kawamura, H. Yanagi, T. Kamiya, H. Hosono, Iron-Based Layered Superconductor: LaOFeP, J. Am. Chem. Soc. 128 (2006) 10012.

[13] Y. Kamihara, T. Watanabe, M. Hirano, H. Hosono, Iron-Based Layered Superconductor La[$O_{1-x}F_x$]FeAs ($x$ = 0.05–0.12) with $T_c$ = 26 K, J. Am. Chem. Soc. 130 (2008) 3296.

[14] H. Takagi, S. Uchida, Y. Tokura, Superconductivity produced by electron doping in $CuO_2$-layered compounds, Phys. Rev. Lett. 62 (1989) 1197.

[15] Y. Tokura, H. Takagi, S. Uchida, A superconducting copper oxide compound with electrons as the charge carriers, Nature 337 (1989) 345.

[16] N.P. Armitage, P. Fournier, R.L. Greene, Progress and perspectives on electron-doped cuprates, Rev. Mod. Phys. 82 (2010) 2421.

[17] J. Yuan, G. He, H. Yang, Y.J. Shi, B.Y. Zhu, K. Jin, Research trends in electron-doped cuprate superconductors, Sci. China-Phys. Mech. Astron. 58 (2015) 11.

[18] P. Fournier, T' and infinite-layer electron-doped cuprates, Phys. C Supercond. 514 (2015) 314.

[19] M. Thinkham, Introduction to superconductivity, 2th ed., Dover Publications Inc, 2004.

[20] M. Imada, A. Fujimori, Y. Tokura, Metal-insulator transitions, Rev. Mod. Phys. 70 (1998) 1039.

[21] S.J. Hagen, J.L. Peng, Z.Y. Li, R.L. Greene, In-plane transport-properties of single-crystal $R_{2-x}Ce_xCuO_{4-y}$ (R = Nd,Sm), Phys. Rev. B 43 (1991) 13606.

[22] F. Gollnik, M. Naito, Doping dependence of normal- and superconducting-state transport properties of $Nd_{2-x}Ce_xCuO_{4\pm y}$ thin films, Phys. Rev. B 58 (1998) 11734.

[23] Y. Onose, Y. Taguchi, K. Ishizaka, Y. Tokura, Charge dynamics in underdoped $Nd_{2-x}Ce_xCuO_4$: Pseudogap and related phenomena, Phys. Rev. B 69 (2004) 024504.

[24] K. Jin, X.H. Zhang, P. Bach, R.L. Greene, Evidence for antiferromagnetic order in $La_{2-x}Ce_xCuO_4$ from angular magnetoresistance measurements, Phys. Rev. B 80 (2009) 012501.

[25] V.P. Jovanović, L. Fruchter, Z.Z. Li, H. Raffy, Anisotropy of the in-plane angular magnetoresistance of electron-doped $Sr_{1-x}La_xCuO_2$ thin films, Phys. Rev. B 81 (2010) 134520.

[26] S. Tanda, M. Honma, T. Nakayama, Critical sheet resistance observed in high-$T_c$ oxide-superconductor $Nd_{2-x}Ce_xCuO_4$ thin-films, Phys. Rev. B 43 (1991) 8725.

[27] S. Tanda, S. Ohzeki, T. Nakayama, Bose-glass-vortex-glass phase-transition and dynamic scaling for high-$T_c$ $Nd_{2-x}Ce_xCuo_4$ thin-films, Phys. Rev. Lett. 69 (1992) 530.

[28] W. Jiang, S.N. Mao, X.X. Xi, X.G. Jiang, J.L. Peng, T. Venkatesan, C.J. Lobb, R.L. Greene, Anomalous Transport Properties in Superconducting $Nd_{1.85}Ce_{0.15}CuO_{4\pm\delta}$, Phys. Rev. Lett. 73 (1994) 1291.

[29] P. Fournier, X. Jiang, W. Jiang, S.N. Mao, T. Venkatesan, C.J. Lobb, R.L. Greene, Thermomagnetic transport properties of $Nd_{1.85}Ce_{0.15}CuO_{4+\delta}$ films: Evidence for two types of charge carriers, Phys. Rev. B 56 (1997) 14149.

[30] K. Jin, B.Y. Zhu, B.X. Wu, J. Vanacken, V.V. Moshchalkov, B. Xu, L.X. Cao, X.G. Qiu, B.R. Zhao, Normal-state transport in electron-doped $La_{2-x}Ce_xCuO_4$ thin films in magnetic fields up to 40 Tesla, Phys. Rev. B 77 (2008) 172503.

[31] S.I. Woods, A.S. Katz, M.C. de Andrade, J. Herrmann, M.B. Maple, R.C. Dynes, Destruction of superconductivity in $Nd_{2-x}Ce_xCuO_{4-\delta}$ thin films by ion irradiation, Phys. Rev. B 58 (1998) 8800.

[32] S.I. Woods, A.S. Katz, S.I. Applebaum, M.C. de Andrade, M.B. Maple, R.C. Dynes, Nature of conduction in disordered $Nd_{2-x}Ce_xCuO_{4-\delta}$ films, Phys. Rev. B 66 (2002) 014538.

[33] K. Jin, J. Yuan, L. Zhao, H. Wu, X.Y. Qi, B.Y. Zhu, L.X. Cao, X.G. Qiu, B. Xu, X.F. Duan, B.R. Zhao, Coexistence of superconductivity and ferromagnetism in a dilute cobalt-doped $La_{1.89}Ce_{0.11}CuO_{4\pm\delta}$ system, Phys. Rev. B 74 (2006) 94518.

[34] B.X. Wu, K. Jin, J. Yuan, H.B. Wang, T. Hatano, B.R. Zhao, B.Y. Zhu, Thickness-induced insufficient oxygen reduction in $La_{2-x}Ce_xCuO_{4\pm\delta}$ thin films, Supercond. Sci. Technol. 22 (2009) 085004.

[35] M.P.A. Fisher, Quantum phase-transitions in disordered 2-dimensional superconductors, Phys. Rev. Lett. 65 (1990) 923.

[36] P. Fournier, J. Higgins, H. Balci, E. Maiser, C.J. Lobb, R.L. Greene, Anomalous saturation of the phase coherence length in underdoped $Pr_{2-x}Ce_xCuO_4$ thin films, Phys. Rev. B 62 (2000) 11993(R).

[37] T. Sekitani, M. Naito, N. Miura, Kondo effect in underdoped n-type superconductors, Phys. Rev. B 67 (2003) 174503.

[38] S. Finkelman, M. Sachs, G. Droulers, N.P. Butch, J. Paglione, P. Bach, R.L. Greene, Y. Dagan, Resistivity at low temperatures in electron-doped cuprate superconductors, Phys. Rev. B 82 (2010) 094508.

[39] W. Chen, B.M. Andersen, P.J. Hirschfeld, Theory of resistivity upturns in metallic cuprates, Phys. Rev. B 80 (2009) 134518.

[40] Y. Dagan, M.C. Barr, W.M. Fisher, R. Beck, T. Dhakal, A. Biswas, R.L. Greene, Origin of the anomalous low temperature upturn in the resistivity of the electron-doped cuprate superconductors, Phys. Rev. Lett. 94 (2005) 057005.

[41] B.X. Wu, K. Jin, J. Yuan, H.B. Wang, T. Hatano, B.R. Zhao, B.Y. Zhu, Preparation of electron-doped $La_{2-x}Ce_xCuO_{4\pm\delta}$ thin films with various Ce doping by dc magnetron sputtering, Phys. C Supercond. 469 (2009) 1945.

[42] A.T. Bollinger, G. Dubuis, J. Yoon, D. Pavuna, J. Misewich, I. Bozovic, Superconductor-insulator transition in $La_{2-x}Sr_xCuO_4$ at the pair quantum resistance, Nature 472 (2011) 458.

[43] X. Leng, J. Garcia-Barriocanal, S. Bose, Y. Lee, A.M. Goldman, Electrostatic control of the evolution from a superconducting phase to an insulating phase in ultrathin $YBa_2Cu_3O_{7-x}$ films, Phys. Rev. Lett. 107 (2011) 027001.

[44] A. Sawa, M. Kawasaki, H. Takagi, Y. Tokura, Electron-doped superconductor $La_{2-x}Ce_xCuO_4$: Preparation of thin films and modified doping range for superconductivity, Phys. Rev. B 66 (2002) 014531.

[45] S.W. Zeng, Z. Huang, W.M. Lv, N.N. Bao, K. Gopinadhan, L.K. Jian, T.S. Herng, Z.Q. Liu, Y.L. Zhao, C.J. Li, H.J.H. Ma, P. Yang, J. Ding, T. Venkatesan, Ariando, Two-dimensional superconductor-insulator quantum phase transitions in an electron-doped cuprate, Phys. Rev. B 92 (2015) 020503(R).

[46] D.-H. Lee, Z. Wang, S. Kivelson, Quantum percolation and plateau transitions in the quantum Hall effect, Phys. Rev. Lett. 70 (1993) 4130.

[47] A. Kapitulnik, N. Mason, S.A. Kivelson, S. Chakravarty, Effects of dissipation on quantum phase transitions, Phys. Rev. B 63 (2001) 125322.

[48] Y. Xing, H.-M. Zhang, H.-L. Fu, H. Liu, Y. Sun, J.-P. Peng, F. Wang, X. Lin, X.-C. Ma, Q.-K. Xue, J. Wang, X.C. Xie, Quantum Griffiths singularity of superconductor-metal transition in Ga thin films, Science 350 (2015) 542.

[49] T. Toyoda, Finite-temperature fermi-liquid theory of electrical conductivity, Phys. Rev. A 39 (1989) 2659.

[50] M. Gurvitch, A.T. Fiory, Resistivity of $La_{1.825}Sr_{0.175}CuO_4$ and $YBa_2Cu_3O_7$ to 1100 K: Absence of saturation and its implications, Phys. Rev. Lett. 59 (1987) 1337.

[51] N. Doiron-Leyraud, P. Auban-Senzier, S.R. de Cotret, C. Bourbonnais, D. Jérome, K. Bechgaard, L. Taillefer, Correlation between linear resistivity and $T_c$ in the Bechgaard salts and the pnictide superconductor $Ba(Fe_{1-x}Co_x)_2As_2$, Phys. Rev. B 80 (2009) 214531.

[52] H. von Löhneysen, A. Rosch, M. Vojta, P. Wolfle, Fermi-liquid instabilities at magnetic quantum phase transitions, Rev. Mod. Phys. 79 (2007) 1015.

[53] R. Daou, N. Doiron-Leyraud, D. LeBoeuf, S.Y. Li, F. Laliberté, O. Cyr-Choinière, Y.J. Jo, L. Balicas, J.-Q. Yan, J.S. Zhou, J.B. Goodenough, L. Taillefer, Linear temperature dependence of resistivity and change in the Fermi surface at the pseudogap critical point of a high-$T_c$ superconductor, Nat. Phys. 5 (2009) 31.

[54] K. Jin, N.P. Butch, K. Kirshenbaum, J. Paglione, R.L. Greene, Link between spin fluctuations and electron pairing in copper oxide superconductors, Nature 476 (2011) 73.

[55] R.H. Liu, G. Wu, T. Wu, D.F. Fang, H. Chen, S.Y. Li, K. Liu, Y.L. Xie, X.F. Wang, R.L. Yang, L. Ding, C. He, D.L. Feng, X.H. Chen, Anomalous transport properties and phase diagram of the FeAs-based $SmFeAsO_{1-x}F_x$ superconductors, Phys. Rev. Lett. 101 (2008) 087001.

[56] N.F. Mott, Conduction in Non-Crystalline Systems IX. Minimum Metallic Conductivity, Philos. Mag. 26 (1972) 1015.

[57] P.L. Bach, S.R. Saha, K. Kirshenbaum, J. Paglione, R.L. Greene, High-temperature resistivity in the iron pnictides and the electron-doped cuprates, Phys. Rev. B 83 (2011) 212506.

[58] H. Wu, L. Zhao, J. Yuan, L.X. Cao, J.P. Zhong, L.J. Gao, B. Xu, P.C. Dai, B.Y. Zhu, X.G. Qiu, B.R. Zhao, Transport properties of electron-doped $La_{2-x}Ce_xCuO_4$ cuprate thin films, Phys. Rev. B 73 (2006) 104512.

[59] K. Jin, L. Zhao, H. Wu, J. Yuan, S.J. Zhu, L.J. Gao, B.Y. Zhu, B. Xu, L.X. Cao, X.G. Qiu, B.R. Zhao, Magnetic cobalt-ion substitution effect in electron-doped $La_{1.89}Ce_{0.11}CuO_4$ superconductor, Phys. C Supercond. 460 (2007) 410.

[60] P. Fournier, P. Mohanty, E. Maiser, S. Darzens, T. Venkatesan, C.J. Lobb, G. Czjzek, R.A. Webb, R.L. Greene, Insulator-metal crossover near optimal doping in $Pr_{2-x}Ce_xCuO_4$: Anomalous normal-state low temperature resistivity, Phys. Rev. Lett. 81 (1998) 4720.

[61] N. Barisic, M.K. Chan, Y. Li, G. Yu, X. Zhao, M. Dressel, A. Smontara, M. Greven, Universal sheet resistance and revised phase diagram of the cuprate high-temperature superconductors, Proc. Natl. Acad. Sci. U. S. A. 110 (2013) 12235.

[62] N.E. Hussey, R.A. Cooper, X. Xu, Y. Wang, I. Mouzopoulou, B. Vignolle, C. Proust, Dichotomy in the $T$-linear resistivity in hole-doped cuprates, Philos. Trans. A Math. Phys. Eng. Sci. 369 (2011) 1626.

[63] L. Taillefer, Scattering and pairing in cuprate superconductors, Annu. Rev. Condens. Matter Phys. 1 (2010) 51.

[64] B.L. Altshuler, A.G. Aronov, D.E. Khmelnitsky, Effects of electron-electron collisions with small energy transfers on quantum localization, J. Phys. C Solid State 15 (1982) 7367.

[65] T. Moriya, K. Ueda, Spin fluctuations and high temperature superconductivity, Adv. Phys. 49 (2000) 555.

[66] A. Rosch, Magnetotransport in nearly antiferromagnetic metals, Phys. Rev. B 62 (2000) 4945.

[67] E. Abrahams, C.M. Varma, Hall effect in the marginal Fermi liquid regime of high-$T_c$ superconductors, Phys. Rev. B 68 (2003) 094502.

[68] V.A. Khodel, J.W. Clark, K.G. Popov, V.R. Shaginyan, Occurrence of flat bands in strongly correlated Fermi systems and high-$T_c$ superconductivity of electron-doped compounds, Jetp Lett. 101 (2015) 413.

[69] M. Shahbazi, C. Bourbonnais, Electrical transport near quantum criticality in low-dimensional organic superconductors, Phys. Rev. B 92 (2015) 195141.

[70] P.S. Weiss, B.N. Narozhny, J. Schmalian, P. Wolfle, Interference of quantum critical excitations and soft diffusive modes in a disordered antiferromagnetic metal, Phys. Rev. B 93 (2016) 045128.

[71] S. Hikami, A.I. Larkin, Y. Nagaoka, Spin-orbit interaction and magnetoresistance in the two dimensional random system, Prog. Theor. Phys. 63 (1980) 707.

[72] O. Matsumoto, A. Utsuki, A. Tsukada, H. Yamamoto, T. Manabe, M. Naito, Superconductivity in undoped T'-$RE_2CuO_4$ with $T_c$ over 30 K, Phys. C Supercond. 468 (2008) 1148.

[73] A.N. Lavrov, H.J. Kang, Y. Kurita, T. Suzuki, S. Komiya, J.W. Lynn, S.-H. Lee, P.C. Dai, Y. Ando, Spin-flop transition and the anisotropic magnetoresistance of $Pr_{1.3-x}La_{0.7}Ce_xCuO_4$: Unexpectedly strong spin-charge coupling in the electron-doped cuprates, Phys. Rev. Lett. 92 (2004) 227003.

[74] P. Fournier, M.-E. Gosselin, S. Savard, J. Renaud, I. Hetel, P. Richard, G. Riou, Fourfold oscillations and anomalous magnetoresistance irreversibility in the nonmetallic regime of $Pr_{1.85}Ce_{0.15}CuO_4$, Phys. Rev. B 69 (2004) 220501(R).

[75] A.I. Ponomarev, L.D. Sabirzyanova, A.A. Ivanov, A.S. Moskvin, Y.D. Panov, Anisotropic low-temperature in-plane magnetoresistance in electron doped $Nd_{2-x}Ce_xCuO_{4+\delta}$, Jetp Lett. 81 (2005) 394.

[76] T. Wu, C.H. Wang, G. Wu, D.F. Fang, J.L. Luo, G.T. Liu, X.H. Chen, Giant anisotropy of the magnetoresistance and the 'spin valve' effect in antiferromagnetic $Nd_{2-x}Ce_xCuO_4$, J. Phys. Condens. Matter 20 (2008) 275226.

[77] Y. Ando, A.N. Lavrov, K. Segawa, Magnetoresistance anomalies in antiferromagnetic $YBa_2Cu_3O_{6+x}$: Fingerprints of charged stripes, Phys. Rev. Lett. 83 (1999) 2813.

[78] Y. Ando, A.N. Lavrov, S. Komiya, Anisotropic magnetoresistance in lightly doped $La_{2-x}Sr_xCuO_4$: Impact of antiphase domain boundaries on the electron transport, Phys. Rev. Lett. 90 (2003) 247003.

[79] X.F. Wang, T. Wu, G. Wu, H. Chen, Y.L. Xie, J.J. Ying, Y.J. Yan, R.H. Liu, X.H. Chen, Anisotropy in the electrical resistivity and susceptibility of superconducting BaFe$_2$As$_2$ single crystals, Phys. Rev. Lett. 102 (2009) 117005.

[80] K. Jin, G. He, X. Zhang, S. Maruyama, S. Yasui, R. Suchoski, J. Shin, Y. Jiang, H.S. Yu, J. Yuan, L. Shan, F.V. Kusmartsev, R.L. Greene, I. Takeuchi, Anomalous magnetoresistance in the spinel superconductor LiTi$_2$O$_4$, Nat. Commun. 6 (2015) 7183.

[81] W. Yu, J.S. Higgins, P. Bach, R.L. Greene, Transport evidence of a magnetic quantum phase transition in electron-doped high-temperature superconductors, Phys. Rev. B 76 (2007) 020503(R).

[82] H. Yu, G. He, Z. Lin, J. Yuan, B. Zhu, Y.-f. Yang, T. Xiang, F.V. Kusmartsev, L. Li, J. Wang, K. Jin, A close look at antiferromagnetsm in multidimensional phase diagram of electron-doped copper oxide, arXiv: 1510.07388 (2015).

[83] I.M. Vishik, F. Mahmood, Z. Alpichshev, N. Gedik, J. Higgins, R.L. Greene, Dynamics of quasiparticles and antiferromagnetic correlations in electron-doped cuprate La$_{2-x}$Ce$_x$CuO$_{4\pm\delta}$, arXiv: 1601.06694 (2016).

[84] R. Xu, A. Husmann, T.F. Rosenbaum, M.L. Saboungi, J.E. Enderby, P.B. Littlewood, Large magnetoresistance in non-magnetic silver chalcogenides, Nature 390 (1997) 57.

[85] T. Sekitani, H. Nakagawa, N. Miura, M. Naito, Negative magneto-resistance of the normal state in Nd$_{2-x}$Ce$_x$CuO$_4$ below $T_c$ and the effect of high magnetic fields, Physica B 294 (2001) 358.

[86] R.A. Cooper, Y. Wang, B. Vignolle, O.J. Lipscombe, S.M. Hayden, Y. Tanabe, T. Adachi, Y. Koike, M. Nohara, H. Takagi, C. Proust, N.E. Hussey, Anomalous Criticality in the Electrical Resistivity of La$_{2-x}$Sr$_x$CuO$_4$, Science 323 (2009) 603.

[87] A.L. Friedman, J.L. Tedesco, P.M. Campbell, J.C. Culbertson, E. Aifer, F.K. Perkins, R.L. Myers-Ward, J.K. Hite, C.R. Eddy, G.G. Jernigan, D.K. Gaskill, Quantum linear magnetoresistance in multi layer epitaxial graphene, Nano Lett. 10 (2010) 3962.

[88] Z.H. Wang, L. Yang, X.J. Li, X.T. Zhao, H.L. Wang, Z.D. Zhang, X.P.A. Gao, Granularity Controlled Nonsaturating Linear Magnetoresistance in Topological Insulator Bi$_2$Te$_3$ Films, Nano Lett. 14 (2014) 6510.

[89] J.Y. Feng, Y. Pang, D.S. Wu, Z.J. Wang, H.M. Weng, J.Q. Li, X. Dai, Z. Fang, Y.G. Shi, L. Lu, Large linear magnetoresistance in Dirac semimetal Cd$_3$As$_2$ with Fermi surfaces close to the Dirac points, Phys. Rev. B 92 (2015) 081306(R)

[90] Y.F. Zhao, H.W. Liu, J.Q. Yan, W. An, J. Liu, X. Zhang, H.C. Wang, Y. Liu, H. Jiang, Q. Li, Y. Wang, X.Z. Li, D. Mandrus, X.C. Xie, M.H. Pan, J. Wang, Anisotropic magnetotransport and exotic longitudinal linear magnetoresistance in WTe$_2$ crystals, Phys. Rev. B 92 (2015) 041104(R).

[91] M.Z. Cieplak, A. Malinowski, S. Guha, M. Berkowski, Localization and interaction effects in strongly underdoped La$_{2-x}$Sr$_x$CuO$_4$, Phys. Rev. Lett. 92 (2004) 187003.

[92] P. Li, F.F. Balakirev, R.L. Greene, High-field Hall resistivity and magnetoresistance of electron-doped Pr$_{2-x}$Ce$_x$CuO$_{4-\delta}$ Phys. Rev. Lett. 99 (2007) 047003.

[93] C. Herring, Effect of random inhomogeneities on electrical and galvanomagnetic measurements, J. Appl. Phys. 31 (1960) 1939.

[94] V. Guttal, D. Stroud, Model for a macroscopically disordered conductor with an exactly linear high-field magnetoresistance, Phys. Rev. B 71 (2005) 201304(R).

[95] S.A. Bulgadaev, F. Kusmartsev, Large linear magnetoresistivity in strongly inhomogeneous planar and layered systems, Phys. Lett. A 342 (2005) 188.

[96] A.A. Abrikosov, Quantum magnetoresistance, Phys. Rev. B 58 (1998) 2788.

[97] J. Fenton, A.J. Schofield, Breakdown of weak-field magnetotransport at a metallic quantum critical point, Phys. Rev. Lett. 95 (2005) 247201.

[98] A.Y. Liu, I.I. Mazin, J. Kortus, Beyond Eliashberg superconductivity in $MgB_2$: anharmonicity, two-phonon scattering, and multiple gaps, Phys. Rev. Lett. 87 (2001) 087005.

[99] D. LeBoeuf, N. Doiron-Leyraud, J. Levallois, R. Daou, J.-B. Bonnemaison, N.E. Hussey, L. Balicas, B.J. Ramshaw, R. Liang, D.A. Bonn, W.N. Hardy, S. Adachi, C. Proust, L. Taillefer, Electron pockets in the Fermi surface of hole-doped high-$T_c$ superconductors, Nature 450 (2007) 533.

[100] P.M.C. Rourke, A.F. Bangura, C. Proust, J. Levallois, N. Doiron-Leyraud, D. LeBoeuf, L. Taillefer, S. Adachi, M.L. Sutherland, N.E. Hussey, Fermi-surface reconstruction and two-carrier model for the Hall effect in $YBa_2Cu_4O_8$, Phys. Rev. B 82 (2010) 020514(R).

[101] J. Chang, R. Daou, C. Proust, D. LeBoeuf, N. Doiron-Leyraud, F. Laliberté, B. Pingault, B.J. Ramshaw, R.X. Liang, D.A. Bonn, W.N. Hardy, H. Takagi, A.B. Antunes, I. Sheikin, K. Behnia, L. Taillefer, Nernst and Seebeck Coefficients of the Cuprate Superconductor $YBa_2Cu_3O_{6.67}$: A Study of Fermi Surface Reconstruction, Phys. Rev. Lett. 104 (2010) 057005.

[102] F. Hunte, J. Jaroszynski, A. Gurevich, D.C. Larbalestier, R. Jin, A.S. Sefat, M.A. McGuire, B.C. Sales, D.K. Christen, D. Mandrus, Two-band superconductivity in $LaFeAsO_{0.89}F_{0.11}$ at very high magnetic fields, Nature 453 (2008) 903.

[103] D.J. Singh, M.-H. Du, Density functional study of $LaFeAsO_{1-x}F_x$: a low carrier density superconductor near itinerant magnetism, Phys. Rev. Lett. 100 (2008) 237003.

[104] J. Jaroszynski, S. Riggs, F. Hunte, A. Gurevich, D. Larbalestier, G. Boebinger, F. Balakirev, A. Migliori, Z. Ren, W. Lu, J. Yang, X. Shen, X. Dong, Z. Zhao, R. Jin, A. Sefat, M. McGuire, B. Sales, D. Christen, D. Mandrus, Comparative high-field magnetotransport of the oxypnictide superconductors $RFeAsO_{1-x}F_x$ (R=La, Nd) and $SmFeAsO_{1-\delta}$, Phys. Rev. B 78 (2008) 064511.

[105] X.L. Dong, K. Jin, D.N. Yuan, H.X. Zhou, J. Yuan, Y.L. Huang, W. Hua, J.L. Sun, P. Zheng, W. Hu, Y.Y. Mao, M.W. Ma, G.M. Zhang, F. Zhou, Z.X. Zhao, $(Li_{0.84}Fe_{0.16})OHFe_{0.98}Se$ superconductor: Ion-exchange synthesis of large single crystal and highly two-dimensional electron properties Phys. Rev. B 92 (2015) 064515.

[106] N.P. Armitage, F. Ronning, D.H. Lu, C. Kim, A. Damascelli, K.M. Shen, D.L. Feng, H. Eisaki, Z.-X. Shen, P.K. Mang, N. Kaneko, M. Greven, Y. Onose, Y. Taguchi, Y. Tokura, Doping dependence of an n-type cuprate superconductor investigated by angle-resolved photoemission spectroscopy, Phys. Rev. Lett. 88 (2002) 257001.

[107] H. Matsui, T. Takahashi, T. Sato, K. Terashima, H. Ding, T. Uefuji, K. Yamada, Evolution of the pseudogap across the magnet-superconductor phase boundary of $Nd_{2-x}Ce_xCuO_4$, Phys. Rev. B 75 (2007) 224514.

[108] P. Richard, M. Neupane, Y.M. Xu, P. Fournier, S. Li, P. Dai, Z. Wang, H. Ding, Competition between antiferromagnetism and superconductivity in the electron-doped cuprates triggered by oxygen reduction, Phys. Rev. Lett. 99 (2007) 157002.

[109] M. Horio, T. Adachi, Y. Mori, A. Takahashi, T. Yoshida, H. Suzuki, L.C.C. Ambolode II, K. Okazaki, K. Ono, H. Kumigashira, H. Anzai, M. Arita, H. Namatame, M. Taniguchi, D. Ootsuki, K. Sawada, M. Takahashi, T. Mizokawa, Y. Koike, A. Fujimori, Suppression of the antiferromagnetic pseudogap in the electron-doped high-temperature superconductor by protect annealing, Nat. Commun. 7 (2016) 10567.

[110] A.F. Santander-Syro, M. Ikeda, T. Yoshida, A. Fujimori, K. Ishizaka, M. Okawa, S. Shin, R.L. Greene, N. Bontemps, Two-Fermi-surface superconducting state and a nodal d-wave energy gap of the electron-doped $Sm_{1.85}Ce_{0.15}CuO_{4-\delta}$ cuprate superconductor, Phys. Rev. Lett. 106 (2011) 197002.

[111] J.W. Harter, L. Maritato, D.E. Shai, E.J. Monkman, Y. Nie, D.G. Schlom, K.M. Shen, Nodeless superconducting phase arising from a strong ($\pi$, $\pi$) antiferromagnetic phase in the infinite-layer electron-doped $Sr_{1-x}La_xCuO_2$ compound, Phys. Rev. Lett. 109 (2012) 267001.

[112] J. Gauthier, S. Gagné, J. Renaud, M.-È. Gosselin, P. Fournier, P. Richard, Different roles of cerium substitution and oxygen reduction in transport in $Pr_{2-x}Ce_xCuO_4$ thin films, Phys. Rev. B 75 (2007) 024424.

[113] Y. Dagan, M.M. Qazilbash, C.P. Hill, V.N. Kulkarni, R.L. Greene, Evidence for a quantum phase transition in $Pr_{2-x}Ce_xCuO_{4-\delta}$ from transport measurements, Phys. Rev. Lett. 92 (2004) 167001.

[114] K. Jin, B.Y. Zhu, B.X. Wu, L.J. Gao, B.R. Zhao, Low-temperature Hall effect in electron-doped superconducting $La_{2-x}Ce_xCuO_4$ thin films, Phys. Rev. B 78 (2008) 174521.

[115] P. Li, R.L. Greene, Normal-state Nernst effect in electron-doped $Pr_{2-x}Ce_xCuO_{4-\delta}$ : Superconducting fluctuations and two-band transport, Phys. Rev. B 76 (2007) 174512.

[116] F.F. Tafti, F. Laliberté, M. Dion, J. Gaudet, P. Fournier, L. Taillefer, Nernst effect in the electron-doped cuprate superconductor Superconducting fluctuations, upper critical field and the origin of the dome, Phys. Rev. B 90 (2014) 024519.

[117] P.C. Li, K. Behnia, R.L. Greene, Evidence for a quantum phase transition in electron-doped $Pr_{2-x}Ce_xCuO_{4-\delta}$ from thermopower measurements, Phys. Rev. B 75 (2007) 020506(R).

[118] Y. Dagan, M.M. Qazilbash, R.L. Greene, Tunneling into the normal state of $Pr_{2-x}Ce_xCuO_4$, Phys. Rev. Lett. 94 (2005) 187003.

[119] A. Zimmers, J.M. Tomczak, R.P.S.M. Lobo, N. Bontemps, C.P. Hill, M.C. Barr, Y. Dagan, R.L. Greene, A.J. Millis, C.C. Homes, Infrared properties of electron-doped cuprates: Tracking normal-state gaps and quantum critical behavior in $Pr_{2-x}Ce_xCuO_4$, Europhys. Lett. 70 (2005) 225.

[120] J. Lin, A. Millis, Theory of low-temperature Hall effect in electron-doped cuprates, Phys. Rev. B 72 (2005) 214506.

[121] C. Kusko, R.S. Markiewicz, M. Lindroos, a.A. Bansil, Fermi surface evolution and collapse of the Mott pseudogap in $Nd_{2-x}Ce_xCuO_{4\pm\delta}$, Phys. Rev. B 66 (2002) 140513(R).

[122] D. Senechal, A.M. Tremblay, Hot spots and pseudogaps for hole- and electron-doped high-temperature superconductors, Phys. Rev. Lett. 92 (2004) 126401.

[123] T. Xiang, H.G. Luo, D.H. Lu, K.M. Shen, Z.X. Shen, Intrinsic electron and hole bands in electron-doped cuprate superconductors, Phys. Rev. B 79 (2009) 014524.

[124] G. Blatter, M.V. Feigelman, V.B. Geshkenbein, A.I. Larkin, V.M. Vinokur, Vortices in high-temperature superconductors, Rev. Mod. Phys. 66 (1994) 1125.

[125] K. Noto, S. Shinzawa, Y. Muto, Hall effect in intrinsic type-II superconductors near lower critical-field, Solid State Commun. 18 (1976) 1081.

[126] A.W. Smith, T.W. Clinton, C.C. Tsuei, C.J. Lobb, Sign reversal of the Hall resistivity in amorphous $Mo_3Si$, Phys. Rev. B 49 (1994) 12927.

[127] G. D'Anna, V. Berseth, L. Forro, A. Erb, E. Walker, Hall anomaly and vortex-lattice melting in superconducting single crystal $YBa_2Cu_3O_{7-\delta}$, Phys. Rev. Lett. 81 (1998) 2530.

[128] L.M. Wang, H.C. Yang, H.E. Horng, Mixed-state Hall effect in $YBa_2Cu_3O_y$/$PrBa_2Cu_3O_y$ superlattices, Phys. Rev. Lett. 78 (1997) 527.

[129] T.W. Clinton, A.W. Smith, Q. Li, J.L. Peng, R.L. Greene, C.J. Lobb, M. Eddy, C.C. Tsuei, Anisotropy, pinning, and the mixed-state Hall effect, Phys. Rev. B 52 (1995) R7046.

[130] W. Göb, W. Lang, J.D. Pedarnig, R. Rössler, D. Bäuerle, Magnetic field and current density dependence of the mixed-state Hall effect in $Bi_2Sr_2CaCu_2O_x$, Phys. C Supercond. 317 (1999) 627.

[131] S.J. Hagen, C.J. Lobb, R.L. Greene, M. Eddy, Flux-flow Hall effect in superconducting $Tl_2Ba_2CaCu_2O_8$ films, Phys. Rev. B 43 (1991) 6246.

[132] A.V. Samoilov, A. Legris, F. Rullieralbenque, P. Lejay, S. Bouffard, Z.G. Ivanov, L.G. Johansson, Mixed-state Hall conductivity in high-$T_c$ superconductors: Direct evidence of its independence on disorder, Phys. Rev. Lett. 74 (1995) 2351.

[133] W.N. Kang, S.H. Yun, J.Z. Wu, D.H. Kim, Scaling behavior and mixed-state hall effect in epitaxial $HgBa_2CaCu_2O_{4+\delta}$ thin films, Phys. Rev. B 55 (1997) 621.

[134] W. Göb, W. Liebich, W. Lang, I. Puica, R. Sobolewski, R. Rössler, J.D. Pedarnig, D. Bäuerle, Double sign reversal of the vortex Hall effect in $YBa_2Cu_3O_{4-\delta}$ thin films in the strong pinning limit of low magnetic fields, Phys. Rev. B 62 (2000) 9780.

[135] B.D. Josephson, Potential differences in the mixed state of type-II superconductors Phys. Lett. 16 (1965) 242.

[136] J. Bardeen, M.J. Stephen, Theory of motion of vortices in superconductors, Phys. Rev. 140 (1965) A1197.

[137] R. Ikeda, Hall-sign dependent on dimensionality of vortex-pinning disorder, Phys. C Supercond. 316 (1999) 189.

[138] N.B. Kopnin, V.M. Vinokur, Effects of pinning on the flux flow Hall resistivity, Phys. Rev. Lett. 83 (1999) 4864.

[139] I. Puica, W. Lang, W. Göb, R. Sobolewski, Hall-effect anomaly near $T_c$ and renormalized superconducting fluctuations in $YBa_2Cu_3O_{7-x}$, Phys. Rev. B 69 (2004) 104513.

[140] B.Y. Zhu, D.Y. Xing, Z.D. Wang, B.R. Zhao, Z.X. Zhao, Sign reversal of the mixed-state Hall resistivity in type-II superconductors, Phys. Rev. B 60 (1999) 3080.

[141] A.T. Dorsey, Vortex motion and the Hall effect in type-II superconductors: A time-dependent Ginzburg-Landau theory approach, Phys. Rev. B 46 (1992) 8376.

[142] N.B. Kopnin, Hall effect in moderately clean superconductors and the transverse force on a moving vortex, Phys. Rev. B 54 (1996) 9475.

[143] N.B. Kopnin, B.I. Ivlev, V.A. Kalatsky, The flux-flow Hall effect in type-II superconductors an explanation of the sign reversal, J. Low Temp. Phys. 90 (1993) 1.

[144] N.B. Kopnin, A.V. Lopatin, Flux-flow Hall effect in clean type-II superconductors, Phys. Rev. B 51 (1995) 15291.

[145] A. van Otterlo, M. Feigel'man, V. Geshkenbein, G. Blatter, Vortex dynamics and the Hall anomaly: A microscopic analysis, Phys. Rev. Lett. 75 (1995) 3736.

[146] J.E. Hirsch, F. Marsiglio, Hole superconductivity in oxides: A two-band model, Phys. Rev. B 43 (1991) 424.

[147] S.J. Hagen, A.W. Smith, M. Rajeswari, J.L. Peng, Z.Y. Li, R.L. Greene, S.N. Mao, X.X. Xi, S. Bhattacharya, Q. Li, C.J. Lobb, Anomalous flux-flow Hall effect $Nd_{1.85}Ce_{0.15}CuO_{4-y}$ and evidence for vortex dynamics, Phys. Rev. B 47 (1993) 1064.

[148] R.J. Troy, A.T. Dorsey, Transport properties and fluctuations in type-II superconductors near $H_{c2}$, Phys. Rev. B 47 (1993) 2715.

[149] T.B. Charikova, N.G. Shelushinina, G.I. Harus, D.S. Petukhov, A.V. Korolev, V.N. Neverov, A.A. Ivanov, Doping effect on the anomalous behavior of the Hall effect in electron-doped superconductor $Nd_{2-x}Ce_xCuO_{4+\delta}$, Phys. C Supercond. 483 (2012) 113.

[150] H.G. Luo, T. Xiang, Superfluid response in electron-doped cuprate superconductors, Phys. Rev. Lett. 94 (2005) 027001.

[151] C.S. Liu, H.G. Luo, W.C. Wu, T. Xiang, Two-band model of Raman scattering on electron-doped high-$T_c$ superconductors, Phys. Rev. B 73 (2006) 174517.

[152] K. Jin, B.X. Wu, B.Y. Zhu, B.R. Zhao, A. Volodin, J. Vanacken, A.V. Silhanek, V.V. Moshchalkov, Sign reversal of the Hall resistance in the mixed-state of $La_{1.89}Ce_{0.11}CuO_4$ and $La_{1.89}Ce_{0.11}(Cu_{0.99}Co_{0.01})O_4$ thin films, Phys. C Supercond. 479 (2012) 53.

[153] K. Jin, W. Hu, B.Y. Zhu, D. Kim, J. Yuan, T. Xiang, M.S. Fuhrer, I. Takeuchi, R.L. Greene, Evolution of electronic states in n-type copper oxide superconductor via electric double layer gating, arXiv: 1506.05727 (2015).

[154] J. Li, G.F. Zhang, W. Hu, Y. Huang, M. Ji, H.C. Sun, X. Zhou, D.Y. An, L.Y. Hao, Q. Zhu, J. Yuan, K. Jin, H.X. Guo, D. Fujita, T. Hatano, K. Yamaura, E. Takayama-Muromachi, H.B. Wang, P.H. Wu, J. Vanacken, V.V. Moshchalkov, High upper critical fields of superconducting $Ca_{10}(Pt_4As_8)(Fe_{1.8}Pt_{0.2}As_2)_5$ whiskers, Appl. Phys. Lett. 106 (2015) 262601.

[155] J. Jaroszynski, F. Hunte, L. Balicas, Y.-j. Jo, I. Raičević, A. Gurevich, D. Larbalestier, F. Balakirev, L. Fang, P. Cheng, Y. Jia, H. Wen, Upper critical fields and thermally-activated transport of $NdFeAsO_{0.7}F_{0.3}$ single crystal, Phys. Rev. B 78 (2008) 174523.

[156] M. Kano, Y. Kohama, D. Graf, F. Balakirev, A. S. Sefat, M. A. Mcguire, B. C. Sales, D. Mandrus, S. W. Tozer, Anisotropy of the upper critical field in a Co-doped $BaFe_2As_2$ single crystal, J. Phys. Soc. Jpn. 78 (2009) 084719.

[157] C. Senatore, R. Flükiger, M. Cantoni, G. Wu, R. Liu, X. Chen, Upper critical fields well above 100 T for the superconductor $SmFeAsO_{0.85}F_{0.15}$ with $T_c$=46K, Phys. Rev. B 78 (2008) 054514.

[158] E.D. Mun, M.M. Altarawneh, C.H. Mielke, V.S. Zapf, R. Hu, S.L. Bud'ko, P.C. Canfield, Anisotropic $H_{c2}$ of $K_{0.8}Fe_{1.76}Se_2$ determined up to 60 T, Phys. Rev. B 83 (2011) 100514(R).

[159] P. Fournier, R.L. Greene, Doping dependence of the upper critical field of electron-doped $Pr_{2-x}Ce_xCuO_4$ thin films, Phys. Rev. B 68 (2003) 094507.

[160] J. Herrmann, M.C. deAndrade, C.C. Almasan, R.P. Dickey, M.B. Maple, Magnetoresistivity of thin films of the electron-doped high-$T_c$ superconductor $Nd_{1.85}Ce_{0.15}CuO_{4\pm\delta}$, Phys. Rev. B 54 (1996) 3610.

[161] S.H. Han, C.C. Almasan, M.C. de Andrade, Y. Dalichaouch, M.B. Maple, Determination of the upper critical field of the electron-doped superconductor $Sm_{1.85}Ce_{0.15}CuO_{4-y}$ from resistive fluctuations, Phys. Rev. B 46 (1992) 14290.

[162] Y. Wang, H. Gao, Anisotropic magnetotransport of superconducting and normal state in an electron-doped $Nd_{1.85}Ce_{0.15}CuO_{4-\delta}$ single crystal, Phys. C Supercond. 470 (2010) 689.

[163] H.H. Wen, W.L. Yang, Z.X. Zhao, Macroscopic phase separation in overdoped high temperature superconductors, Phys. C Supercond. 341 (2000) 1735.

[164] H. Balci, C.P. Hill, M.M. Qazilbash, R.L. Greene, Nernst effect in electron-doped $Pr_{2-x}Ce_xCuO_4$, Phys. Rev. B 68 (2003) 054520.

[165] M.S. Kim, J.A. Skinta, T.R. Lemberger, A. Tsukada, M. Naito, Magnetic penetration depth measurements of $Pr_{2-x}Ce_xCuO_{4-\delta}$ films on buffered substrates: Evidence for a nodeless gap, Phys. Rev. Lett. 91 (2003) 087001.

[166] A. Gurevich, Enhancement of the upper critical field by nonmagnetic impurities in dirty two-gap superconductors, Phys. Rev. B 67 (2003) 184515.

[167] Y. Wang, L. Li, N.P. Ong, Nernst effect in high-$T_c$ superconductors, Phys. Rev. B 73 (2006) 024510.

[168] A. Pourret, P. Spathis, H. Aubin, K. Behnia, Nernst effect as a probe of superconducting fluctuations in disordered thin films, New J. Phys. 11 (2009) 055071.

[169] O. Cyr-Choiniere, R. Daou, F. Laliberte, D. LeBoeuf, N. Doiron-Leyraud, J. Chang, J.-Q. Yan, J.-G. Cheng, J.-S. Zhou, J.B. Goodenough, S. Pyon, T. Takayama, H. Takagi, Y. Tanaka, L. Taillefer, Enhancement of the Nernst effect by stripe order in a high-$T_c$ superconductor, Nature 458 (2009) 743.

[170] A.V. Ettingshausen, W. Nernst, Ueber das Auftreten electromotorischer Kräfte in Metallplatten, welche von einem Wärmestrome durchflossen werden und sich im magnetischen Felde befinden, Annalen der Physik Und Chemie 265(10) (1886) 343.

[171] K. Behnia, The Nernst effect and the boundaries of the Fermi liquid picture, J. Phys. Condens Matter 21 (2009) 113101.

[172] V.J. Emery, S.A. Kivelson, Importance of phase fluctuations in superconductors with small superfluid density, Nature 374 (1995) 434.

[173] I. Ussishkin, S.L. Sondhi, D.A. Huse, Gaussian superconducting fluctuations, thermal transport, and the nernst effect, Phys. Rev. Lett. 89 (2002) 287001.

[174] E.H. Sondheimer, The theory of the galvanomagnetic and thermomagnetic effects in metals, Proc. R. Soc. London, Ser. A 193 (1948) 484.

[175] Z.A. Xu, N.P. Ong, Y. Wang, T. Kakeshita, S. Uchida, Vortex-like excitations and the onset of superconducting phase fluctuation in underdoped $La_{2-x}Sr_xCuO_4$, Nature 406 (2000) 486.

[176] M.S. Nam, A. Ardavan, S.J. Blundell, J.A. Schlueter, Fluctuating superconductivity in organic molecular metals close to the Mott transition, Nature 449 (2007) 584.

[177] W. Wu, I.J. Lee, P.M. Chaikin, Giant nernst effect and lock-in currents at magic angles in $(TMTSF)_2PF_6$, Phys. Rev. Lett. 91 (2003) 056601.

[178] R. Bel, K. Behnia, Y. Nakajima, K. Izawa, Y. Matsuda, H. Shishido, R. Settai, Y. Onuki, Giant Nernst effect in $CeCoIn_5$, Phys. Rev. Lett. 92 (2004) 217002.

[179] I. Sheikin, H. Jin, R. Bel, K. Behnia, C. Proust, J. Flouquet, Y. Matsuda, D. Aoki, Y. Onuki, Evidence for a new magnetic field scale in $CeCoIn_5$, Phys. Rev. Lett. 96 (2006) 077207

[180] A.H. Wilson, The theory of metals, Second ed., Cambridge University Press, 1953.

[181] J.A. Clayhold, A.W. Linnen, F. Chen, C.W. Chu, Normal-state Nernst effect in a $Tl_2Ba_2CaCu_2O_{8+\delta}$ epitaxial film, Phys. Rev. B 50 (1994) 4252.

[182] Y. Wang, Z.A. Xu, T. Kakeshita, S. Uchida, S. Ono, Y. Ando, N.P. Ong, Onset of the vortexlike Nernst signal above $T_c$ in $La_{2-x}Sr_xCuO_4$ and $Bi_2Sr_{2-y}La_yCuO_6$, Phys. Rev. B 64 (2001) 224519.

[183] R. Bel, K. Behnia, H. Berger, Ambipolar Nernst effect in $NbSe_2$, Phys. Rev. Lett. 91 (2003) 066602.

[184] Y.J. Uemura, G.M. Luke, B.J. Sternlieb, J.H. Brewer, J.F. Carolan, W.N. Hardy, R. Kadono, J.R. Kempton, R.F. Kiefl, S.R. Kreitzman, P. Mulhern, T.M. Riseman, D.L. Williams, B.X. Yang, S. Uchida, H. Takagi, J. Gopalakrishnan, A.W. Sleight, M.A. Subramanian, C.L. Chien, M.Z. Cieplak, G. Xiao, V.Y. Lee, B.W. Statt, C.E. Stronach, W.J. Kossler, X.H. Yu, Universal correlations between $T_c$ and $n_s/m$ (carrier density over effective mass) in high-$T_c$ cuprate superconductors, Phys. Rev. Lett. 62 (1989) 2317.

[185] B.B. Goodman, Type II superconductors, Rep. Prog. Phys. 29 (1966) 445.

[186] N.P. Ong, Y. Wang, Vorticity, phase stiffness and the cuprate phase diagram, Phys. C Supercond. 408 (2004) 11.

[187] L. Li, J.G. Checkelsky, S. Komiya, Y. Ando, N.P. Ong, Low-temperature vortex liquid in $La_{2-x}Sr_xCuO_4$, Nat. Phys. 3 (2007) 311.

[188] A. Larkin, A. Varlamov, Theory of fluctuations in superconductors, Rev. ed., Oxford Science Publication, 2005.

[189] J. Chang, N. Doiron-Leyraud, O. Cyr-Choinière, G. Grissonnanche, F. Laliberté, E. Hassinger, J.-P. Reid, R. Daou, S. Pyon, T. Takayama, H. Takagi, L. Taillefer, Decrease of upper critical field with underdoping in cuprate superconductors, Nat. Phys. 8 (2012) 751.

[190] A. Pourret, H. Aubin, J. Lesueur, C.A. Marrache-Kikuchi, L. Bergé, L. Dumoulin, K. Behnia, Length scale for the superconducting Nernst signal above $T_c$ in $Nb_{0.15}Si_{0.85}$, Phys. Rev. B 76 (2007) 214504.

[191] B. Leridon, T.K. Ng, C.M. Varma, Josephson effect for superconductors lacking time-reversal and inversion symmetries, Phys. Rev. Lett. 99 (2007) 027002.

[192] V.V. Moshchalkov, L. Trappeniers, J. Vanacken, Doped $CuO_2$ planes in high $T_c$ cuprates: 2D or not 2D?, J. Low Temp. Phys. 117 (1999) 1283.

[193] N.W. Ashcroft, N.D. Mermin, Solid State Physics, Philadelphia, PA: Saunders, 1976.

[194] R.D. Barnard, Thermoelectricity in metals and alloys London, Taylor & Francis, New York, Halsted Press, 1972.

[195] F. Laliberte, J. Chang, N. Doiron-Leyraud, E. Hassinger, R. Daou, M. Rondeau, B.J. Ramshaw, R. Liang, D.A. Bonn, W.N. Hardy, S. Pyon, T. Takayama, H. Takagi, I. Sheikin, L. Malone, C. Proust, K. Behnia, L. Taillefer, Fermi-surface reconstruction by stripe order in cuprate superconductors, Nat. Commun. 2 (2011) 432.

[196] W. Jiang, X.Q. Xu, S.J. Hagen, J.L. Peng, Z.Y. Li, R.L. Greene, Anisotropic normal-state magnetothermopower of superconducting $Nd_{1.85}Ce_{0.15}CuO_4$ crystals, Phys. Rev. B 48 (1993) 657.

[197] X.Q. Xu, S.N. Mao, W. Jiang, J.L. Peng, R.L. Greene, Oxygen dependence of the transport properties of $Nd_{1.78}Ce_{0.22}CuO_{4\pm\delta}$, Phys. Rev. B 53 (1996) 871.

[198] R.C. Budhani, M.C. Sullivan, C.J. Lobb, R.L. Greene, Thermopower and Hall conductivity in the magnetic-field-driven normal state of $Pr_{2-x}Ce_xCuO_{4-\delta}$ superconductors, Phys. Rev. B 65 (2002) 100517(R).

[199] D. Shoenberg, Magnetic oscillations in metals, Cambridge University Press, Cambridge, UK, 1984.

[200] N. Doiron-Leyraud, C. Proust, D. LeBoeuf, J. Levallois, J.-B. Bonnemaison, R. Liang, D.A. Bonn, W.N. Hardy, L. Taillefer, Quantum oscillations and the Fermi surface in an underdoped high-$T_c$ superconductor, Nature 447 (2007) 565.

[201] A. Audouard, C. Jaudet, D. Vignolles, R. Liang, D.A. Bonn, W.N. Hardy, L. Taillefer, C. Proust, Multiple quantum oscillations in the de Haas-van Alphen spectra of the underdoped high-temperature superconductor $YBa_2Cu_3O_{6.5}$, Phys. Rev. Lett. 103 (2009) 157003.

[202] S.E. Sebastian, N. Harrison, E. Palm, T.P. Murphy, C.H. Mielke, R. Liang, D.A. Bonn, W.N. Hardy, G.G. Lonzarich, A multi-component Fermi surface in the vortex state of an underdoped high-$T_c$ superconductor, Nature 454 (2008) 200.

[203] S.C. Riggs, O. Vafek, J.B. Kemper, J.B. Betts, A. Migliori, F.F. Balakirev, W.N. Hardy, R. Liang, D.A. Bonn, G.S. Boebinger, Heat capacity through the magnetic-field-induced resistive transition in an underdoped high-temperature superconductor, Nat. Phys. 7 (2011) 332.

[204] G. Grissonnanche, O. Cyr-Choinière, F. Laliberté, S.R. de Cotret, A. Juneau-Fecteau, S. Dufour-Beauséjour, M.-È. Delage, D. LeBoeuf, J. Chang, B.J. Ramshaw, D.A. Bonn, W.N. Hardy, R. Liang, S. Adachi, N.E. Hussey, B. Vignolle, C. Proust, M. Sutherland, S. Krämer, J.-H. Park, D. Graf, N.

Doiron-Leyraud, L. Taillefer, Direct measurement of the upper critical field in cuprate superconductors, Nat. Commun. 5 (2014) 3280.

[205] N. Barišić, S. Badoux, M.K. Chan, C. Dorow, W. Tabis, B. Vignolle, G. Yu, J. Béard, X. Zhao, C. Proust, M. Greven, Universal quantum oscillations in the underdoped cuprate superconductors, Nat. Phys. 9 (2013) 761.

[206] B. Vignolle, A. Carrington, R.A. Cooper, M.M.J. French, A.P. Mackenzie, C. Jaudet, D. Vignolles, C. Proust, N.E. Hussey, Quantum oscillations in an overdoped high-$T_c$ superconductor, Nature 455 (2008) 952.

[207] T. Helm, M.V. Kartsovnik, M. Bartkowiak, N. Bittner, M. Lambacher, A. Erb, J. Wosnitza, R. Gross, Evolution of the Fermi surface of the electron-doped high-temperature superconductor $Nd_{2-x}Ce_xCuO_4$ revealed by Shubnikov–de Haas oscillations, Phys. Rev. Lett. 103 (2009) 157002.

[208] N. Breznay, R.D. McDonald, Y. Krockenberger, K.A. Modic, Z. Zhu, I.M. Hayes, N.L. Nair, T. Helm, H. Irie, H. Yamamoto, J.G. Analytis, Quantum oscillations suggest hidden quantum phase transition in the cuprate superconductor $Pr_2CuO_{4\pm\delta}$
arXiv: 1510.04268 (2015).

[209] S. Sachdev, Quantum phase transition, Cambridge University Press, Cambridge, UK, 1999.

[210] N.P. Butch, K. Jin, K. Kirshenbaum, R.L. Greene, J. Paglione, Quantum critical scaling at the edge of Fermi liquid stability in a cuprate superconductor, Proc. Natl. Acad. Sci. U. S. A. 109 (2012) 8440.

[211] S. Nakamae, K. Behnia, N. Mangkorntong, M. Nohara, H. Takagi, S.J.C. Yates, N.E. Hussey, Electronic ground state of heavily overdoped nonsuperconducting $La_{2-x}Sr_xCuO_4$, Phys. Rev. B 68 (2003) 100502(R).

[212] D.v.d. Marel, H.J.A. Molegraaf, J. Zaanen, Z. Nussinov, F. Carbone, A. Damascelli, H. Eisaki, M. Greven, P. Kes, M. Li, Quantum critical behaviour in a high-$T_c$ superconductor, Nature 425 (2003) 271.

[213] S.D. Wilson, S. Li, P. Dai, W. Bao, J.-H. Chung, H.J. Kang, S.-H. Lee, S. Komiya, Y. Ando, Q. Si, Evolution of low-energy spin dynamics in the electron-doped high-transition-temperature superconductor $Pr_{0.88}LaCe_{0.12}CuO_{4-\delta}$, Phys. Rev. B 74 (2006) 144514.

[214] B. Keimer, S.A. Kivelson, M.R. Norman, S. Uchida, J. Zaanen, From quantum matter to high-temperature superconductivity in copper oxides, Nature 518 (2015) 179.

[215] T. Helm, M.V. Kartsovnik, C. Proust, B. Vignolle, C. Putzke, E. Kampert, I. Sheikin, E.-S. Choi, J.S. Brooks, N. Bittner, W. Biberacher, A. Erb, J. Wosnitza, R. Gross, Correlation between Fermi surface transformations and superconductivity in the electron-doped high-$T_c$ superconductor $Nd_{2-x}Ce_xCuO_4$, Phys. Rev. B 92 (2015) 094501.

[216] K. Yamada, K. Kurahashi, T. Uefuji, M. Fujita, S. Park, S.-H. Lee, Y. Endoh, Commensurate spin dynamics in the superconducting state of an electron-doped cuprate superconductor, Phys. Rev. Lett. 90 (2003) 137004.

[217] M. Matsuda, Y. Endoh, K. Yamada, H. Kojima, I. Tanaka, R.J. Birgeneau, M.A. Kastner, G. Shirane, Magnetic order, spin correlations, and superconductivity in single-crystal $Nd_{1.85}Ce_{0.15}CuO_{4+\delta}$, Phys. Rev. B 45 (1992) 12548.

[218] T. Uefuji, K. Kurahashi, M. Fujita, M. Matsuda, K. Yamada, Electron-doping effect on magnetic order and superconductivity in $Nd_{2-x}Ce_xCuO_4$ single crystal, Phys. C Supercond. 378 (2002) 273.

[219] E.M. Motoyama, G. Yu, I.M. Vishik, O.P. Vajk, P.K. Mang, M. Greven, Spin correlations in the electron-doped high-transition-temperature superconductor $Nd_{2-x}Ce_xCuO_{4\pm\delta}$, Nature 445 (2007) 186.

[220] P.K. Mang, O.P. Vajk, A. Arvanitaki, J.W. Lynn, M. Greven, Spin correlations and magnetic order in nonsuperconducting $Nd_{2-x}Ce_xCuO_{4\pm\delta}$, Phys. Rev. Lett. 93 (2004) 027002.

[221] S.D. Wilson, S. Li, H. Woo, P. Dai, H.A. Mook, C.D. Frost, S. Komiya, Y. Ando, High-Energy Spin Excitations in the Electron-Doped Superconductor $Pr_{0.88}LaCe_{0.12}CuO_{4-\delta}$ with $T_c$=21 K, Phys. Rev. Lett. 96 (2006) 157001.

[222] K. Ishii, M. Fujita, T. Sasaki, M. Minola, G. Dellea, C. Mazzoli, K. Kummer, G. Ghiringhelli, L. Braicovich, T. Tohyama, K. Tsutsumi, K. Sato, R. Kajimoto, K. Ikeuchi, K. Yamada, M. Yoshida, M. Kurooka, J. Mizuki, High-energy spin and charge excitations in electron-doped copper oxide superconductors, Nat. Commun. 5 (2014) 3714.

[223] M. Fujita, M. Matsuda, S.-H. Lee, M. Nakagawa, K. Yamada, Low-energy spin fluctuations in the ground states of electron-doped $Pr_{1-x}LaCe_xCuO_{4+\delta}$ cuprate superconductors, Phys. Rev. Lett. 101 (2008) 107003.

[224] H. Saadaoui, Z. Salman, H. Luetkens, T. Prokscha, A. Suter, W.A. MacFarlane, Y. Jiang, K. Jin, R.L. Greene, E. Morenzoni, R.F. Kiefl, The phase diagram of electron-doped $La_{2-x}Ce_xCuO_{4-\delta}$, Nat. Commun. 6 (2015) 6041.

[225] H.Y. Hwang, Y. Iwasa, M. Kawasaki, B. Keimer, N. Nagaosa, Y. Tokura, Emergent phenomena at oxide interfaces, Nat. Mater. 11 (2012) 103-113.

[226] H. Koinuma, I. Takeuchi, Combinatorial solid-state chemistry of inorganic materials, Nat. Mater. 3 (2004) 429-438.

Figure captions

Fig. 1. The crystal structures of (a) hole-doped, (b) electron-doped and (c) infinite-layer cuprates. Here RE is one of the rare-earth ions, including Nd, Pr, La, Sm and Eu.

Fig. 2. (a) The illustration and (b) the real image of typical Hall-bar to measure both longitudinal resistivity $\rho_{xx}$ and Hall resistivity $\rho_{xy}$. The black area in figure (a) is film patterned by lithography.

Fig. 3. The low temperature metal-insulator transitions tuned by different parameters. Temperature dependence of resistivity for (a) different doping $Nd_{2-x}Ce_xCuO_4$ [21], (b) $Nd_{2-x}Ce_xCuO_4$ films with disorder controlled by annealing process [28], (c) different magnetic field at $x$ = 0.12 $La_{2-x}Ce_xCuO_4$ film [30], (d) ion-irradiated $Nd_{2-x}Ce_xCuO_4$ films [31].

Fig. 4. $\rho_{ab}$ versus T for $Pr_{2-x}Ce_xCuO_4$ thin films of different doping at B = 0 T (dashed lines), 8.7 T (thin lines), and 12 T (thick lines) [60].

Fig. 5. Schematic phase diagram for superconducting films. Distinct zero temperature superconductor-insulator transitions occur at both critical disorder $\Delta_c$ and critical magnetic field $B_c$ [35].

Fig. 6. (a) Resistivity as a function of the scaling variable $[c_0(B-B_c)/T^{1/zv}]$ for $Nd_{2-x}Ce_xCuO_4$, where $B_c$ =2.9 T and $vz$=1.2 are used [27]. (b) Scaling with respect to the single variable $u = |x - x_c|T^{-1/zv}$ with $zv$ = 1.5 for $La_{2-x}Sr_xCuO_4$ [42]. (c) Isotherms of $R(x)$ at temperatures from 2 to 22 K for $YBa_2Cu_3O_{7-\delta}$. Inset in (c): finite size scaling analysis of $R(x)$ with $zv$ = 2.2 [43].

Fig. 7. Temperature dependence of the resistivity for $La_{2-x}Sr_xCuO_4$ and $YBa_2Cu_3O_7$. Data for $V_3Si$ and Cu are shown for comparison [50].

Fig. 8. Linear resistivity at different temperature region in cuprates. (a) Resistivity at T<10 K and B=12 T for the $Pr_{2-x}Ce_xCuO_4$ samples of $x$ = 0.17. The inset shows a magnified view of the subkelvin range [60]. (b) Resistivity at 0 T, 8.7 T and 12 T and Hall coefficient of the overdoped $Pr_{2-x}Ce_xCuO_4$ film, $x$=0.17 [60]. (c) The normalized resistivities as a function of temperature for three samples show linear dependence above $T^*\approx$280K for $HgBa_2CuO_{4+\delta}$ [61]. (d) The resistivity exhibits a quadratic temperature dependence between $T'\approx$ 90 K and $T^{**}\approx$ 170 K for $HgBa_2CuO_{4+\delta}$. This is also seen from the plot of $d\rho/d(T^2)$ (inset) [61].

Fig. 9. Temperature dependence of the normal-state resistivity $\rho(T)$ of (a) $x$ = 0.15 and (b) x= 0.16 of $La_{2-x}Ce_xCuO_4$ films at 7.5 and 7 T; (c) $x$ = 0.19 and (d) x= 0.21 at zero field [54].

Fig. 10. Relation between the superconducting transition temperature and the scattering rate in $La_{2-x}Ce_xCuO_4$ and $Pr_{2-x}Ce_xCuO_4$ [54].

Fig. 11. In-plane resistivity in magnetic fields as a function of log$T$ for (a) (La, Ce)$_2$CuO$_4$, (b) (Pr, Ce)$_2$CuO$_4$ and (c) (Nd, Ce)$_2$CuO$_4$ thin films. The insets show their linear-scale replotted curves of the zero-field data [37].

Fig. 12. Magnetoresistance is tuned by different parameters. (a) The field dependence of the in-plane magnetoresistivity of La$_{2-x}$Ce$_x$CuO$_4$ with $x$ = 0.06, 0.08, and 0.10 at 35 K [24]. (b) Magnetoresistance at 60 K as a function of oxygen content in optimal doping Nd$_{2-x}$Ce$_x$CuO$_4$ [29]. (c) The ab-plane resistivity of Pr$_{2-x}$Ce$_x$CuO$_4$ films vs. magnetic field applied perpendicular to the ab-plane with $x$ = 0.15 (left) and $x$ = 0.16 (right) [40].

Fig. 13. The magnetoresistance isotherms in La$_{2-x}$Ce$_x$CuO$_4$ thin film with $x$=0.12 (a) and optimal doped Nd$_{2-x}$Ce$_x$CuO$_4$ thin film [30]. (b), respectively. The insets show enlarged $\rho(B)$ curves for $x$=0.15 [22].

Fig. 14. The magnetoresistance isotherms in (a) La$_{2-x}$Ce$_x$CuO$_4$ [30] and (b) Pr$_{2-x}$Ce$_x$CuO$_4$ thin films [38].

Fig. 15. Field-induced transition from noncollinear to collinear spin arrangement in Pr$_2$CuO$_4$ [73]. (a) Zero-filed noncollinear spin structure. Only Cu spins are shown. Collinear spin-flop states induced by magnetic fields applied (b) along the Cu-Cu direction, (c) tilted from [010], (d) parallel to [010].

Fig. 16. The in-plane angular magnetoresistance in electron-doped cuprates. (a) Twofold AMR in La$_{2-x}$Ce$_x$CuO$_4$ [24] and (b) fourfold AMR in Pr$_{2-x}$Ce$_x$CuO$_4$ [81] with different doping.

Fig. 17. In-plane magnetoresistance versus magnetic field for Pr$_{2-x}$Ce$_x$CuO$_4$ films with $x$ = 0.17. The inset shows the magnetoresistance in a different temperature range from the main panel [92].

Fig. 18. The evolution of electronic structure measured by ARPES: (a) and (b) in Nd$_{2-x}$Ce$_x$CuO$_4$ various Ce doped [106, 107], (c) in Pr$_{1.3-x}$La$_{0.7}$Ce$_x$CuO$_4$ with different oxygen contents [109].

Fig. 19. The temperature dependence the Hall coefficient for different parameters. (a) the various Ce doping in Pr$_{2-x}$Ce$_x$CuO$_4$ from $x$=0.11 - 0.19 [113]; (b) various oxygen contents for Pr$_{2-x}$Ce$_x$CuO$_4$ at $x$ = 0.17, where the oxygen content increases from sample 1 to sample 14 [112]; (c) B= 14 T of La$_{2-x}$Ce$_x$CuO$_4$ thin films with $x$ from 0.06 to 0.15 [24]; (d) different Co concentrations for La$_{1.89}$Ce$_{0.11}$(Cu$_{1-x}$Co$_x$)O$_4$ [59].

Fig. 20. Superfluid density versus $T/T_c$ for n-type cuprates. $\rho_{s,1}$ and $\rho_{s,2}$ corresponding to the superfluid densities of electrons and holes, respectively [150].

Fig. 21. The Hall resistivity $\rho_{xy}$ versus the magnetic field perpendicular to the ab-plane of (a) Pr$_{1.85}$Ce$_{0.15}$CuO$_4$ ultrathin films [153] and (b) La$_{2-x}$Ce$_x$CuO$_4$: Co thin films at different

temperatures [152].

Fig. 22. (a) Comparison of Nernst effect and resistivity in terms of $H_{c2}$ for $Pr_{1.85}Ce_{0.15}CuO_4$ thin films. The dashed lines show the method to extract $H_{c2}$ [164]. (b) Magnetic field derivative of the resistivity $d\rho_{xx}/dH$ versus $H$ of $La_{1.85}Ce_{0.15}CuO_4$ thin films. Label A equals the maximum of $d\rho_{xx}/dH$ ($T_{conset}$=16 K). The y-axis is plotted on logarithmic scale [114].

Fig. 23. The upper critical field $H_{c2}$ of $La_{1.85}Ce_{0.15}CuO_4$ and $Pr_{1.85}Ce_{0.15}CuO_4$ for different doping levels (a), and of $Ca_{10}(Pt_4As_8)(Fe_{1.8}Pt_{0.2}As_2)_5$ whiskers (b). The data are extracted from Refs. [114, 154].

Fig. 24. The illustration for thermal transport measurement of Nernst signal ($N = -\frac{V_y}{\Delta T} = \frac{E_y}{\nabla_x T}$) under the perpendicular magnetic field and thermopower ($S = \frac{V_x}{\Delta T} = -\frac{E_x}{\nabla_x T}$) out magnetic field.

Fig. 25. The large Nernst signal at normal state exists in both (a) $Pr_{2-x}Ce_xCuO_4$ [115] and (b) $La_{2-x}Ce_xCuO_4$.

Fig. 26. Schematic phase diagram of high-$T_c$ superconductors with temperature $T$ versus doping $x$ [172].

Fig. 27. (a) Nernst signal versus temperature in underdoped $Pr_{2-x}Ce_xCuO_4$ thin film at $x = 0.13$ and $\mu_0 H$=2 T. $H_{c2}(0)\approx 7$ T and $T_c$= 11.8 K. The solid line is the real part of ac susceptibility under zero field [115]. (b) Temperature dependence of the Nernst coefficient, $\upsilon(T)$, for different Sr doping in $La_{1.8-x}Eu_{0.2}Sr_xCuO_4$ [169].

Fig. 28. (a) The thermopower [117] and (b) Hall coefficient [113] at low temperature in electron-doped cuprates $Pr_{2-x}Ce_xCuO_{4\pm\delta}$. Both the abrupt change of thermopower in (a) and the abrupt change of Hall coefficient in (b) around $x = 0.16$ imply the occurrence of a quantum phase transition.

Fig. 29. Quantum oscillation and topology of Fermi surface in the hole-doped cuprates. (a) Quantum oscillations of in-plane resistance in under-doped cuprate $YBa_2Cu_3O_{6.5}$ [200]; (b) the Fermi arc of under-doped cuprate $Ca_{2-x}Na_xCuO_2Cl_2$ [200]; (c) fast quantum oscillations in overdoped cuprate $Tl_2Ba_2CuO_6$ [206]; (d) the large pocket on the Fermi surface for overdoped cuprate $Tl_2Ba_2CuO_6$ [200].

Fig. 30. Quantum oscillation and topology of Fermi surface in the electron-doped cuprates $Nd_{2-x}Ce_xCuO_4$ [207]. (a) Slow quantum oscillations of $c$-axis resistivity in the optimal and slightly over-doped samples with $x = 0.15$ and $x = 0.16$; (b) fast quantum oscillations in overdoped with $x = 0.17$; (c) corresponding fast Fourier transform spectra of the oscillatory resistivities with different doping; (d) reconstructed Fermi surface consisting of one electron pocket and two hole pockets; (e) single component Fermi surface of the over-doped sample with *x* = 0.17.

Fig. 31. Generic phase diagram in the vicinity of a continuous quantum phase transition [52]. The horizontal axis represents the control parameter *r* used to tune the system through the QPT. Dashed lines indicate the boundaries of the quantum critical region. Lower crossover lines are given by $T \propto |\gamma|^{\nu z}$; the high-temperature crossover to nonuniversal (lattice) physics occurs when the correlation length is no longer large to microscopic length scales. The solid line marks the finite-temperature boundary between the ordered and disordered phases. Close to this line, the critical behavior is classical.

Fig. 32. Quantum criticality at the edge of Fermi liquid in electron-doped cuprates $La_{2-x}Ce_xCuO_{4\pm\delta}$ [210]. (a) The multidimensional phase diagram (*x*, *B*, *T*) near the QCP $x_c$. As the magnetic field increases, the QCP moves to the lower doping . (b) A strong increase of the quasiparticle–quasiparticle scattering coefficient $A_2$ (from fits of $\rho = \rho_0 + A_2 T^2$) as a function of magnetic field provides evidence for a field-tuned quantum critical point. Inset: taken in the zero-temperature limit for three Ce concentrations, all of the data fit to one divergent function $A_2 = A_0(\Delta B/B_c)^{-\alpha}$, with critical exponent α=0.38±0.01. (c) and (d) The resistivity *Δρ* data divided by $A_2T^2$ can be fitted very well by the scaling *ΔBγ/T* with suitable exponent *γ* for *x* = 0.15 and *x* = 0.17. The exponent *γ* is 0.4 for *x* = 0.15 and 1 for *x* = 0.17, respectively.

Fig. 33. Phase diagram. (a) Temperature versus hole doping level for the copper oxides, indicating where various phases occur [214]. The $T_{S,\,onset}$ (dotted green line), $T_{C,\,onset}$ and $T_{SC,\,onset}$ (dotted red line for both) refer to the onset temperatures of spin-, charge and superconducting fluctuations, while *T\** indicates the temperature where the crossover to the pseudogap regime occurs. The blue and green regions indicate fully developed antiferromagnetic order and d-wave superconducting order, respectively. The red striped area indicates the presence of fully developed charge order setting in at $T_{CDW}$. $T_{SDW}$ represents the same for incommensurate spin density wave order. Quantum critical points for superconductivity and charge order are indicated by the arrows. (b) Temperature–doping (*T–x*) phase diagram of $La_{2-x}Ce_xCuO_4$ [54]. The superconductivity (yellow), $\rho \propto T$ (red) and Fermi-liquid regimes (blue) terminate at one critical doping, $x_c$. The antiferromagnetic (or spin-density-wave) regime (circles) is estimated from previous in-plane angular magnetoresistance measurements. A QCP associated with a spin-density-wave Fermi surface reconstruction is estimated to occur near *x*= 0.14 (indicated as $x_{FS}$).

Fig. 34. The phase diagram of $La_{2-x}Ce_xCuO_{4\pm\delta}$ achieved by μSR and the boundary of AFM locates in the under-doped regime [224]. The magnetic phase boundary measured with LE-μSR is the brown band.

Fig. 35. The common features of electron-doped cuprate superconductors sorted out from the transport measurements.

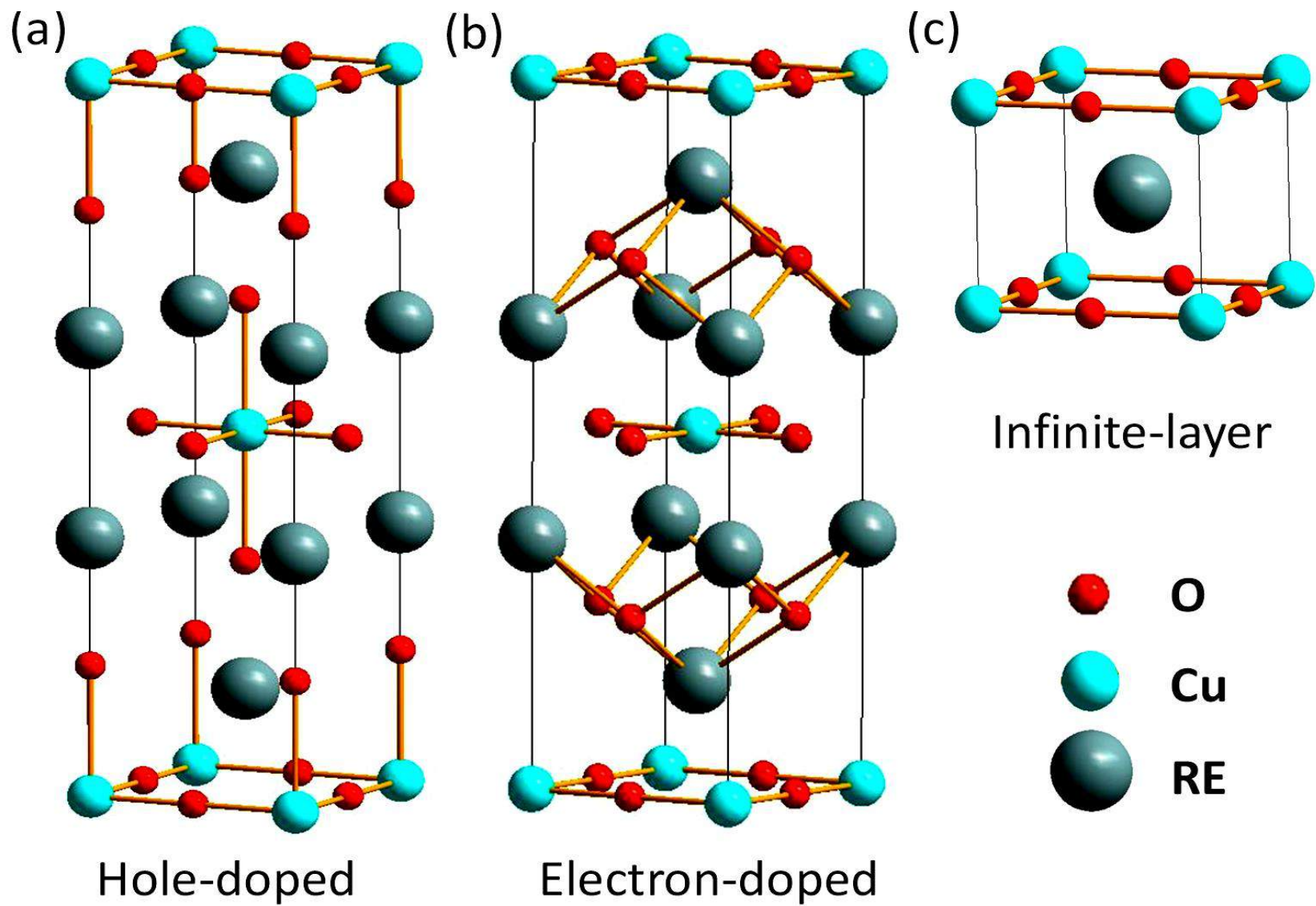

Figure 1

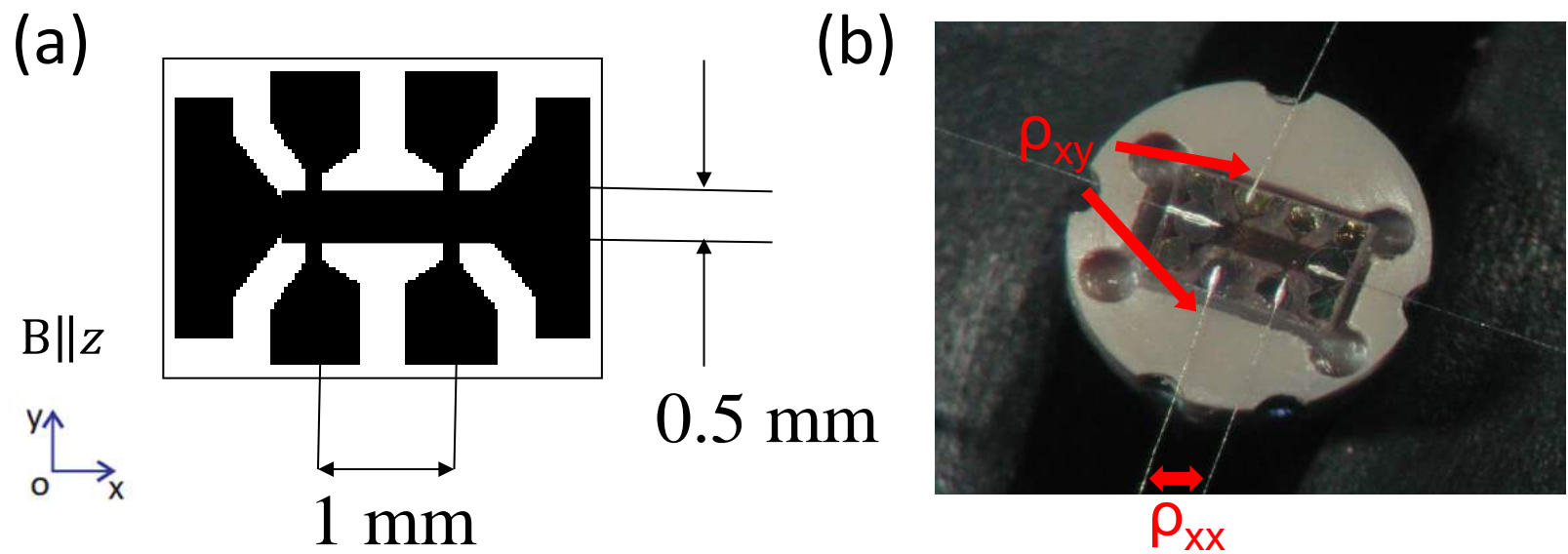

Figure 2

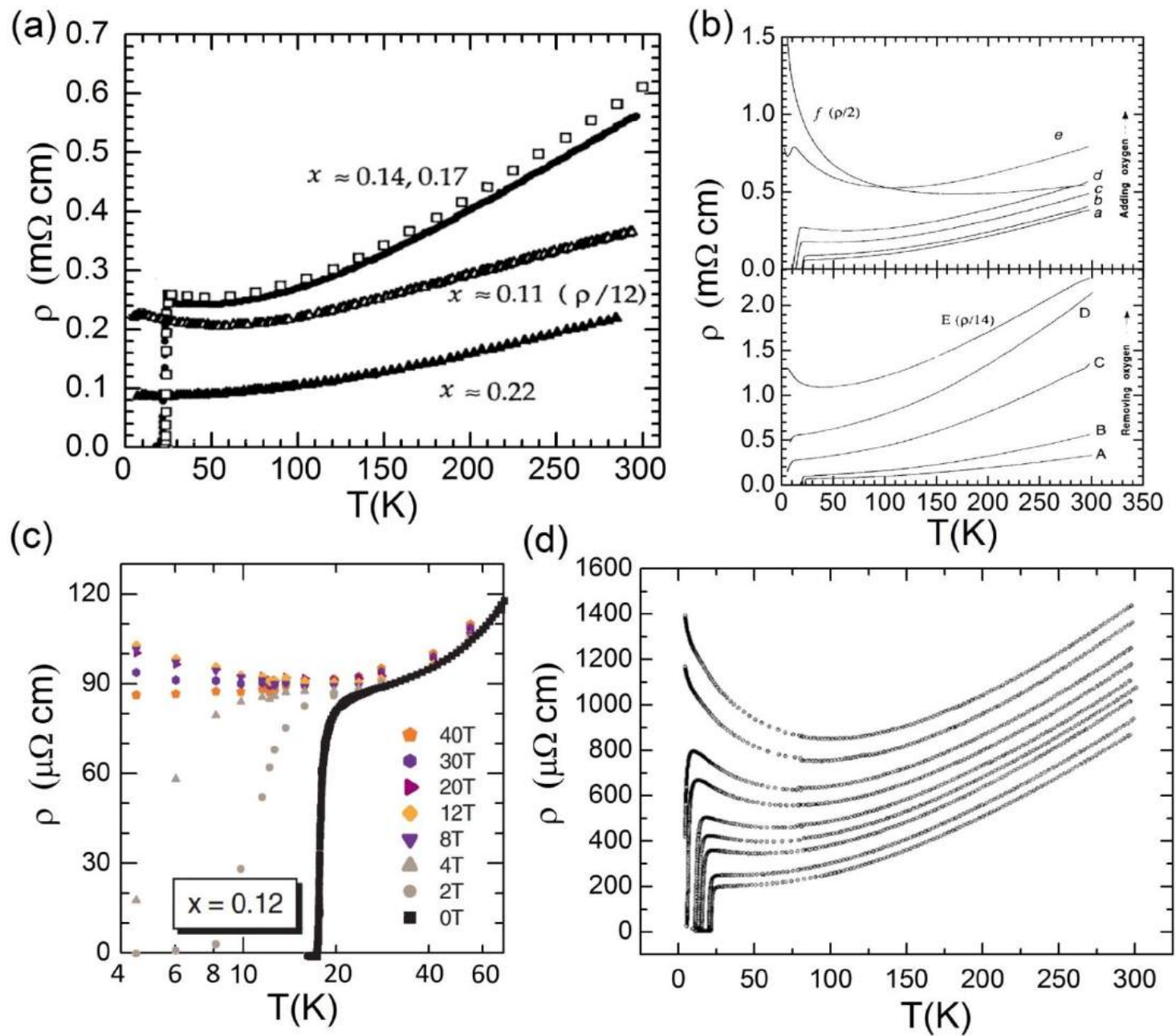

Figure 3

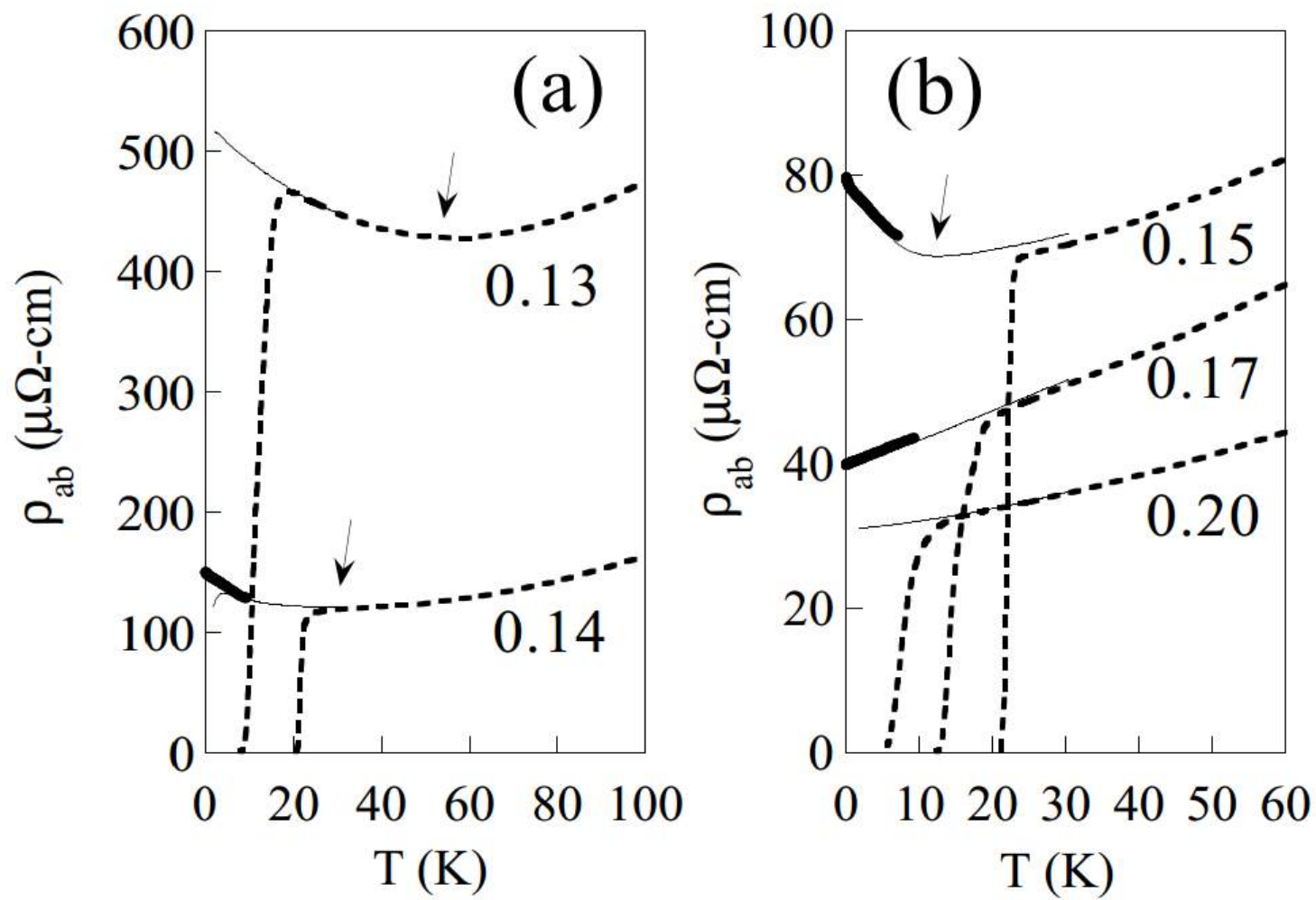

Figure 4

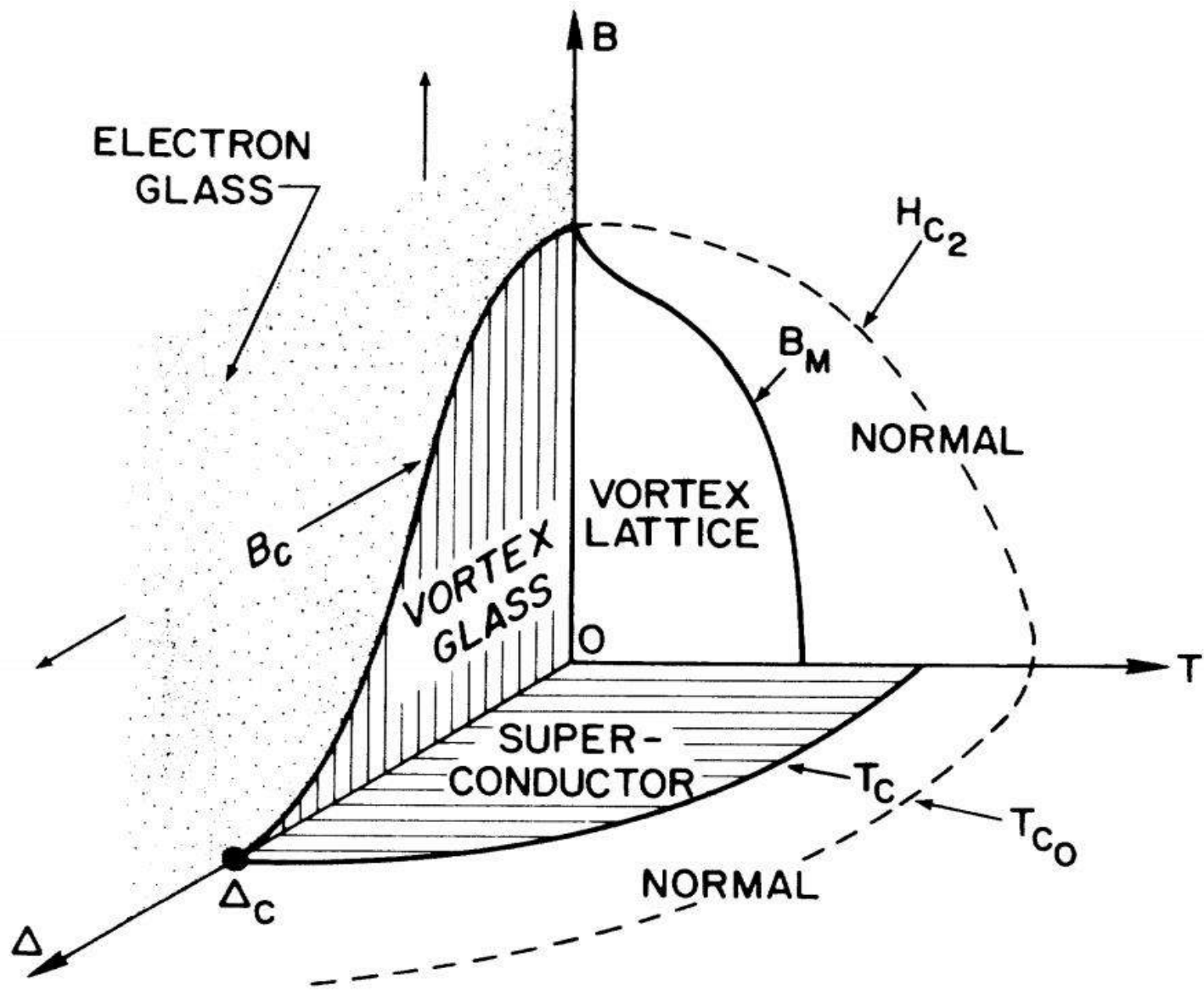

Figure 5

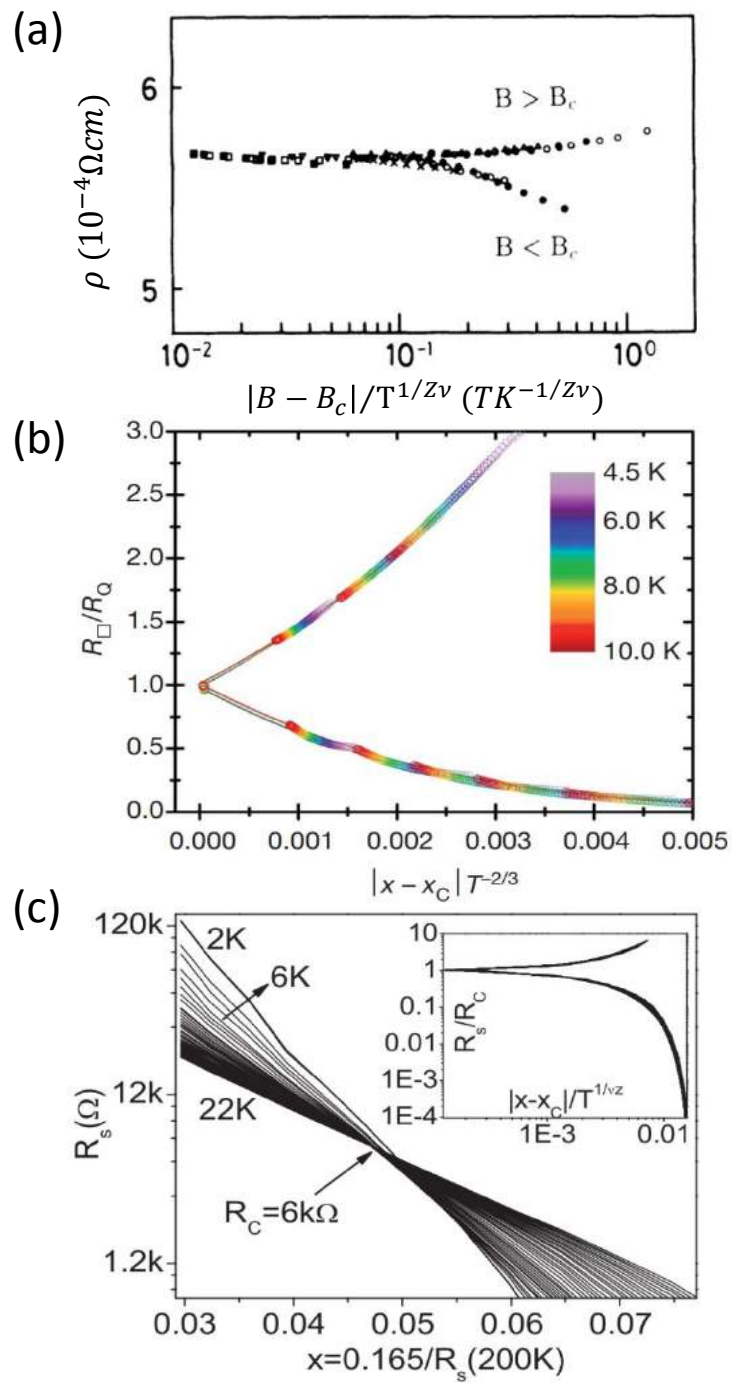

Figure 6

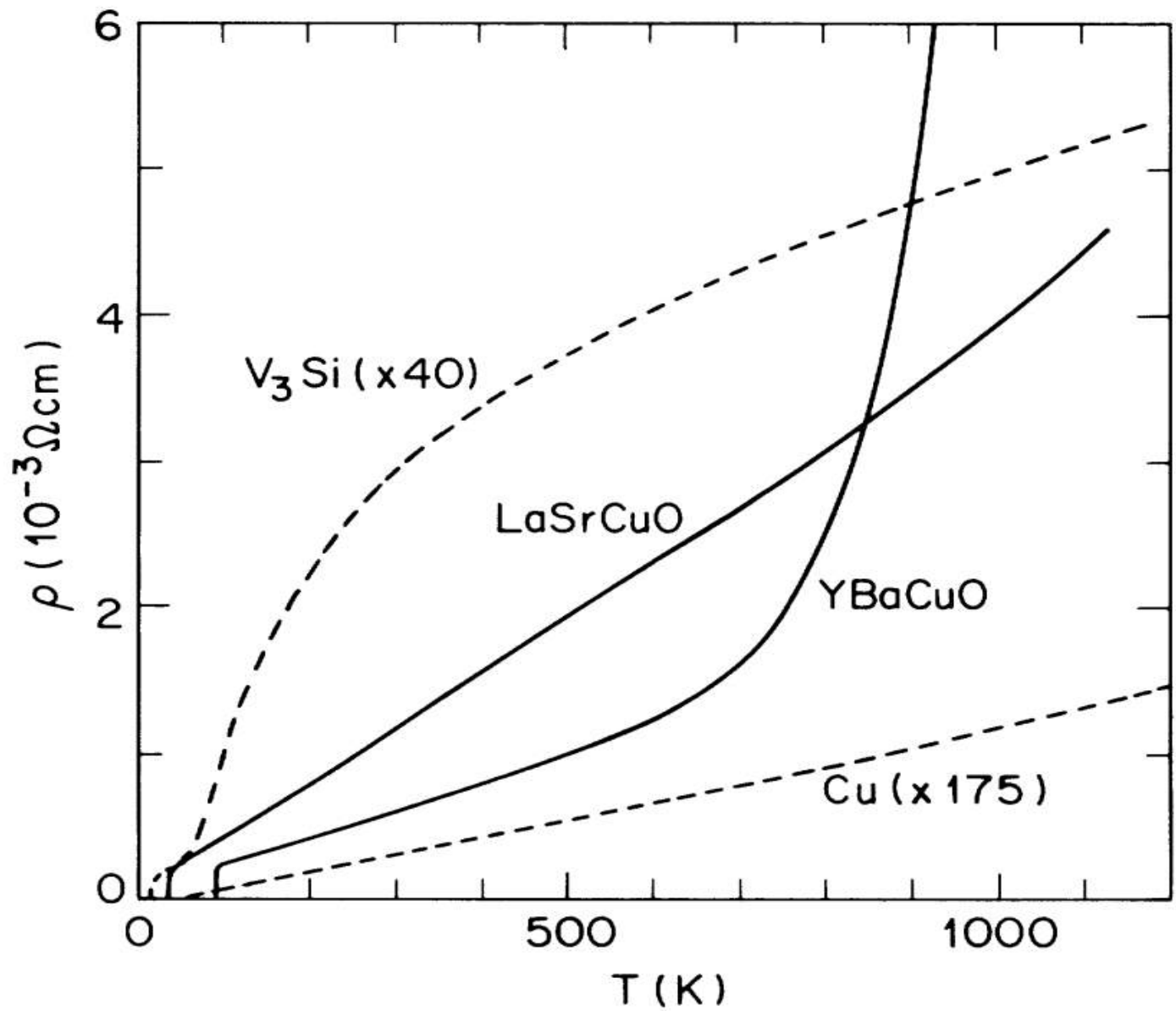

Figure 7

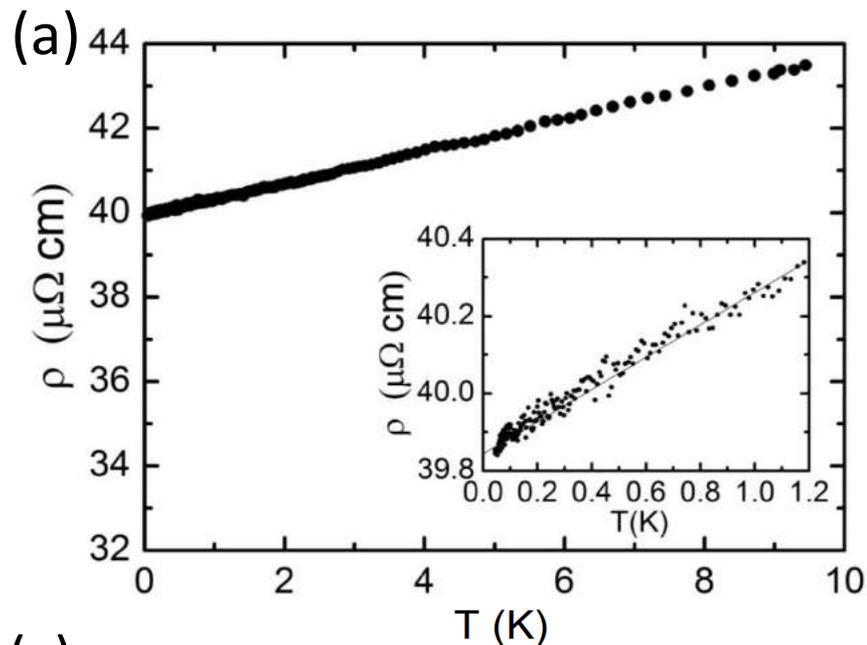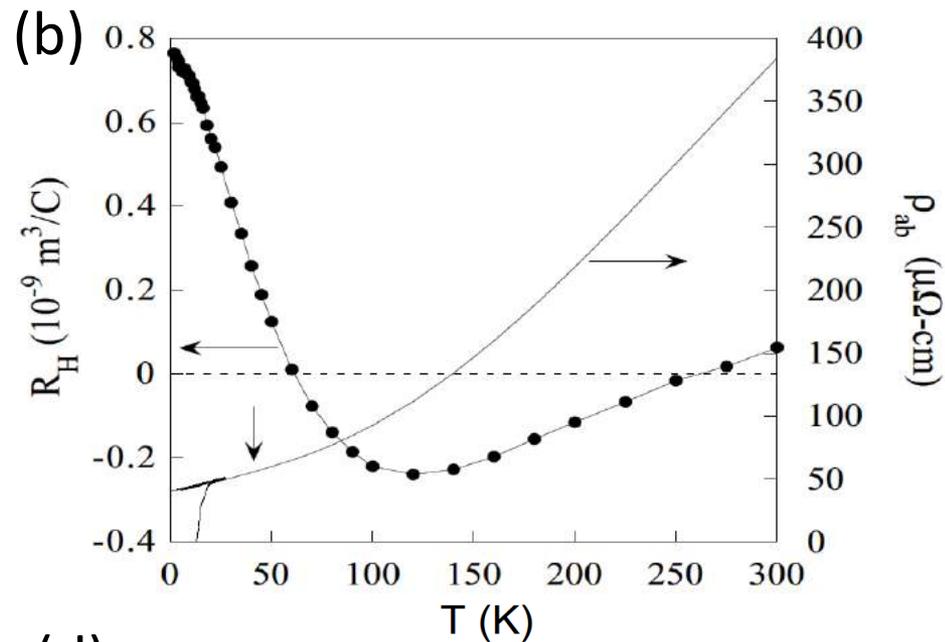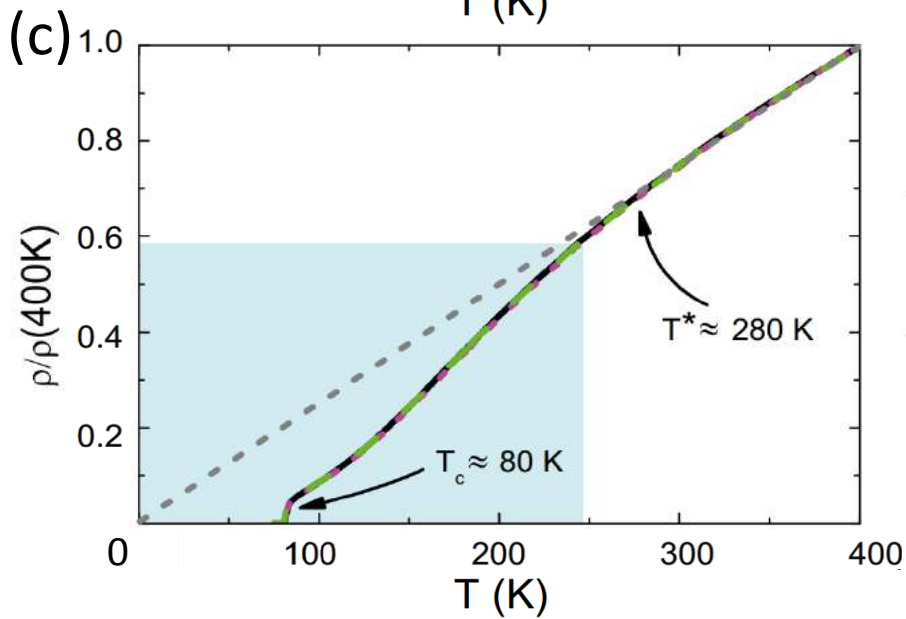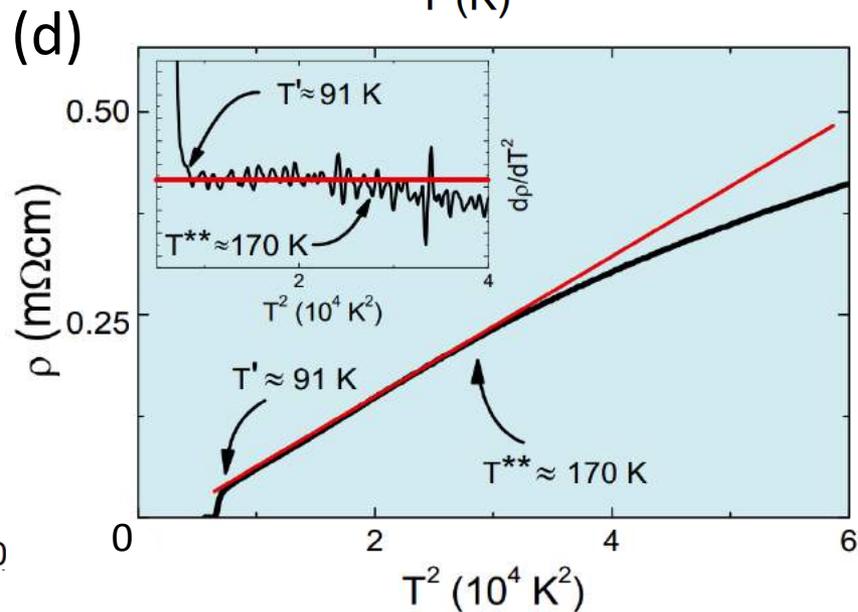

Figure 8

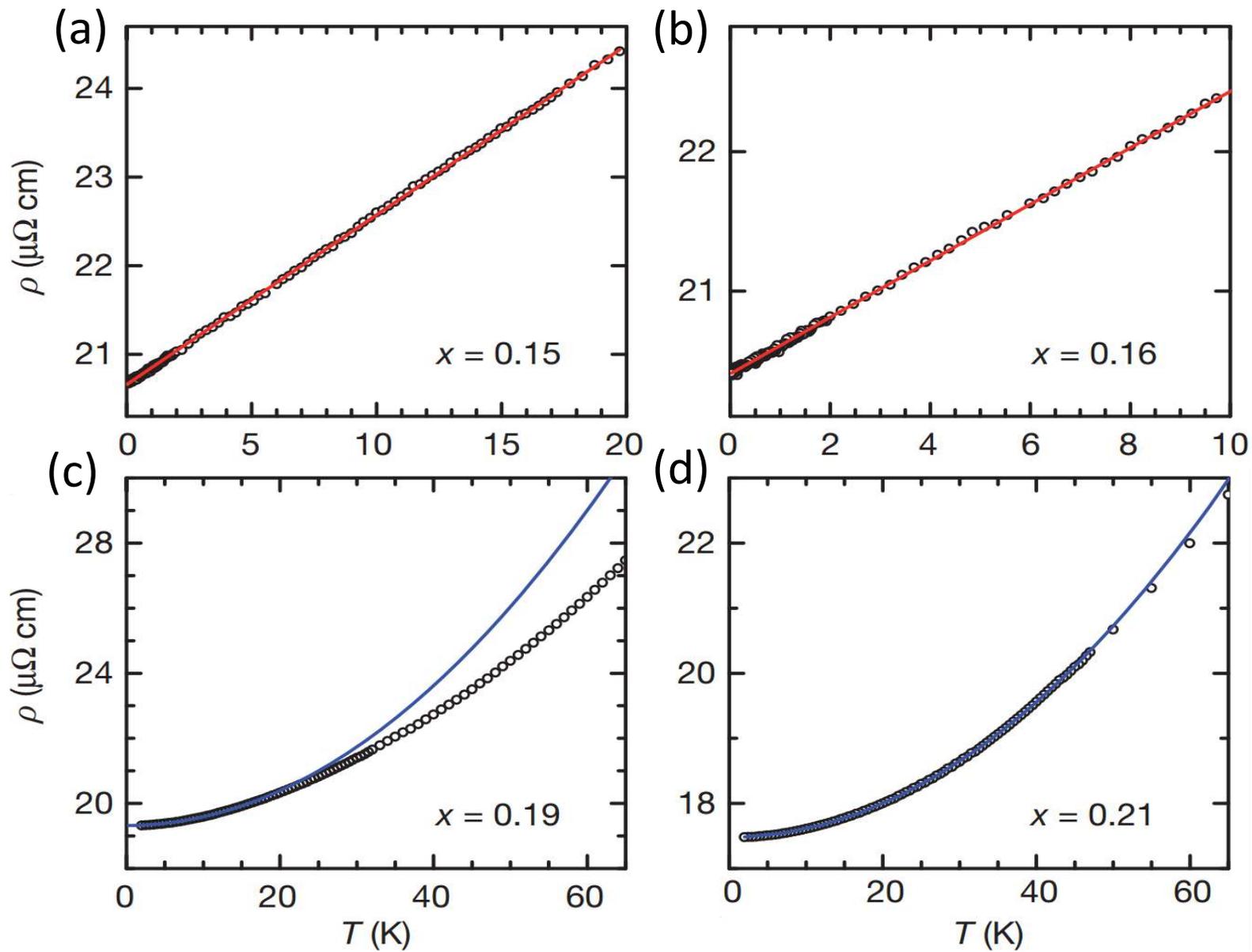

Figure 9

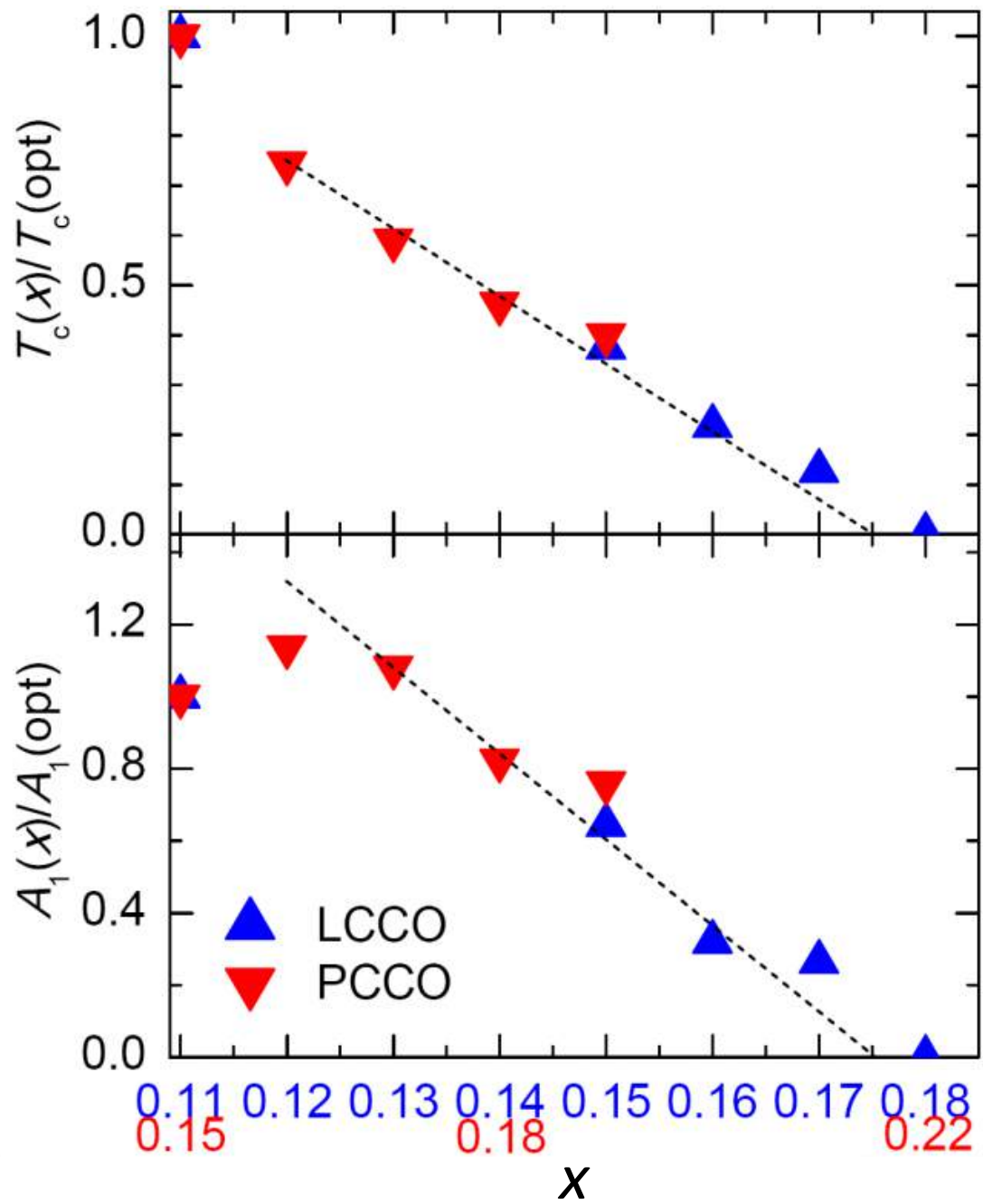

Figure 10

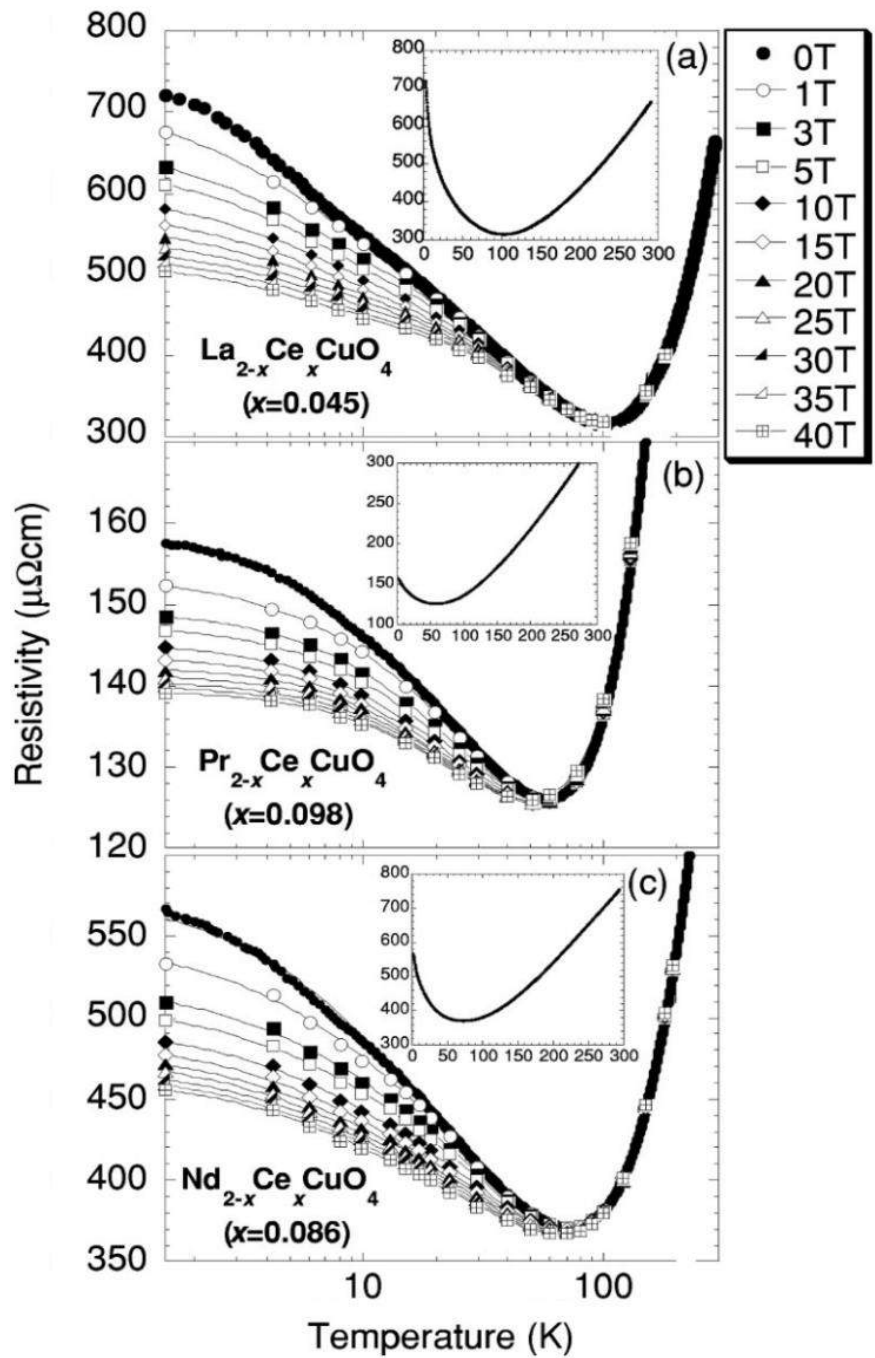

Figure 11

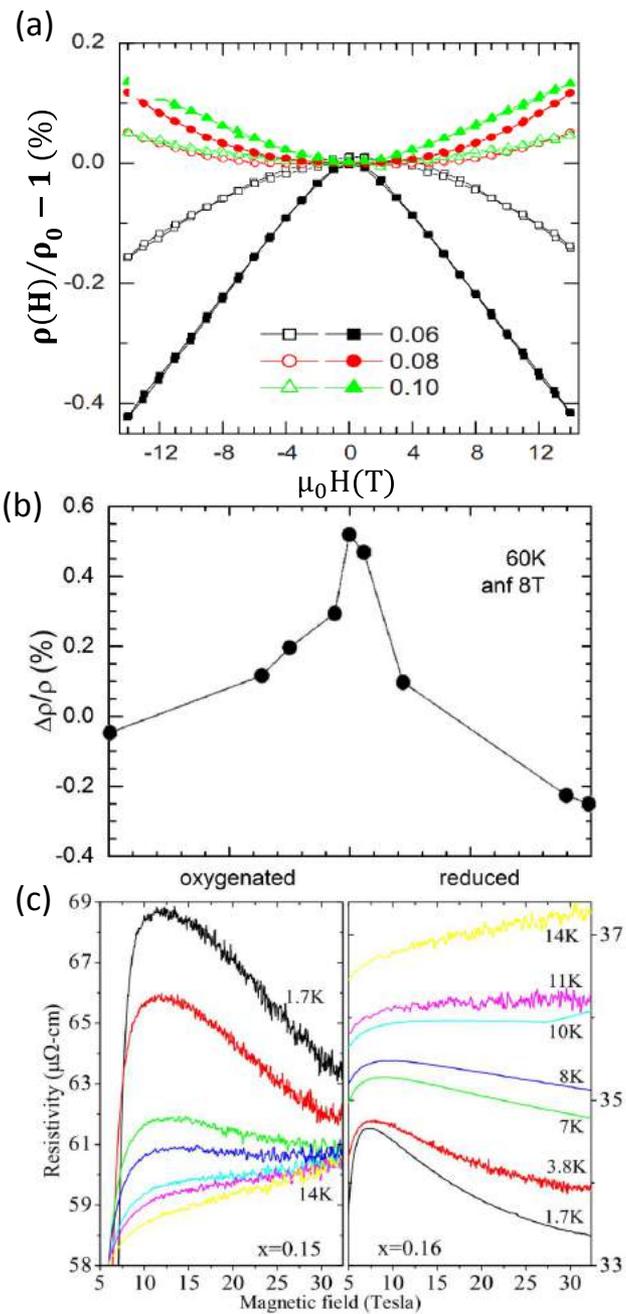

Figure 12

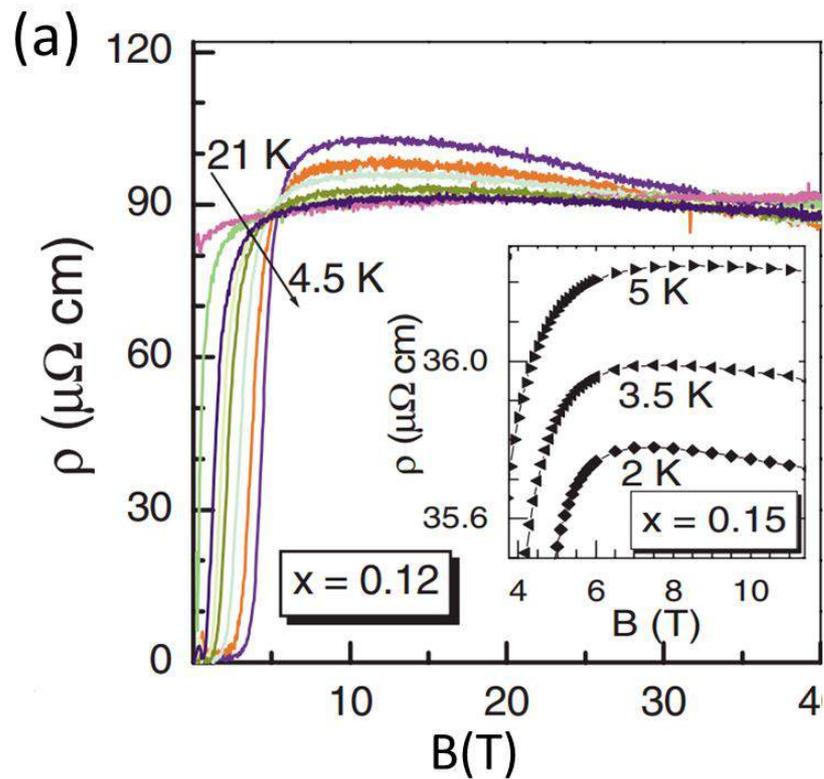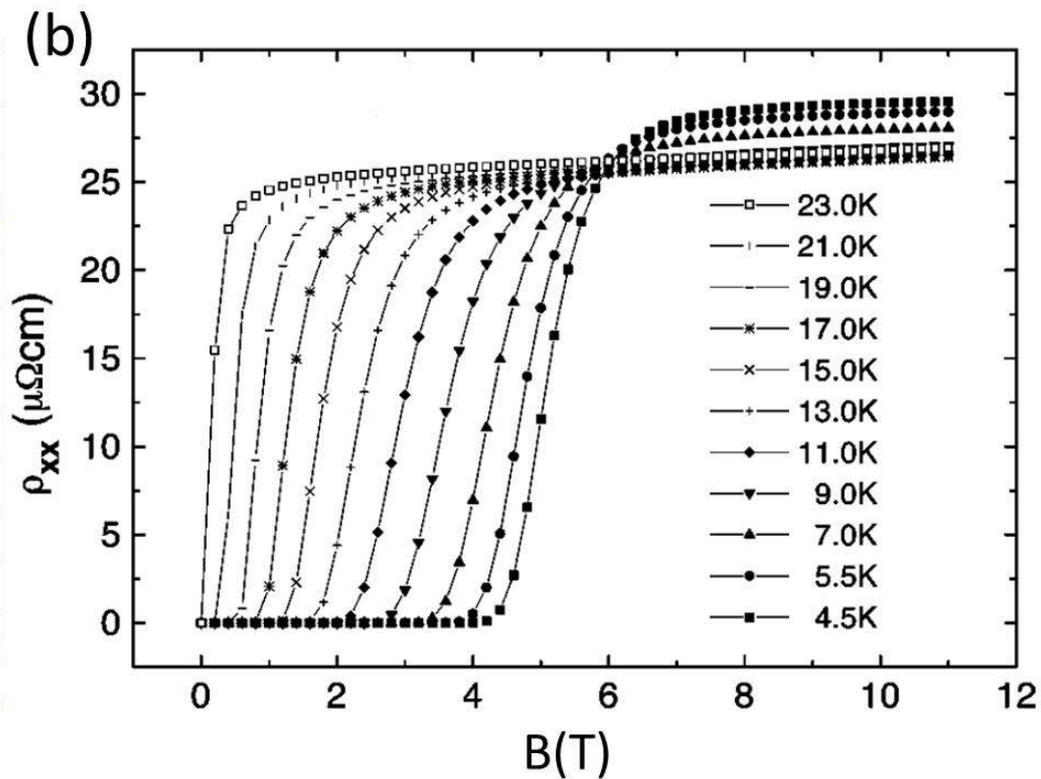

Figure 13

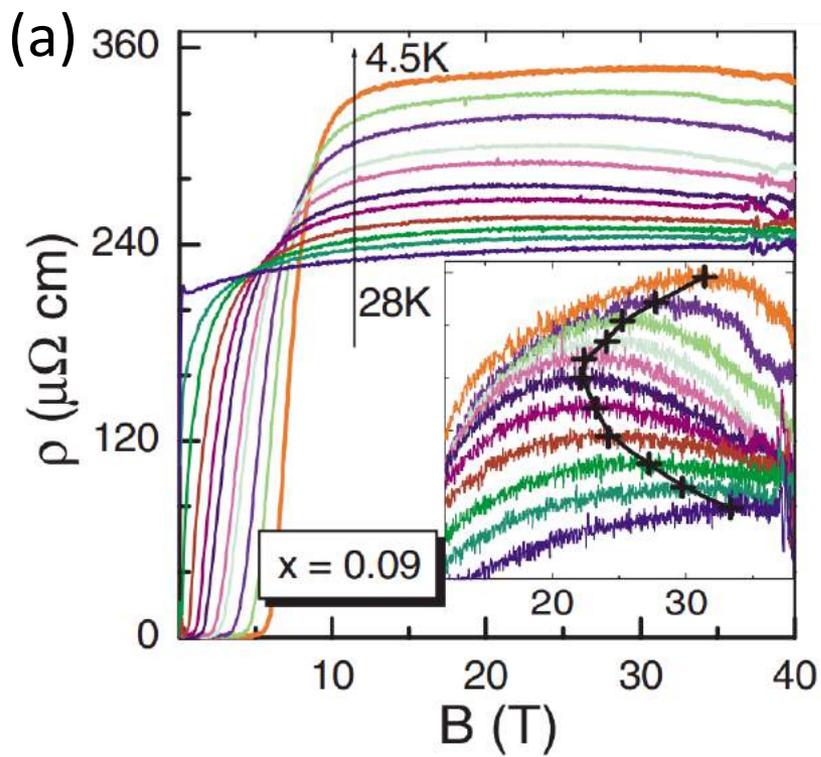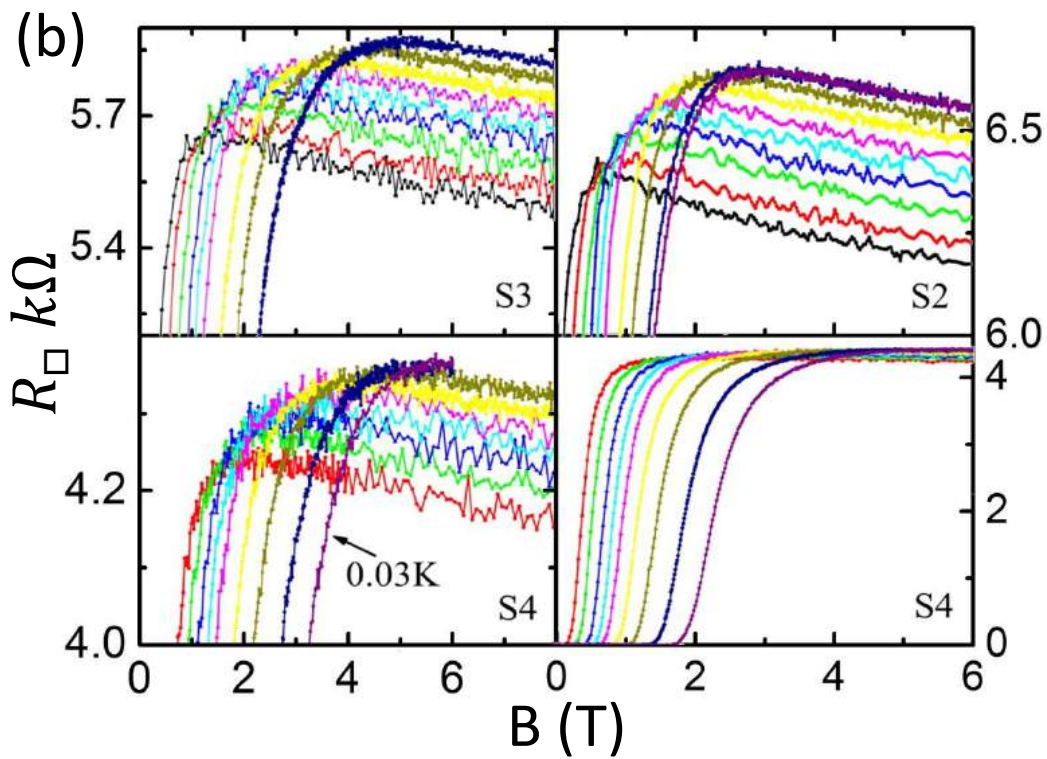

Figure 14

Figure 15

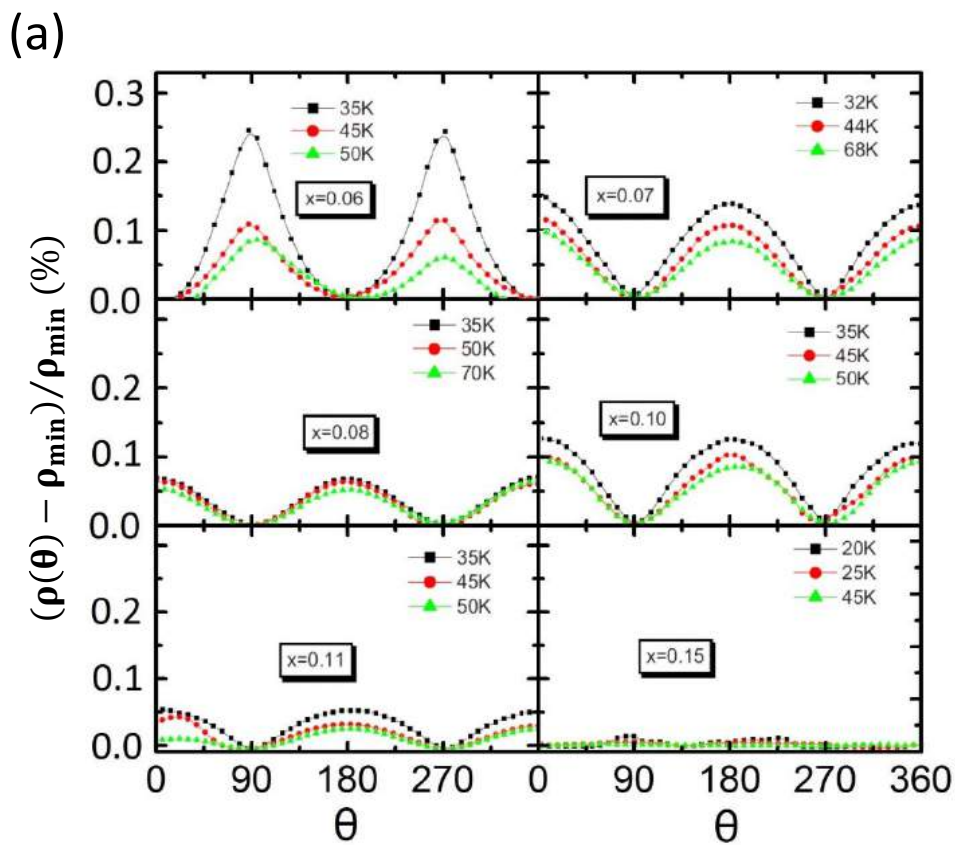 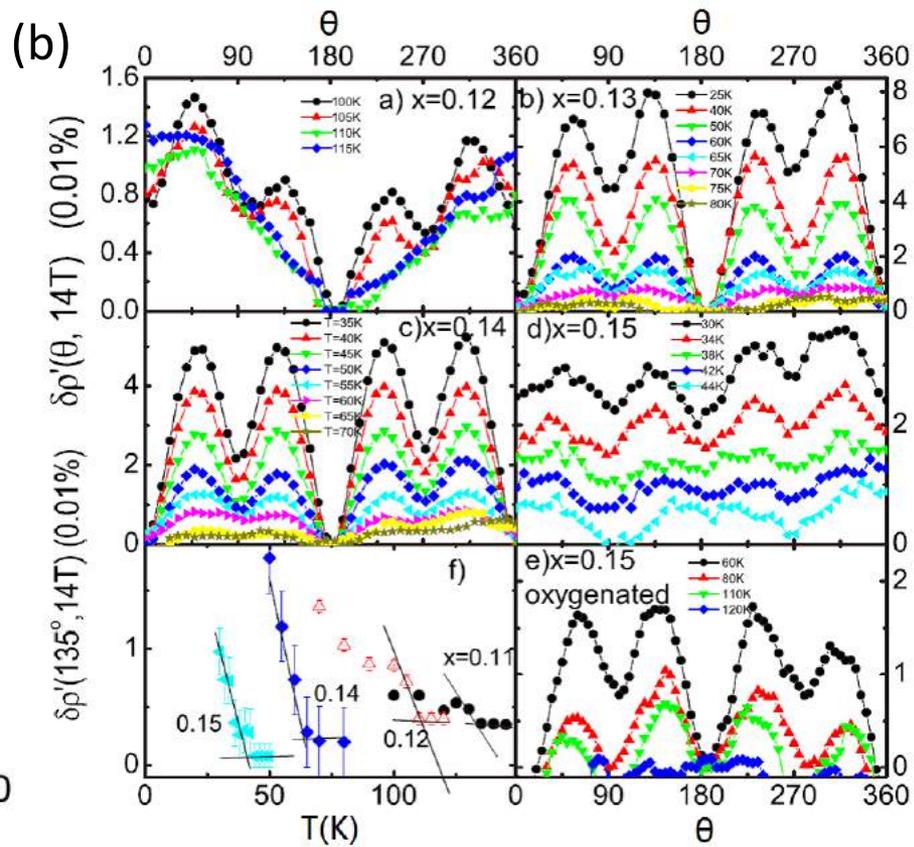

Figure 16

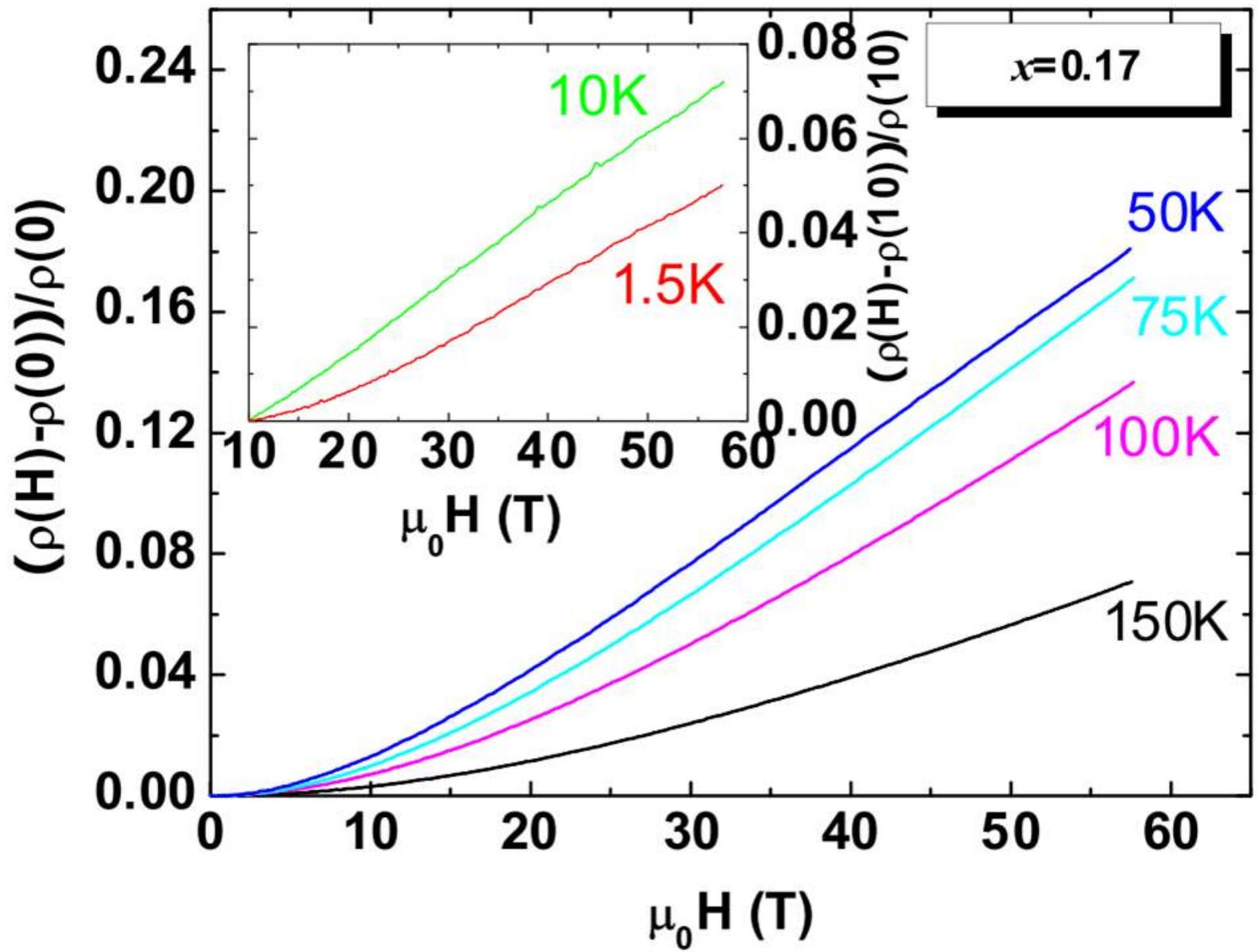

Figure 17

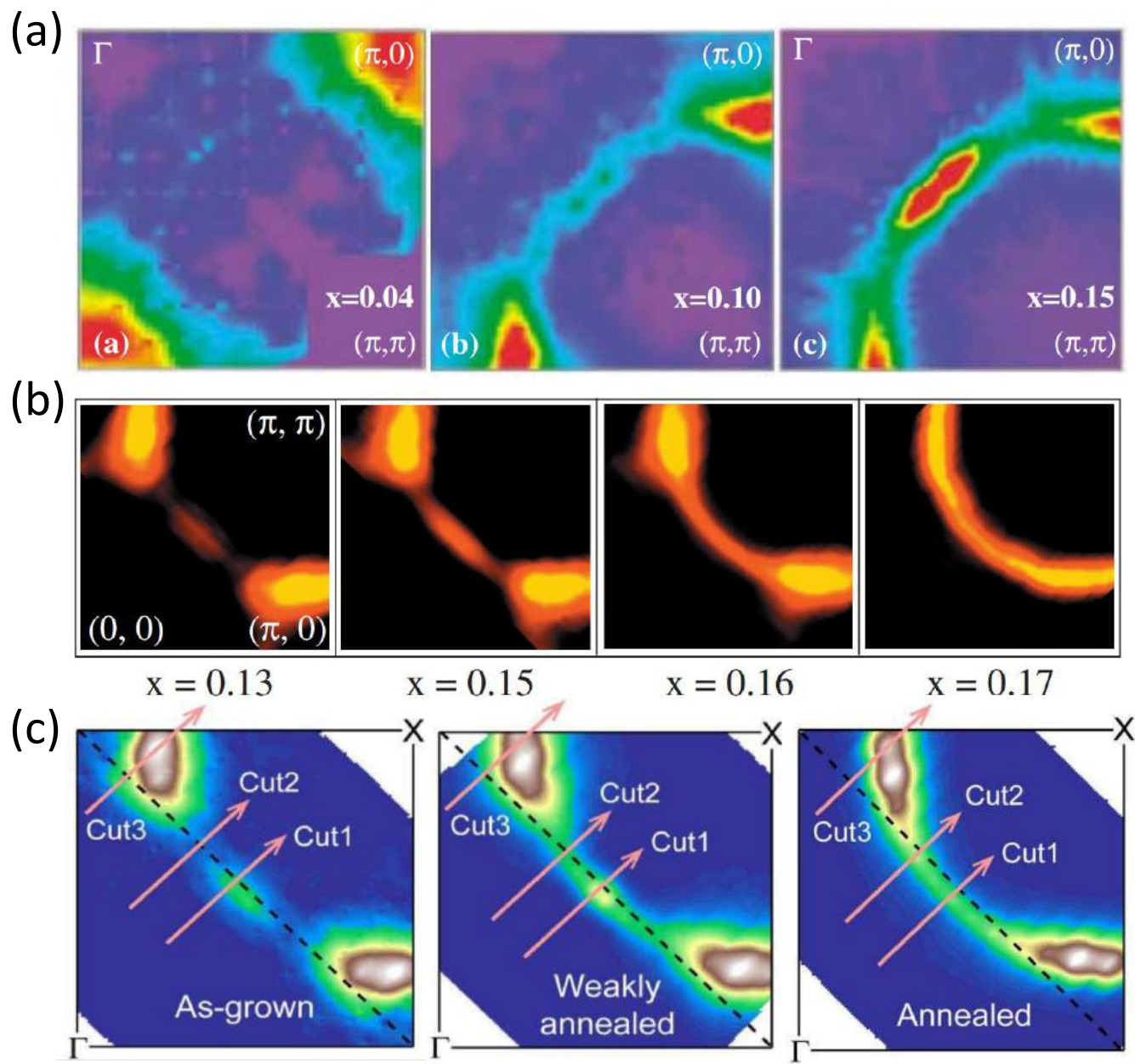

Figure 18

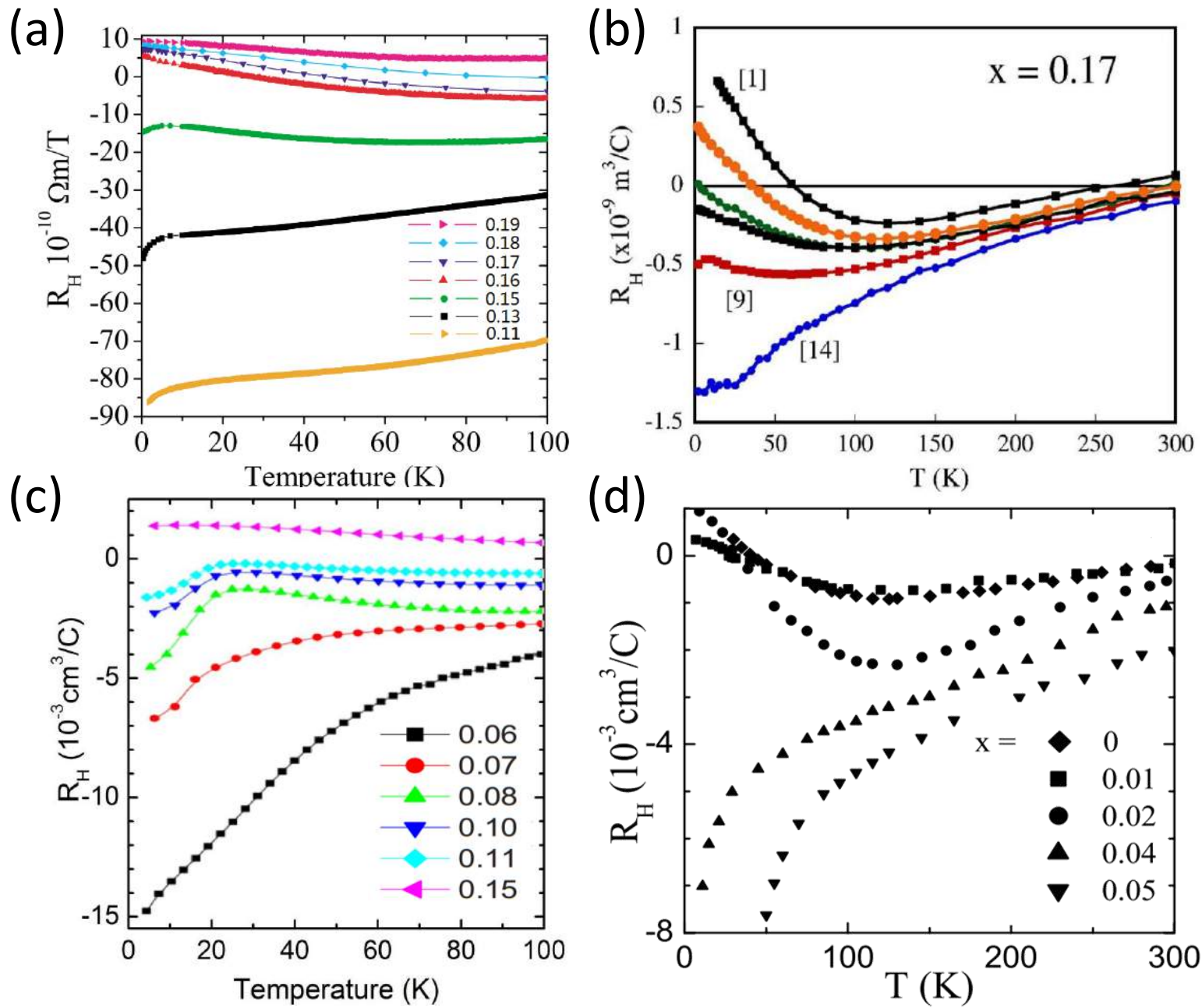

Figure 19

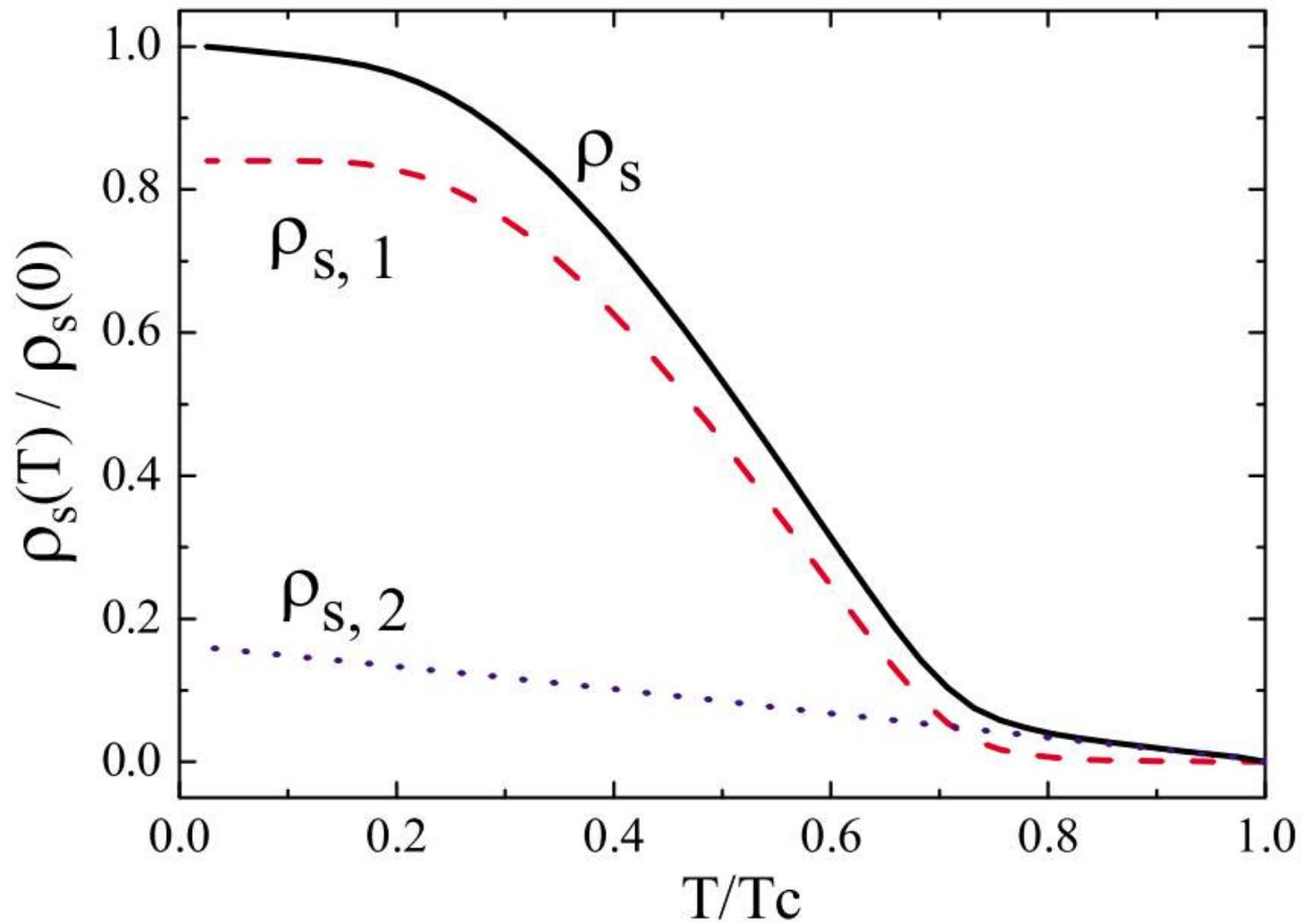

Figure 20

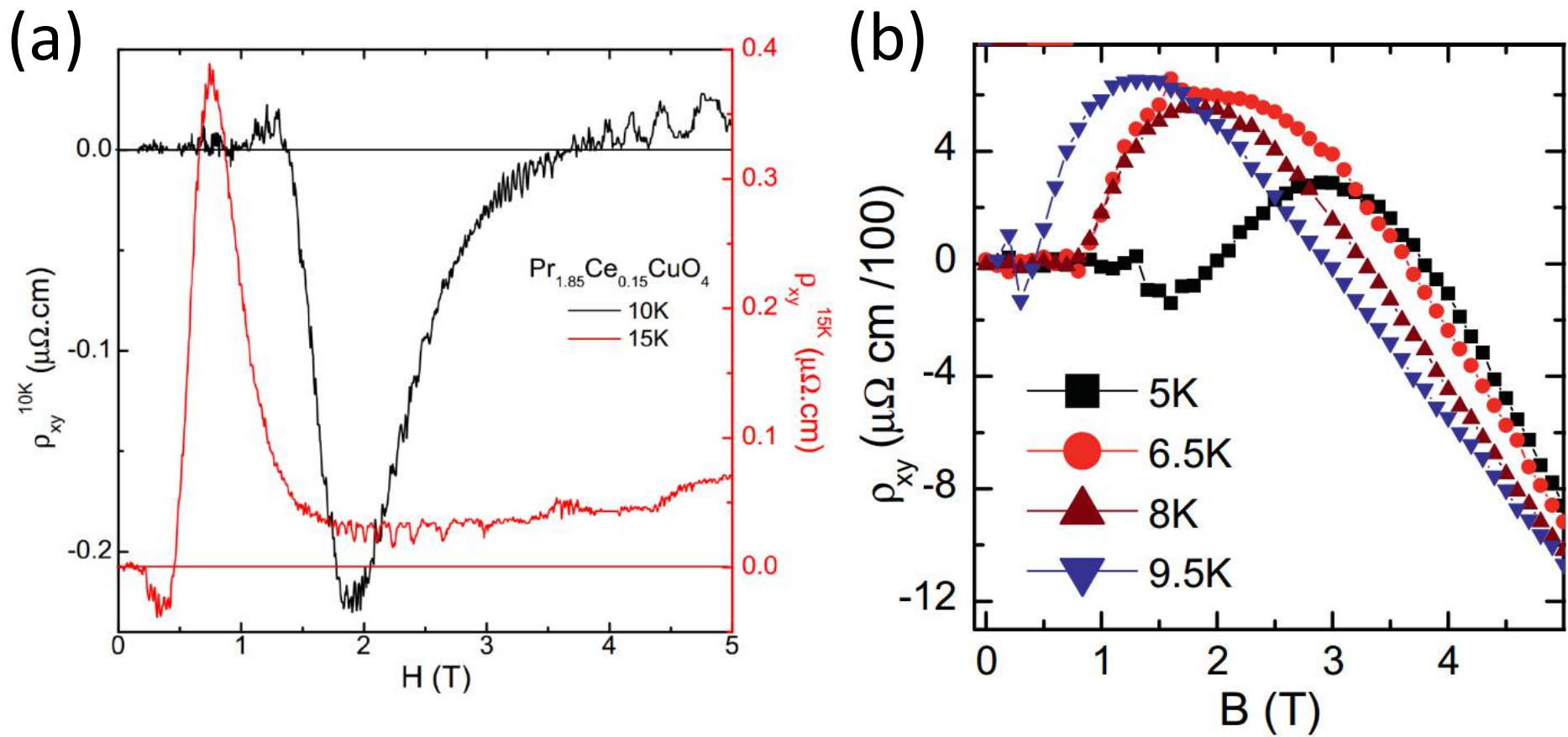

Figure 21

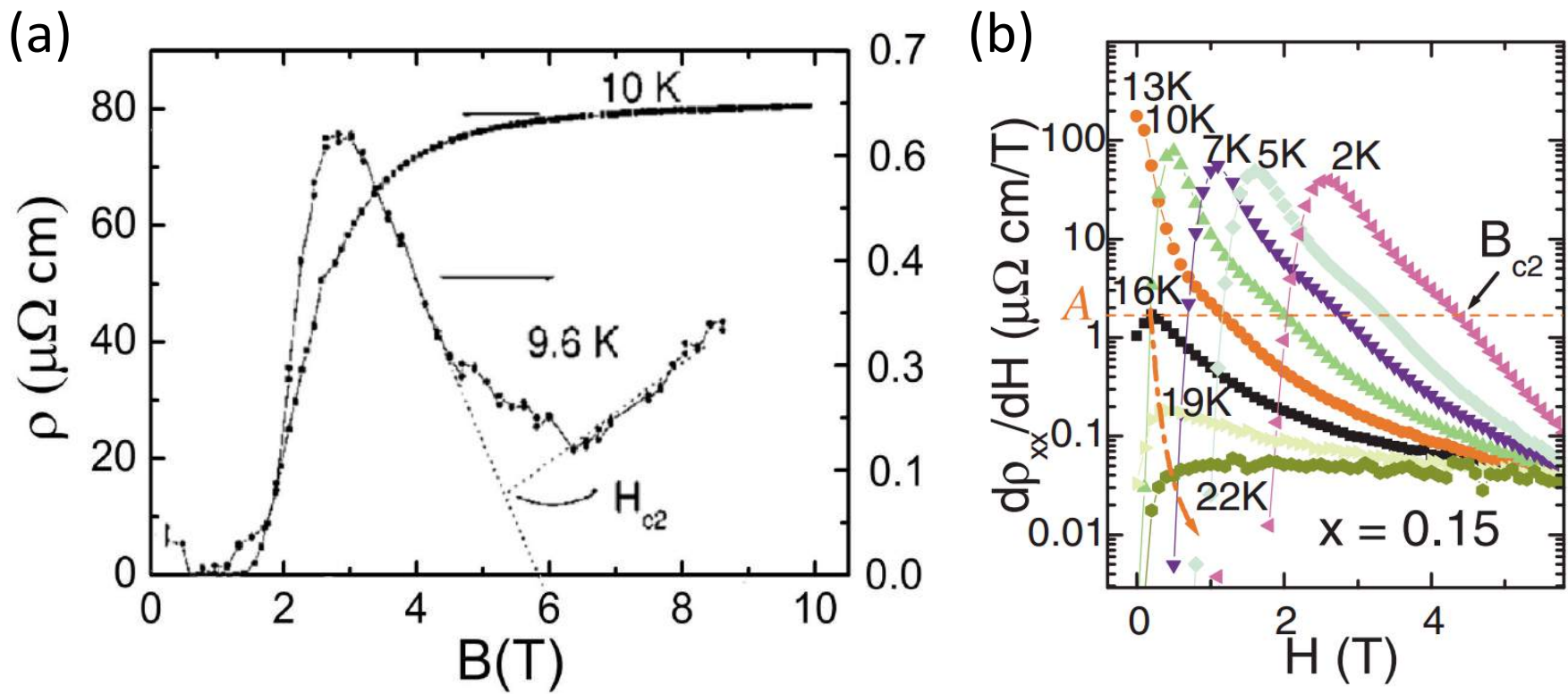

Figure 22

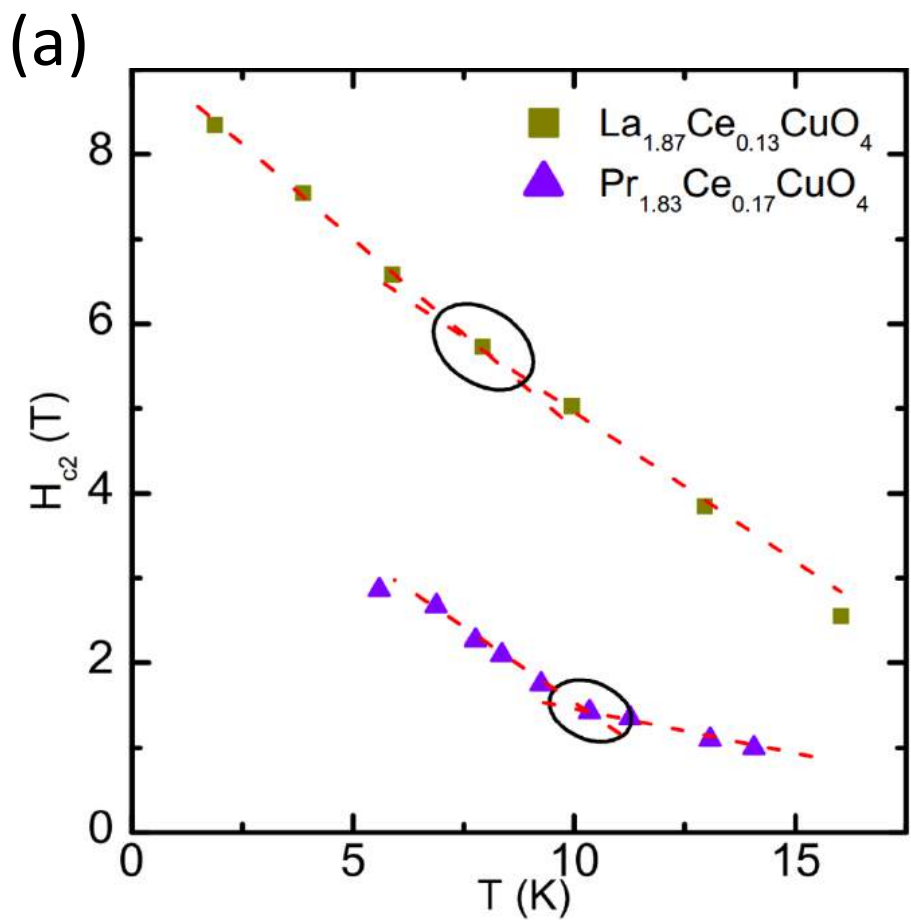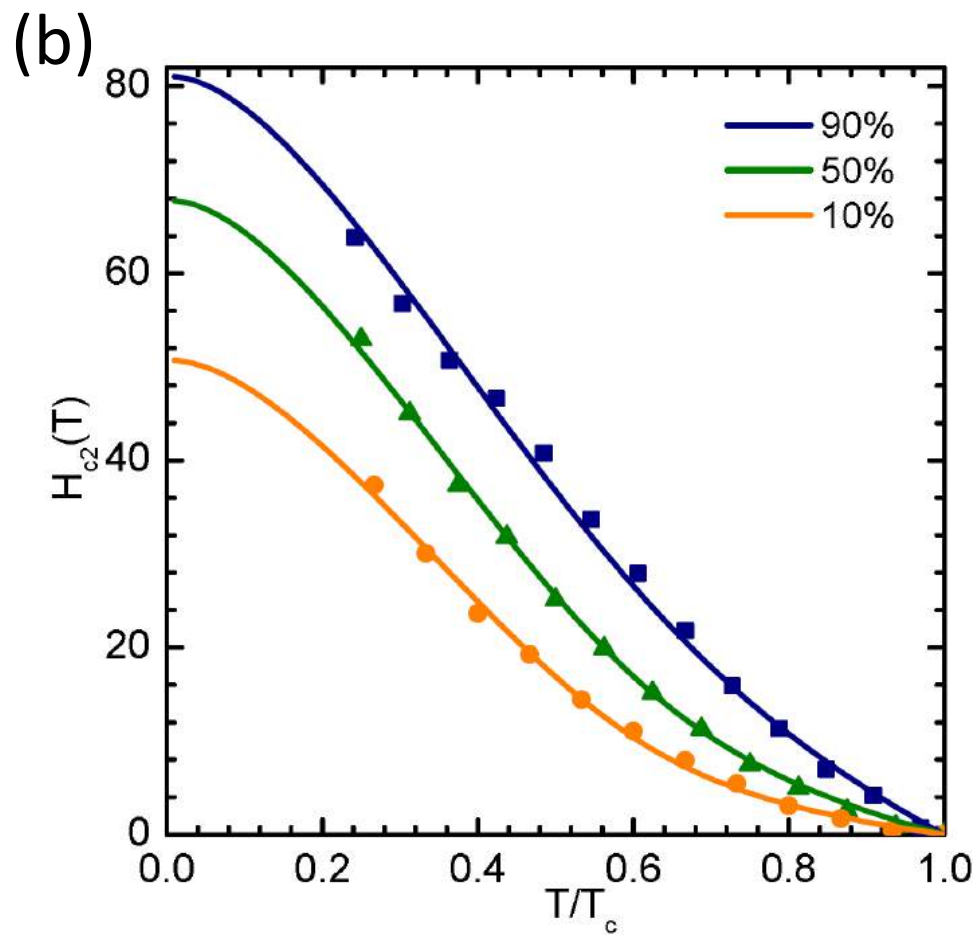

Figure 23

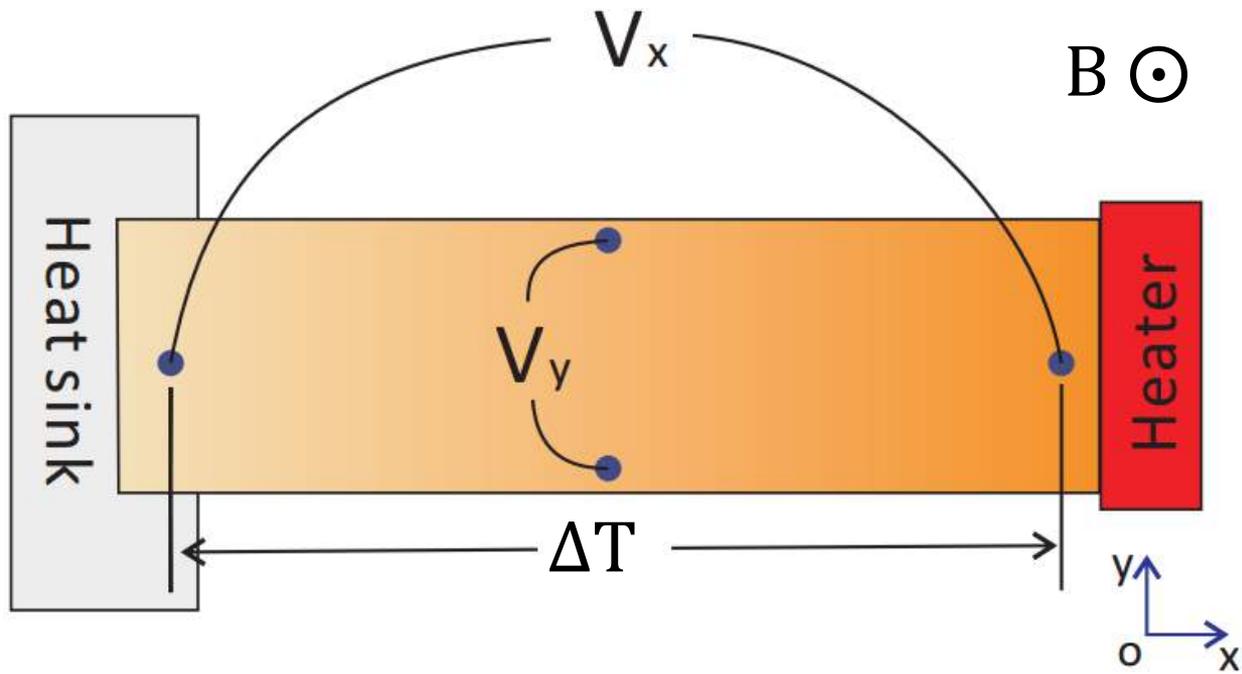

Figure 24

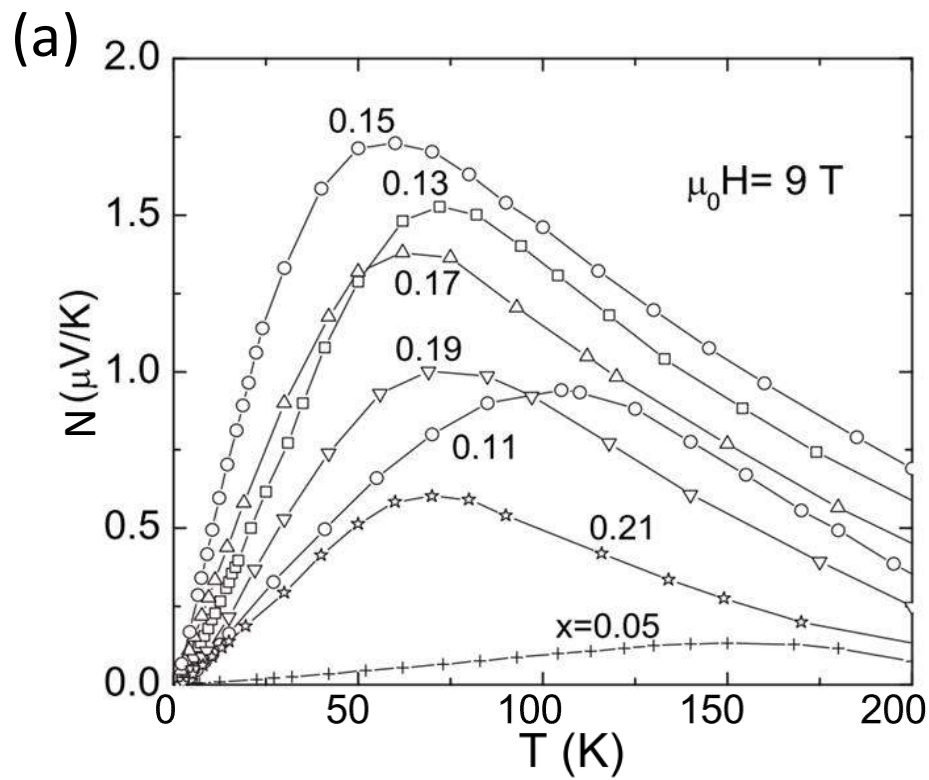 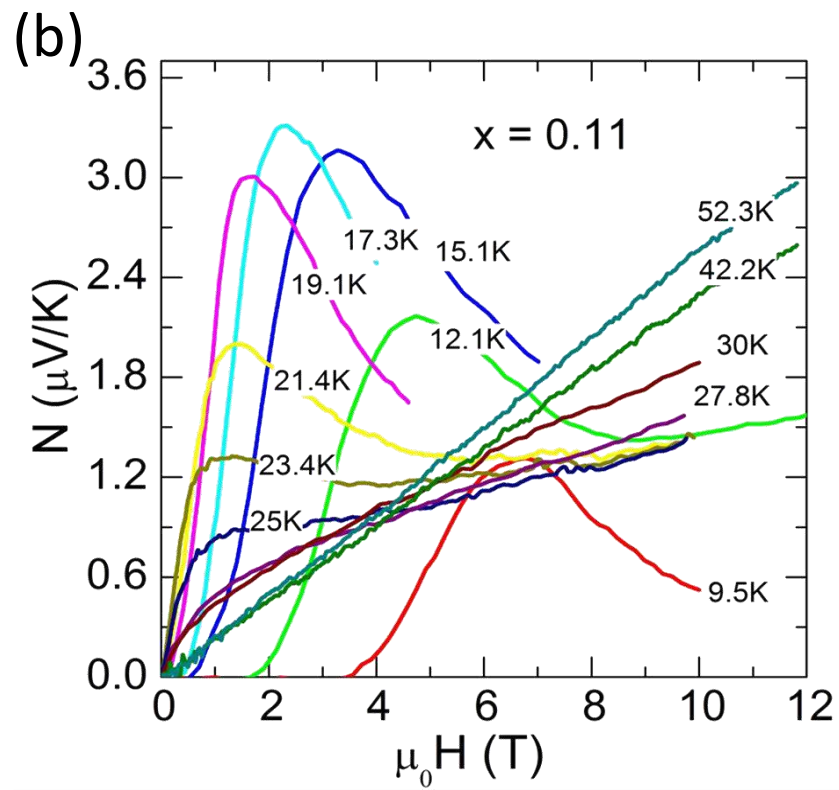

Figure 25

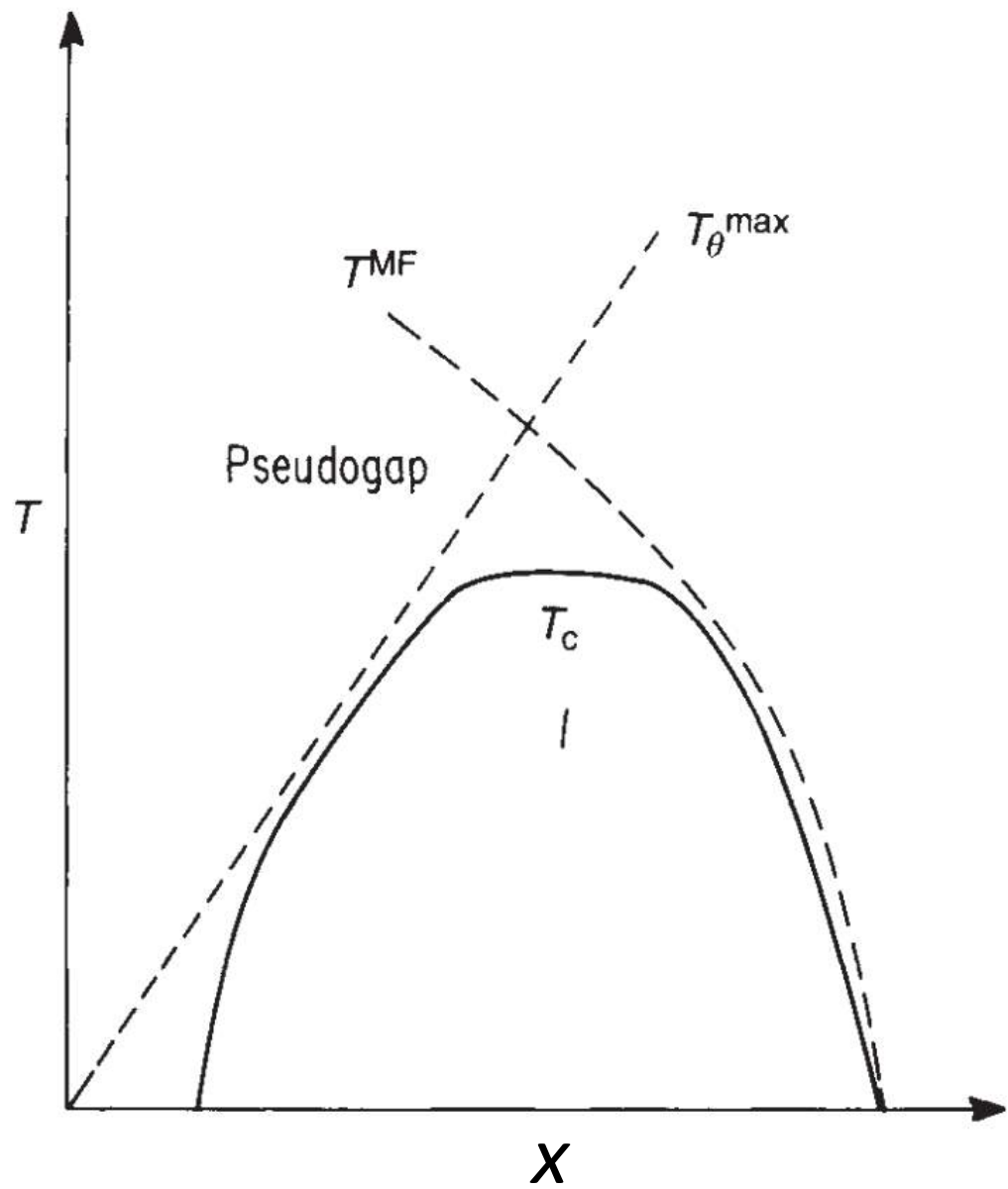

Figure 26

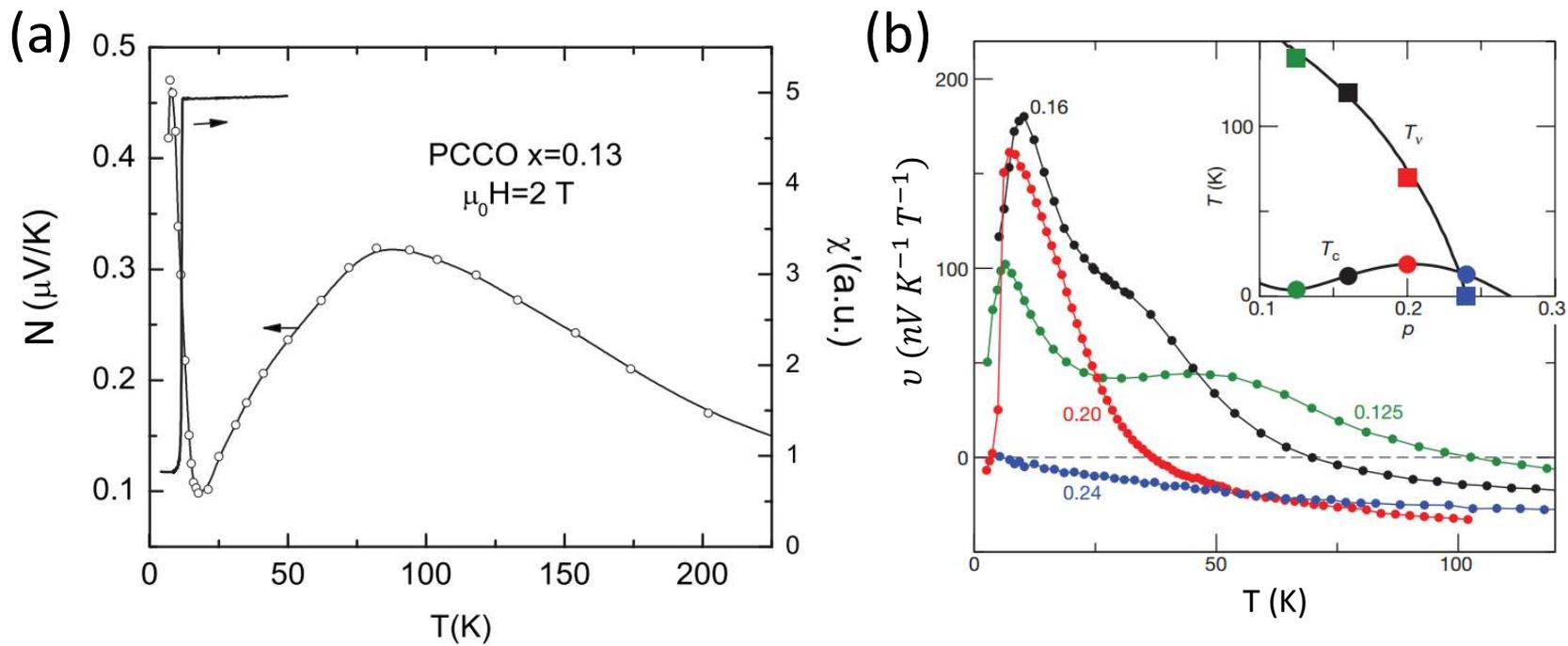

Figure 27

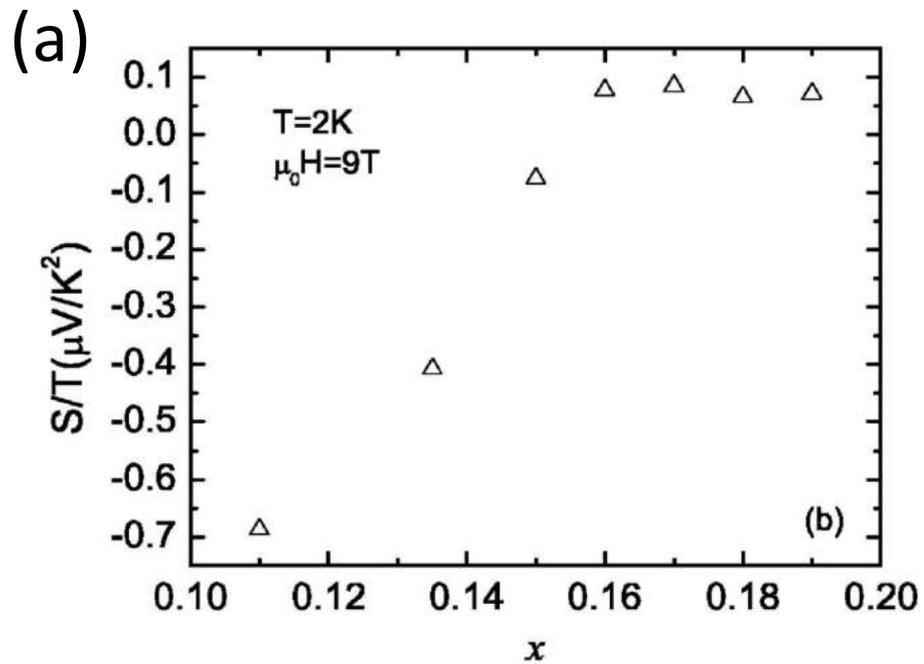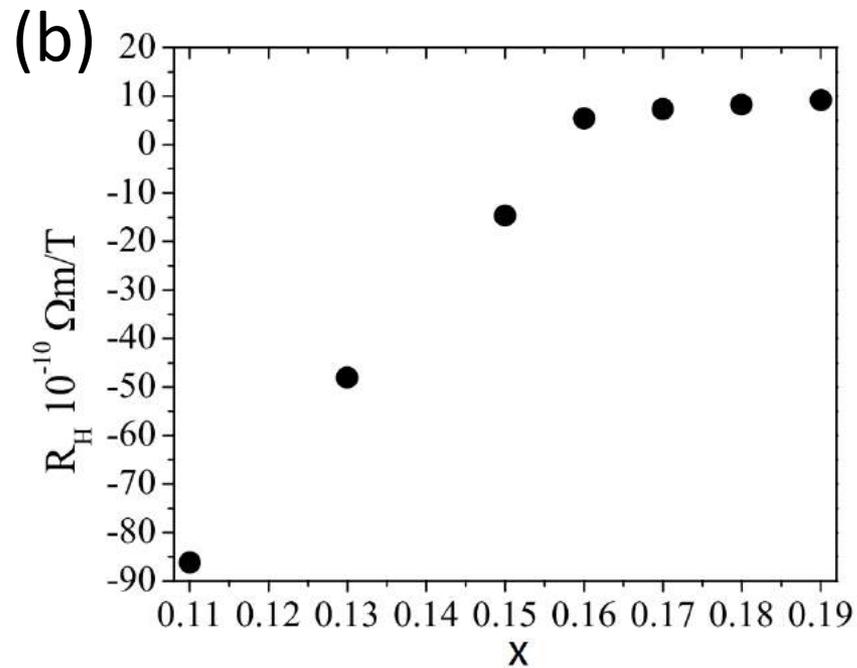

Figure 28

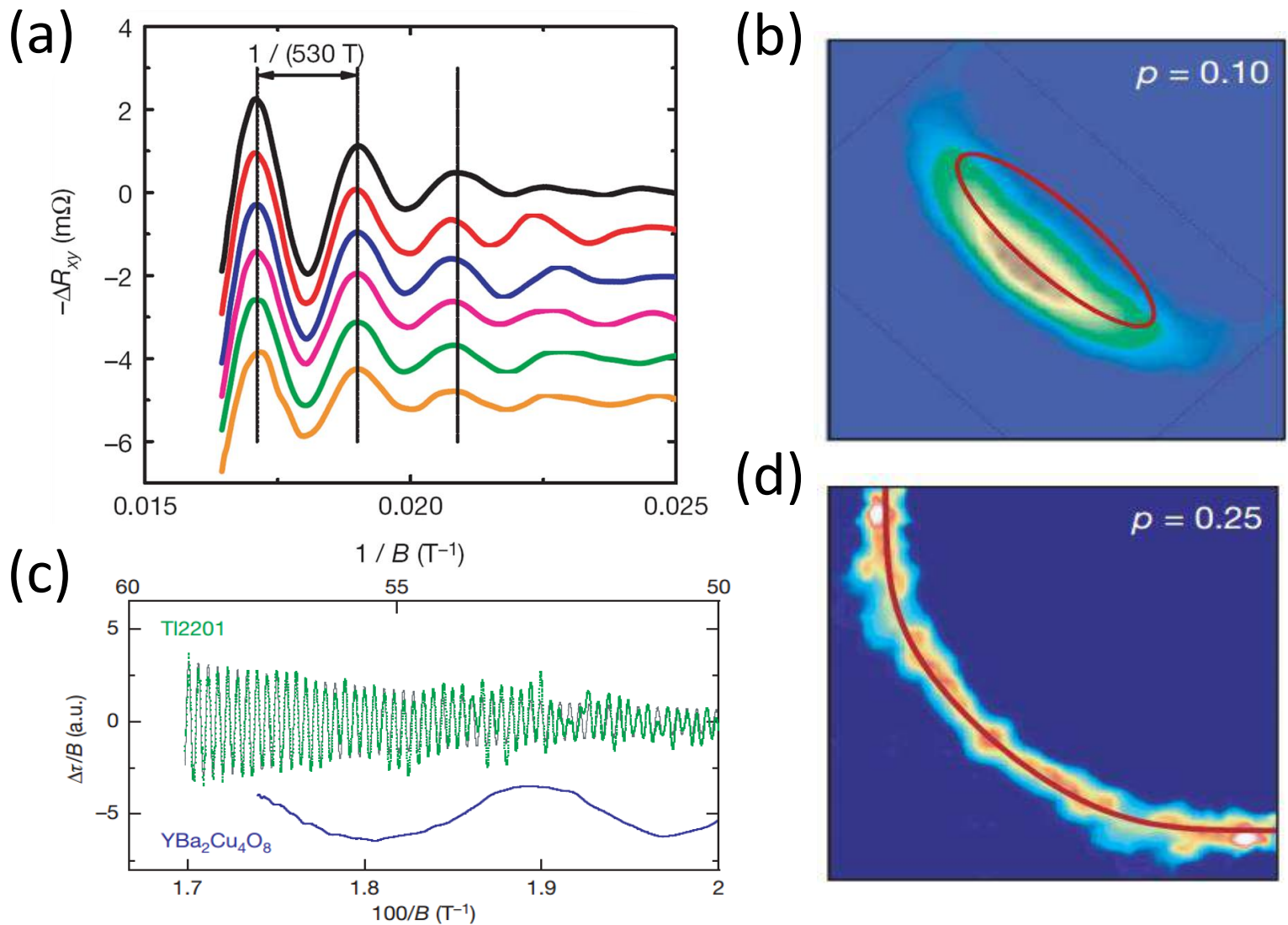

Figure 29

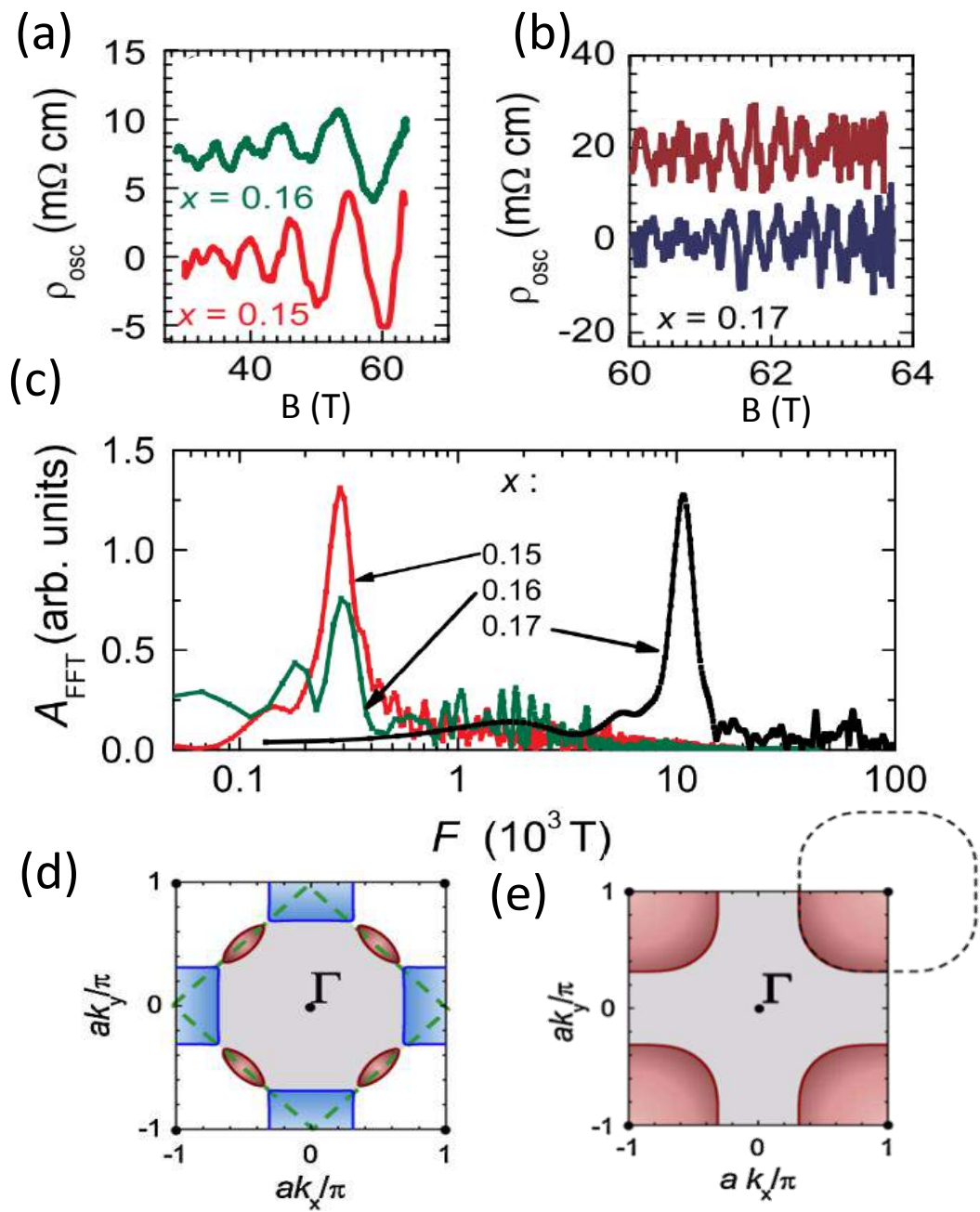

Figure 30

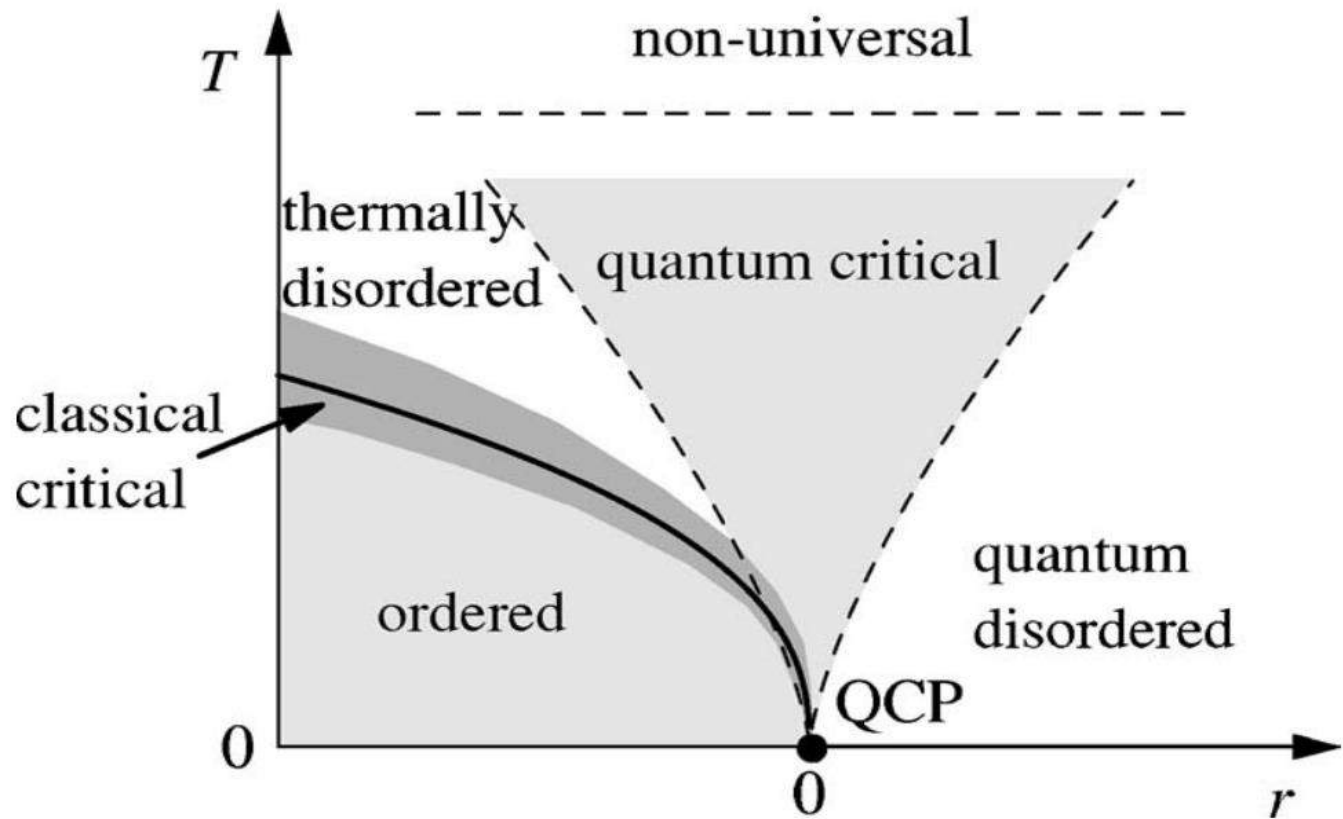

Figure 31

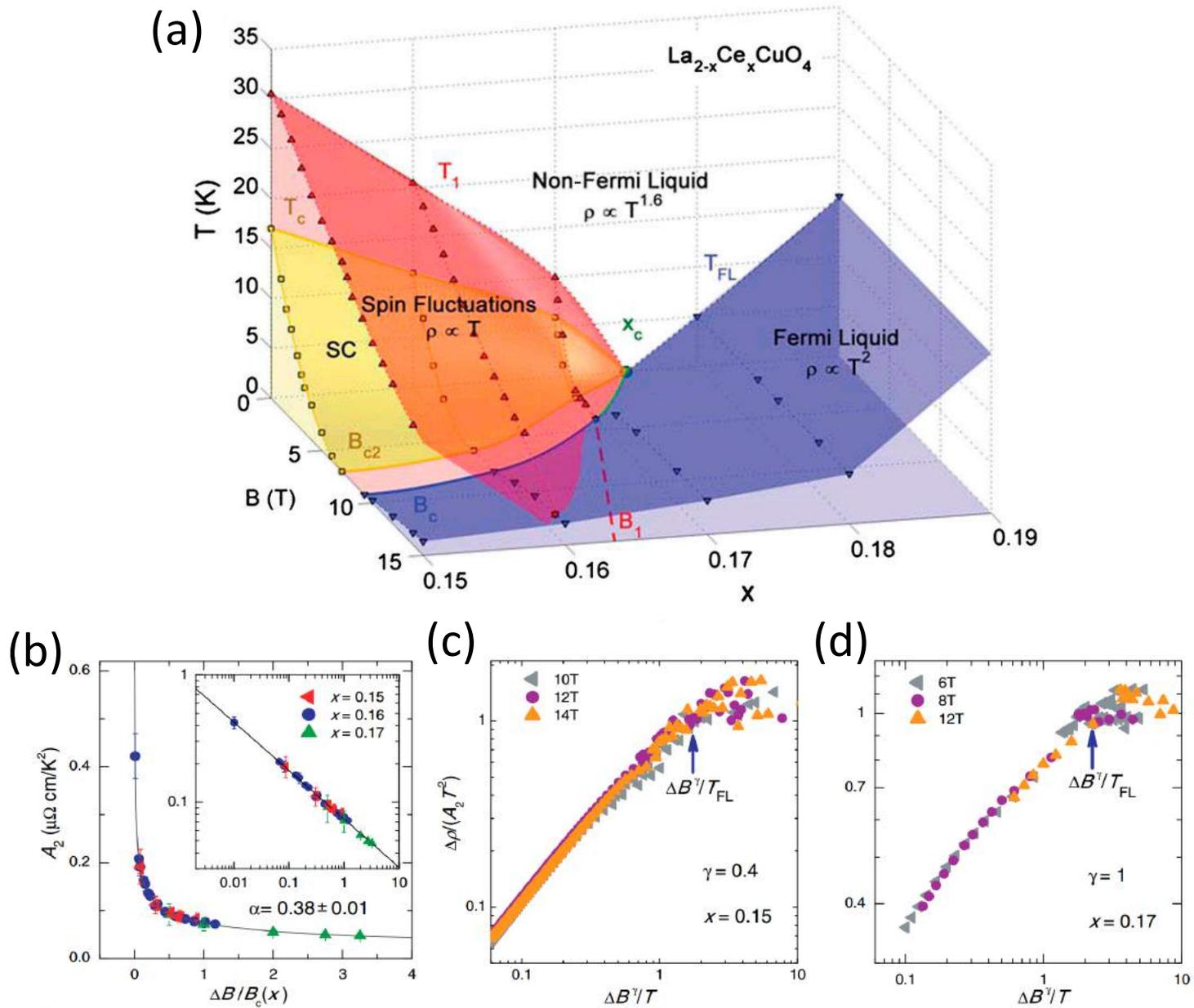

Figure 32

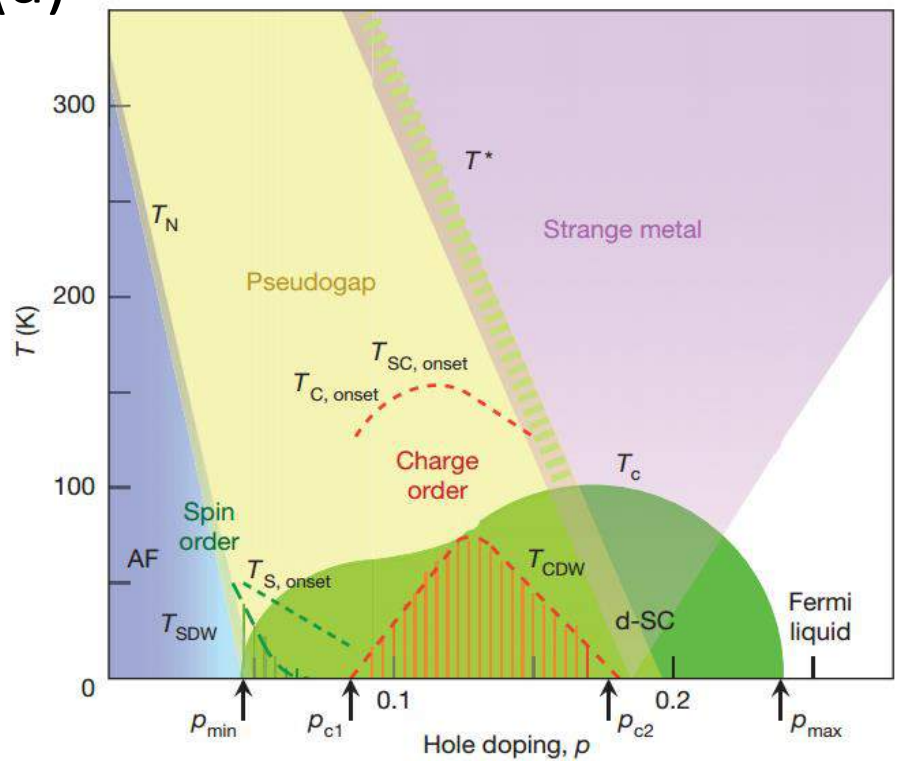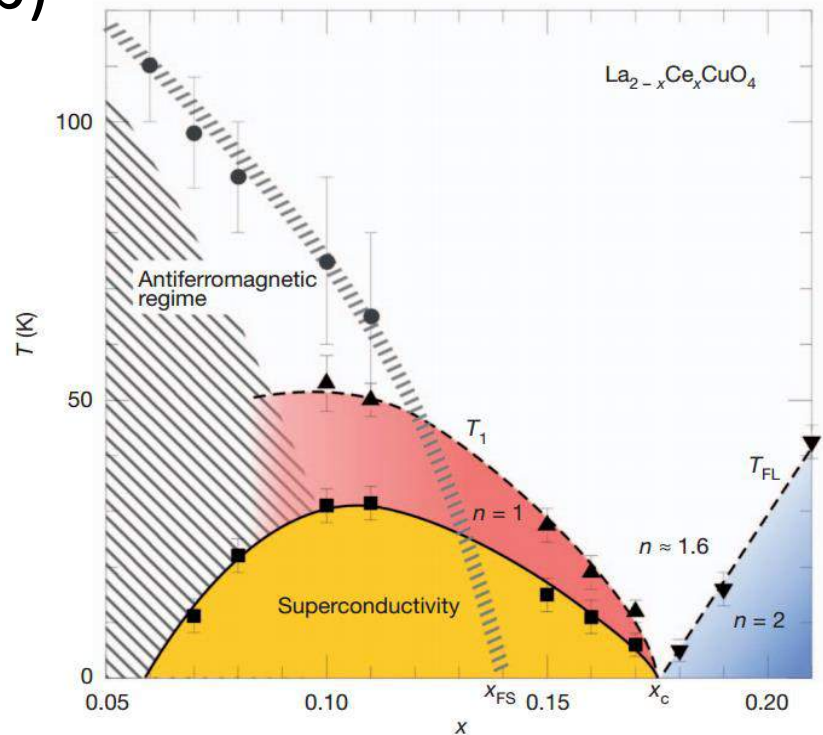

Figure 33

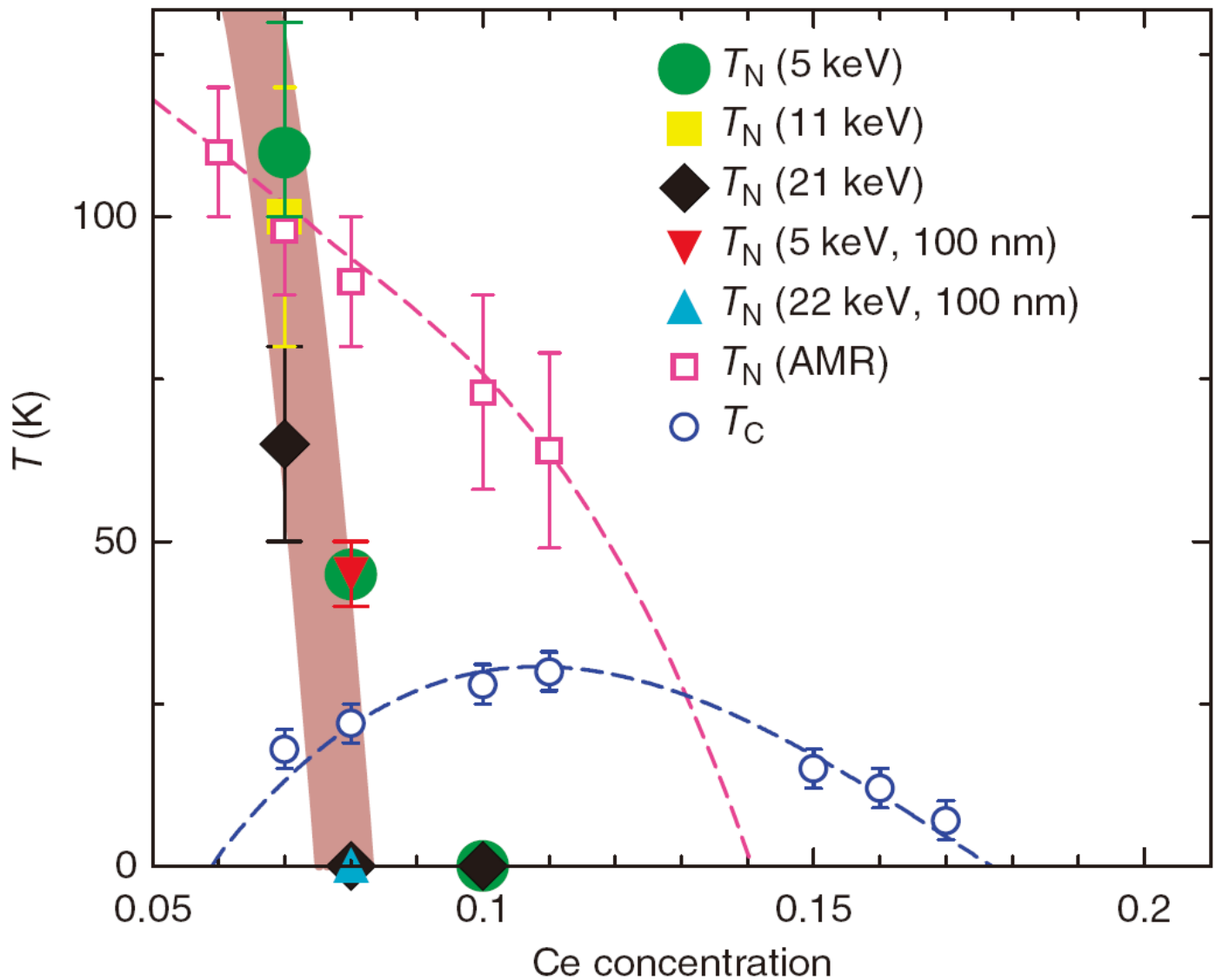

Figure 34

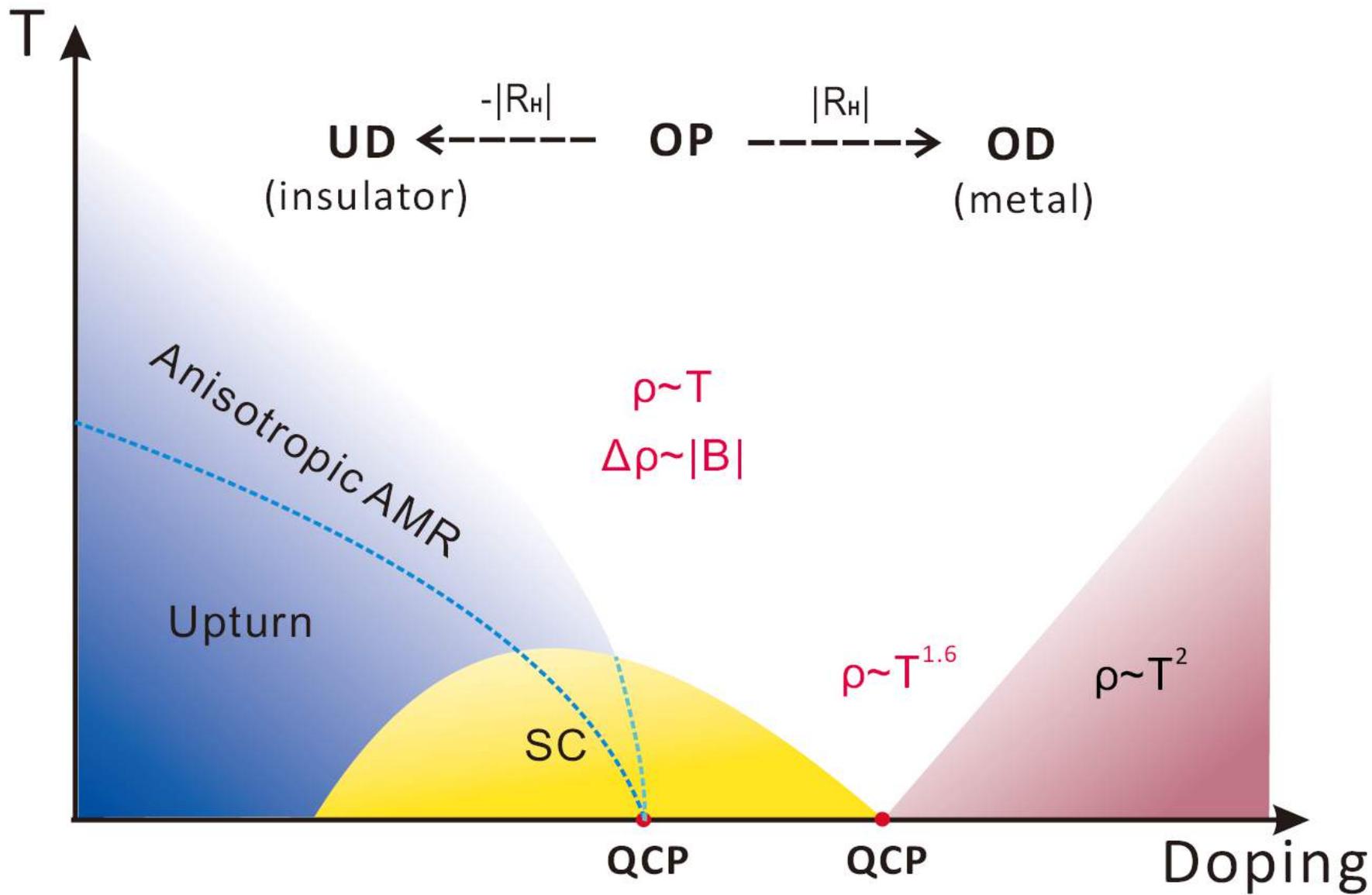

Figure 35